\documentclass[noshowpacs,prx,aps,superscriptaddress,floatfix,twocolumn,nofootinbib]{revtex4-2} 
\usepackage[utf8]{inputenc}
\usepackage{lineno,mathtools,url}
\usepackage{amsmath}
\usepackage{amssymb,mathrsfs}
\usepackage{graphicx}
\usepackage[colorlinks=true]{hyperref}
\usepackage[dvipsnames]{xcolor}
\usepackage{dsfont}
\usepackage{mathbbol} 
\usepackage[normalem]{ulem}

\definecolor{royalblue}{HTML}{4169E1}
\definecolor{WildStrawberry}{HTML}{FF43A4}
\definecolor{crimson}{HTML}{DC143C}

\DeclareMathAlphabet{\mathpzc}{OT1}{pzc}{m}{it}
\DeclareMathOperator{\diag}{diag}
\DeclareMathOperator{\tr}{tr}
\DeclareMathOperator{\arsinh}{arsinh}
\newcommand{\sss}{\scriptscriptstyle}

\newcommand{\NP}[1]{{\color{RedViolet} #1}}

\def\cals{\mathcal{S}}
\def\calst{\mathcal{S}_{\mathscr{T}}}
\def\calsr{\mathcal{S}_{\mathscr{R}}}
\DeclareMathOperator{\arcosh}{arcosh}

\newcommand{\nota}{\phi}
\newcommand{\rmres}{\textrm{res}}
\begin{document}

\title{Bipartite and tripartite entanglement in a Bose-Einstein acoustic black hole}

\author{Mathieu~Isoard}
\affiliation{Université Paris-Saclay, CNRS, LPTMS, 91405, Orsay, France}
\affiliation{Physikalisches Institut, Albert-Ludwigs-Universit\"at Freiburg, D-79104 Freiburg, Germany.}

\author{Nadia~Milazzo}
\affiliation{Université Paris-Saclay, CNRS, LPTMS, 91405, Orsay, France}
\affiliation{Institut f\"ur theoretische Physik, Universit\"at T\"{u}bingen, 72076 T\"ubingen, Germany}

\author{Nicolas~Pavloff}
\affiliation{Université Paris-Saclay, CNRS, LPTMS, 91405, Orsay, France}

\author{Olivier~Giraud}
\affiliation{Université Paris-Saclay, CNRS, LPTMS, 91405, Orsay, France}

\date{\today}

\begin{abstract}
We investigate quantum entanglement in an analogue black hole realized in the flow of a Bose-Einstein condensate. The system is described by a three-mode Gaussian state and we construct the corresponding covariance matrix at zero and finite temperature. We study associated bipartite and tripartite entanglement measures and discuss their experimental observation. We identify a simple optical setup equivalent to the analogue Bose-Einstein black hole which suggests a new way of determining the Hawking temperature and grey-body factor of the system.  
\end{abstract}

\maketitle

\section{Introduction} 
\label{sec: intro}

Analogue gravity aims at providing  platforms making it
possible to conduct laboratory studies of phenomena at the interface
between general relativity and quantum physics, such as Hawking
radiation \cite{Unruh_1981} and black hole superradiance \cite{torres2017}, for which in the gravitational context 
direct observation is not possible or 
no complete theory exists.  It has also been suggested that analogue models can bring some
insight on the information loss paradox
\cite{Chen2017,liberati_information_2019}. The concept
has now broadened so as to include experimental tests of physical effects
of relevance in cosmological scenarii, such as dynamical Casimir
effect, Kibble-Zurek mechanism, Zakharov oscillations, Hubble
friction... see, e.g., \cite{jacquet_next_2020} and references
therein.


In order to reach meaningful results based
on the study of an analogue model, it is important
to  precisely
characterize the experimental system supporting
the analysis 
and to correctly circumscribe the phenomenon under scrutiny.
%
The present work aims at following this line of research in the case
of an analogue of event horizon realized in a Bose-Einstein condensed
(BEC) ultracold atomic vapor. The use of a BEC as an analogue model has
been first suggested by Garay {\it et al.} \cite{Garay2000}, followed by
many others. This motivated Steinhauer
and his group to develop and then ameliorate an
experimental setup making it possible to realise an acoustic horizon
in a quasi one-dimensional BEC
\cite{Lahav2010,steinhauer2014,steinhauer_2016,de_nova_2019,kolobov_2021}. Particular attention has been devoted to the study of the analogous Hawking radiation, which corresponds to the emission of a pair of quasi-particles consisting of a ''Hawking quantum'' and a ''Partner''.
Concomitantly, the theoretical study of this system by means of a Bogoliubov
decomposition has been first suggested in  \cite{Leonhardt2003},
then gradually refined \cite{Macher2009,Recati2009,Larre2012} until a
point where a detailed comparison with experiments has been possible
\cite{Isoard2020}. In this line, a crucial question is the quantum
nature of the Hawking radiation observed in
  \cite{steinhauer_2016,de_nova_2019}: is the phenomenon simply
triggered by noise or does it correspond to spontaneous quantum
emission as in Hawking's original scenario ? A natural test of the latter
consists in demonstrating entanglement of the Hawking pair.
Indeed, experimental observation of correlated pairs of excitations does not suffice to demonstrate the quantum nature of the Hawking process, since the phenomenon also exists, e.g., in the non-quantum setting of water waves \cite{euve2016}. 
Also, as can be inferred from the quantitative results presented in Ref.~\cite{Isoard2020},  
in BEC systems the corresponding signal is robust with respect to temperature: its observation therefore does not rule out the possibility that the analogue Hawking radiation is mostly triggered by thermal and not quantum fluctuations. By contrast, the presence 
of entanglement between the Hawking quantum and its Partner 
demonstrates the presence of quantum effects. Additionally, a quantitative measure of entanglement is necessary for evaluating the respective impact of quantum and thermal effects. However, it has not always been checked whether the measures 
used up to now in the literature provide good quantitative estimates of entanglement in the system. An important goal
of the present work is to identify which, amongst different measures of entanglement, 
enable a quantitative, monotone, experimentally relevant determination of the degree of bipartite entanglement in a finite temperature BEC analogue of black hole.

Several theoretical works have addressed the issue of entanglement in analogue gravity systems. Most of them
\cite{deNova2014,Busch2014b,Busch2014a,Finazzi2014,Boiron2015,nova2015,Fabbri2018,Coutant2018} discuss qualitative measures such as the Peres-Horodecki or Cauchy-Schwarz criteria, which indicate if a state is entangled or not but -- as shown below -- do not provide good estimates of its degree of entanglement. In the present work we follow Refs.~\cite{Giovanazzi2011, Horstmann_2011,bruschi_2013,Jacquet2020}
and focus on {\it quantitative} measures.
It is important to take into account the specificities of BEC physics in order to conduct the corresponding theoretical analysis. In particular, dispersive effects and the lack of Lorentz invariance complexify the standard Hawking quantum/Partner picture by introducing new propagation channels; accordingly the system is described by a three-mode Gaussian state.
Its detailed description makes it possible to quantify its bipartite and also tripartite entanglement. We advocate for the use of a measure of entanglement based on the Gaussian contangle,
and we show that this is an experimentally accessible quantity 
which can provide a signature of the quantum nature of Hawking radiation.
We also show how entanglement can be localized in our system
in an effective two-mode state, which makes it possible to propose a simple and appealing equivalent optical model. This description suggests a new definition of the analogue Hawking temperature and of the associated grey-body factor, in closer agreement with the gravitational paradigm.
Another interesting outcome of this construction is the understanding that genuine tripartite entanglement may occur between the three modes, although two of them are not entangled.

The paper is organized as follows. In \autoref{sec:ABH} we present the theoretical description of an acoustic black hole realized in an ultracold atomic vapor. In \autoref{bog} we review the basics of Bogoliubov transformations and apply it to our situation. The description of Gaussian states appearing in the scattering processes involved in black hole analogues is discussed in \autoref{sec:3mode}. Section \ref{sec:entanglement} is dedicated to the investigation of two-mode and three-mode entanglement in the Gaussian states we are considering here.
The case of a finite-temperature setting is examined in \autoref{sec:finiteT}, where we also provide a proof of principle of the measurability of the quantities we use for assessing the degree of entanglement. Concluding remarks are presented in \autoref{ccl}. Some technical points are given in the appendices. In Appendix \ref{appBogol} we recall some properties of Bogoliubov transformations. In Appendix \ref{technicalpoints} we give some useful explicit expressions of the elements of the covariance matrix. In Appendix \ref{sec:lowwS} we recall the low frequency behavior of the coefficients describing the scattering of linear waves by the acoustic horizon. Appendix \ref{app:tripartite_eigenvalues} details the construction making it possible to localize entanglement in our system. In Appendix \ref{app:app_C} we establish a formula making it possible to compute the Gaussian contangle at finite temperature.

\section{Analogue black hole in BECs}\label{sec:ABH}

We consider a stationary flow of a one-dimensional (1D) BEC which is upstream subsonic and downstream supersonic. This "transonic" configuration mimicks a black hole since acoustic excitations generated in the downstream supersonic region are 
dragged by the flow, and not detected in the upstream region. 

\subsection{The background flow}\label{sec:background}

The complex quantum field $\hat\Psi$
describing the bosonic gas is decomposed into a classical part $\Phi$ (describing the stationary flow of the condensate) supplemented by small quantum fluctuations (described by an operator $\hat\psi$) according to
\begin{equation}\label{abh1}
    \hat\Psi(x,t)=\exp(-i\mu t/\hbar)
    \left(\Phi(x)+\hat\psi(x,t)\right) \; ,
\end{equation}
where $\mu$ is the chemical potential \cite{pitaevskii2016}. The function $\Phi$ is solution of a classical Gross-Pitaevskii equation, with the addition of an external potential $U(x)$ used to implement the transonic flow:
\begin{equation}\label{abh2}
    \mu \Phi(x) = -\frac{\hbar^2}{2 m}  \Phi_{x x} + \left(g |\Phi|^2 + U(x)\right) \Phi\; ,
\end{equation}
where $g>0$ is a nonlinear coefficient accounting for repulsion between the atoms in a mean-field approach. The operator $\hat\psi$ describes the quantum fluctuations on top of the background $\Phi$.

The experimental implementation of the 1D configuration \eqref{abh2} is obtained by a tight transverse confinement of a guided BEC. In the large density limit the transverse degrees of freedom cannot be discarded and the 1D reduction fails.
Also, the so-called Bogoliubov decomposition \eqref{abh1} implies a long-range coherence (off-diagonal long-range order, see e.g.~\cite{pitaevskii2016}) 
which, in 1D, is destroyed by phase fluctuations.
Nonetheless a description of the system relying on Eqs.~\eqref{abh1} and \eqref{abh2} can be ascribed a domain of applicability in the so-called 1D mean field regime \cite{menotti2002}. For a Bose gas with s-wave scattering length $a$ transversely confined by a harmonic
trap of angular frequency $\omega_\perp$, this regime corresponds to the range of densities
\begin{equation}\label{1dmf}
    \frac{m a^2 \omega_\perp}{\hbar} \ll n_{\rm typ} a \ll 1\; ,
\end{equation}
where $n_{\rm typ}$ is a typical
order of magnitude of the linear density $n(x)=|\Phi(x)|^2$. For $^{87}$Rb or $^{23}$Na atoms, the domain of validity \eqref{1dmf} ranges over 4 orders of magnitude in density\footnote{A more detailed discussion of the domain of applicability of the Bogoliubov decomposition \eqref{abh1} can be found, e.g., in \cite{Fabbri2018}.} and in this case $g=2\hbar\omega_\perp a$ \cite{olshanii1998}.

\begin{figure}
\includegraphics[width=\linewidth]{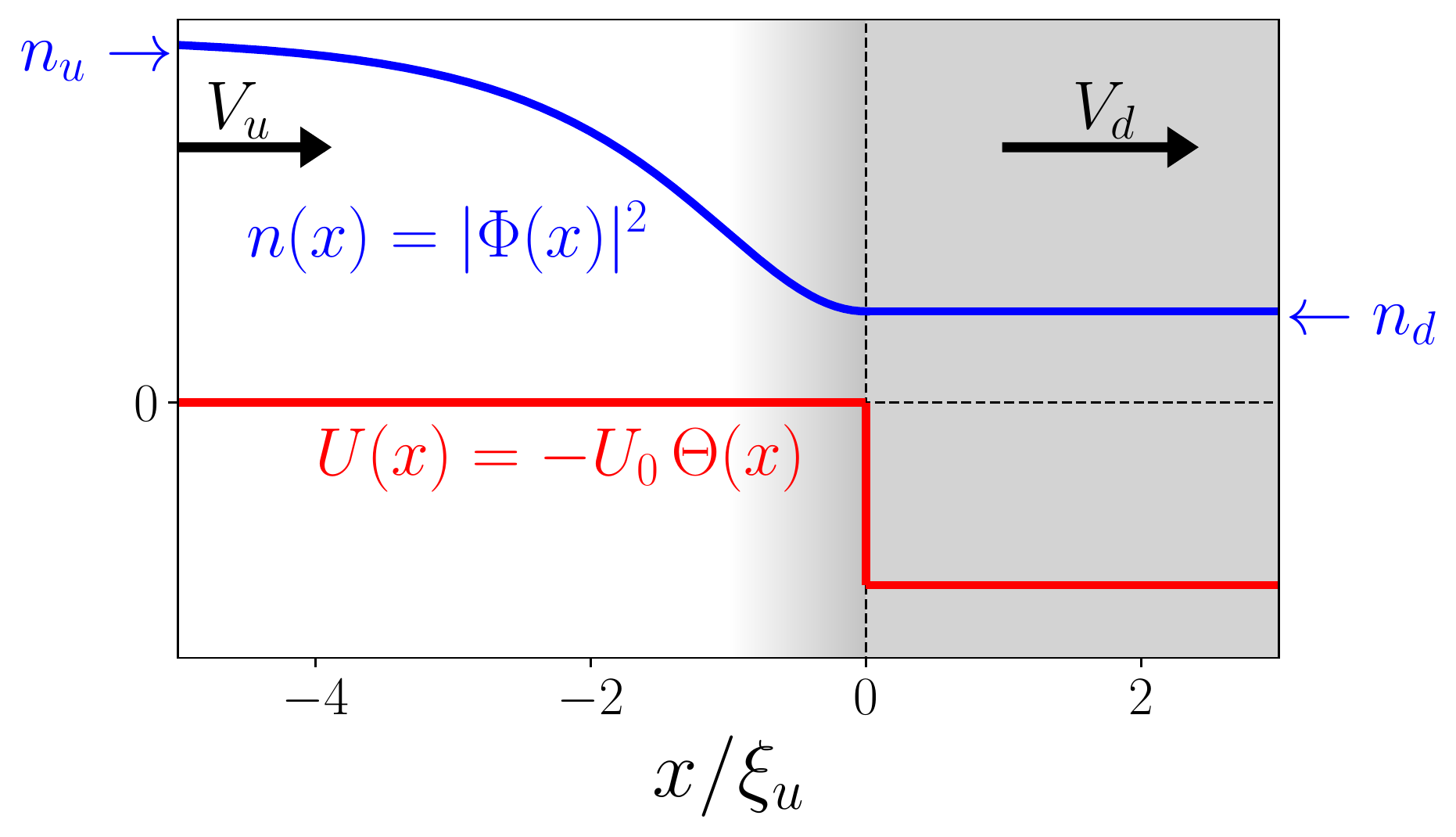}
\caption{Waterfall configuration. The flow is directed from left to right. The downstream classical field $\Phi(x>0)$ is exactly a plane wave of density $n_d$ and velocity $V_d$. $\Phi(x<0)$ is the half profile of a dark soliton  which is asymptotically a plane wave of density $n_u$ and velocity $V_u$, see Eqs.~\eqref{abh2b}. The far upstream flow is subsonic, with a velocity $0<V_u<c_u$ and the downstream flow is supersonic with a velocity $V_d>c_d>0$, where $c_{\alpha}=(g n_{\alpha}/m)^{1/2}$ is the speed of sound in region $\alpha=u$ (upstream) or $d$ (downstream). The shaded region $x > 0$ corresponds to the interior of the analogue black hole; the gradient of grey around $x\lesssim 0$ depicts the (ill-defined, see text) position of the horizon. The coordinate $x$ is plotted in units of the upper healing length $\xi_u=\hbar/(m c_u)$.}
\label{fig:waterfall}
\end{figure}

Several configurations realizing an analogue black hole have be proposed in the past \cite{balbinot2008,carusotto2008,Macher2009,Recati2009,Zapata2011,Larre2012,Parola_2017}. The approach we use in this work is valid in a general setting, but for the sake of illustration we will present numerical results for the so-called "waterfall configuration" \cite{Larre2012} which has been experimentally realized in \cite{steinhauer_2016,de_nova_2019} and which has been shown to lead to a
significant violation of the Cauchy-Schwarz criterion
in \cite{Fabbri2018}. In this configuration $U(x)=-U_0 \Theta(x)$, where $U_0>0$ and $\Theta$ is the Heaviside step function. The corresponding solution of Eq.~\eqref{abh2} is a plane wave flow of density $n_d$ and velocity $V_d>0$ in the downstream region ($x>0$) and half a dark soliton in the upstream region ($x<0$) with asymptotic density $n_u$ and velocity $V_u>0$, meaning that
\begin{equation}\label{abh2b}
    \begin{split}
        \Phi(x>0)=& \sqrt{n_d} \exp\left(i m V_d \, x/\hbar\right) \exp(i\,\beta_d) \; , \\
        \Phi(x\to-\infty)=& \sqrt{n_u} \exp\left(i m V_u\, x/\hbar\right) \exp(i\,\beta_u)\; ,
    \end{split}
\end{equation}
where $\beta_u$ and $\beta_d$ are constant phase factors. This setting is illustrated in Fig.~\ref{fig:waterfall} (see details in \cite{Larre2012}).

In the following we will loosely state that the horizon is located at $x=0$.
However, it is important to note that, in any dispersive analog model, the location of the horizon is ill-defined, as it depends on frequency.  A commonly accepted  way to circumvent this difficulty is to take the zero-frequency value: the analog horizon is then the point where the velocity of the flow $v(x)=\frac{\hbar}{m}{\rm Im}(\Phi^* \Phi_x)/n(x)$ is equal to the local sound velocity $c(x)$. This definition makes sense because it has been shown that the characteristics of analogue Hawking radiation are governed by long wave-length physics, see, e.g., \cite{unruh1995,Brout1995,corley1996}.
However, in the BEC context, the definition of a local sound velocity $c(x)$ is only legitimate in regions where the density varies over a length scale large compared to the healing length. This is not the case  around $x\lesssim 0$ for the waterfall configuration and this forbids a rigorous definition of an horizon. 
Nevertheless, the system still emits a spontaneous analogue Hawking radiation, because the feature that triggers this process is the mismatch between the left subsonic asymptotic flow and the right supersonic one (this is at the heart of the Bogoliubov transform discussed in Sec.~\ref{bog}). 
One may however wonder if the concept of Hawking temperature is still meaningful in the absence of a proper location of the horizon, since, strictly speaking, the widely used semi-classical result \eqref{surf_grav} which defines the Hawking temperature as the analogue surface gravity is not valid here\footnote{Note that this issue is also encountered in profiles smoother than that of the waterfall, see, e.g., \cite{Macher2009}.}. The solution lies in the study of the low frequency behavior of the spectrum of the analogue Hawking radiation which is thermal-like. This makes it possible to determine an effective Hawking temperature, see e.g., \cite{Recati2009,Macher2009,Larre2012}. We will come to this point in more details in Secs. \ref{sec.3mgs} and \ref{sec:entanglement_localization}.

\subsection{Elementary excitations}
\label{sec:ScatteringProcess}

Since the far upstream and downstream background flows are uniform, the elementary excitations which form a basis set for the quantum operator $\hat{\psi}$ are plane waves in these two regions, with dispersion relations of Bogoliubov type (see, e.g., \cite{pitaevskii2016}):
\begin{equation}\label{abh3}
    (\omega-q\, V_{\alpha})^2=\omega_{{\sss\rm B},\alpha}^2(q)\; , \quad \alpha=u \;\mbox{or}\; d\; ,
\end{equation}
where $V_u$ and $V_d$ are the upstream and downstream velocities, and $\omega_{{\sss\rm B},\alpha}$ is the Bogoliubov dispersion relation
\begin{equation}
\label{eq:Bogoliubov_dr}
    \omega_{{\sss\rm B},\alpha}(q)=c_{\alpha} q \sqrt{1+\xi_{\alpha}^2q^2/4}\; , 
\end{equation}
$c_\alpha=(g n_\alpha/m)^{1/2}$ being the speed of sound and $\xi_\alpha=\hbar/(mc_\alpha)$ the "healing length", in the far upstream region if $\alpha=u$ and in the downstream region if $\alpha=d$.
The left-hand side of Eq.~\eqref{abh3} includes a Doppler shift caused by the velocity $V_\alpha$ of the background.

\begin{figure*}[!t]
\includegraphics[width=\linewidth]{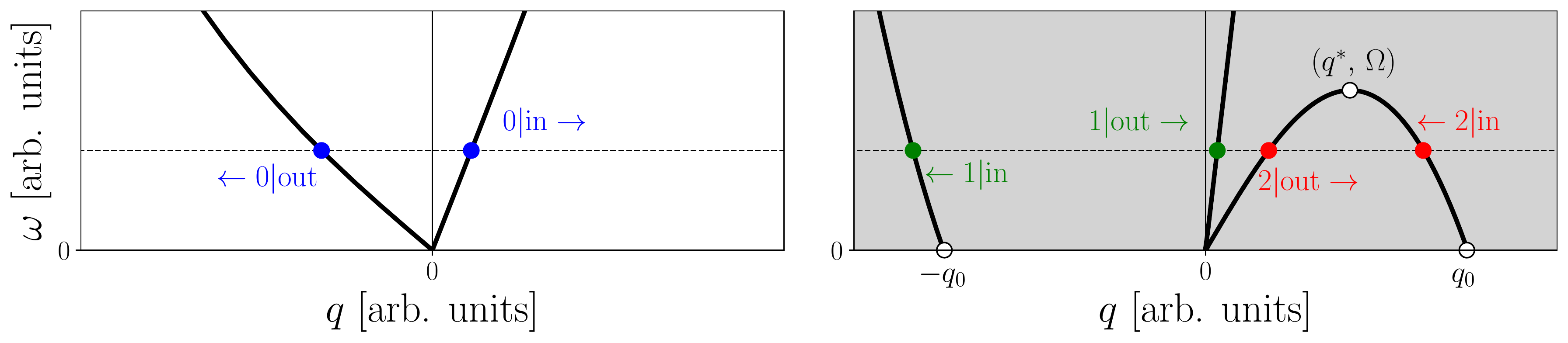}
\caption{Graphical representation of the positive
  frequency part of the dispersion relation \eqref{abh3} in the far
  upstream (left plot) and downstream (right plot) regions. The
  downstream region (grey background) is the interior of the analogue
  black hole while the upstream region (white background) is the
  exterior. In the upstream region, to any given $\omega$ (represented by
  a horizontal dashed line) correspond two channels of propagation
  denoted as $0|{\rm in}$ and $0|{\rm out}$. In the downstream region
  there are four or two channels, depending if $\omega$ is
  smaller or larger than $\Omega$. The arrows indicate the direction
  of propagation of the corresponding waves, and the channels are
  labelled 1 or 2, with an additional "in" (or "out") indicating if
  the wave propagates towards (or away from) the horizon.}
\label{fig:dispersion}
\end{figure*}

It will be useful in the following to define the quantities
\begin{equation}
\label{eq:Mach_nb}
    m_\alpha = \frac{V_\alpha}{c_\alpha}, \quad \alpha=u \;\mbox{or}\; d\; ,
\end{equation}
known as the upstream ($\alpha = u$) and downstream ($\alpha = d$) Mach numbers. It was shown in \cite{Larre2012} that the waterfall configuration, which we use below to exemplify our results, is uniquely characterized once the value of $m_u$, say, is fixed. In particular the parameters of the flow are related by the following relations
\begin{equation}\label{water_eq}
\frac{V_d}{V_u}=\frac{n_u}{n_d}=
\frac{1}{m_u^2}=m_d=\left(\frac{\xi_d}{\xi_u}\right)^2 
=\left(\frac{c_u}{c_d}\right)^2.
\end{equation}
The flow being upstream subsonic ($V_u<c_u$, i.e., $m_u<1$) and downstream supersonic ($V_d>c_d$, i.e., $m_d>1$), the graphs of the corresponding dispersion relations are of different types, as illustrated in Fig.~\ref{fig:dispersion}.
In the upstream region the spectrum has two branches which we label as $0|{\rm in}$
and $0|{\rm out}$. In the downstream supersonic region there are four branches: $1|{\rm in}$, $1|{\rm out}$, $2|{\rm in}$ and $2|{\rm out}$, the last two branches being limited to $\omega\in[0,\Omega]$, where $\Omega$ is the frequency at which these two branches coalesce, and whose value is given by
\begin{equation}\label{threshold}
\begin{split}
& \Omega=q^* V_d-\omega_{{\sss\rm B},d}(q^*) \quad\mbox{with}\\
& q^*\xi_{d}=\left(-2+\frac{m_d^2}{2}+
\frac{m_d}{2}
\sqrt{8+m_d^2}\right)^{\frac{1}{2}}.
\end{split}
\end{equation}
For future convenience (see Sec.~\ref{sec:finiteT}) we define functions $q_{0|{\rm in}}(\omega)$, $q_{1|{\rm in}}(\omega)$ and $q_{2|{\rm in}}(\omega)$ as
the reciprocal of the Bogoliubov dispersion relation \eqref{abh3} along some of these branches; $q_{0|{\rm in}}(\omega)$ and $q_{1|{\rm in}}(\omega)$ are defined for $\omega>0$ and $q_{2|{\rm in}}(\omega)$ only for $\omega\in[0,\Omega]$. A number of previous works  \cite{Zapata2011,Larre2012,deNova2014,Fabbri2018,Isoard2020}  followed the convention introduced in \cite{Recati2009}, in which indices $u$, $d1$, and $d2$ are employed instead of the indices 0, 1 and 2 we use here. We changed convention in order to simplify the manipulation of the matrix notation introduced below.

The particular transonic configuration we consider corresponds, for angular frequencies $\omega$  lower than the threshold $\Omega$, to a specific scattering process of elementary excitations onto the analogue event horizon. For instance, a wave issued from the interior region along the channel identified as $1|{\rm in}$ in Fig.~\ref{fig:dispersion} is transmitted to the exterior along the $0|{\rm out}$ channel and reflected back along the $1|{\rm out}$ and $2|{\rm out}$ channels. The corresponding (complex) transmission and reflection amplitudes are denoted as $S_{10}(\omega)$, $S_{11}(\omega)$ and $S_{12}(\omega)$, respectively. They are obtained by imposing matching conditions at $x=0$, as explained in Ref.~\cite{Larre2012}.
The quantum boson operator corresponding to this whole process is denoted as $\hat{b}_1(\omega)$. Similarly, a wave incident along the $0|{\rm in}$ channel is transmitted towards the interior of the black hole along channels $1|{\rm out}$ (amplitude $S_{01}(\omega)$) and $2|{\rm out}$ (amplitude $S_{02}(\omega)$) and reflected along $0|{\rm out}$ (amplitude $S_{00}(\omega)$); the corresponding quantum mode is associated with operator $\hat{b}_0(\omega)$ \footnote{In the terminology we use, it is important to make a distinction between the "quantum modes" and the "propagation channels": a mode corresponds to a whole process typically involving one or several incoming channels and one or several outgoing channels.}. A third mode describes the scattering of a wave issued from the channel $2|{\rm in}$ onto the outgoing channels $0|{\rm out}$, $1|{\rm out}$ and $2|{\rm out}$.
The channels labelled $2|{\rm in}$ and $2|{\rm out}$ are particular, in the sense that they have a negative norm, i.e., a negative energy in the rest frame of the fluid
\cite{blaizot1986,Fetter_EFermi_1999,Barcelo2010,robertson2012}. As a result, the mode initiated by the incoming channel $2|{\rm in}$ should be quantized using an operator $\hat{b}^\dagger_2(\omega)$, i.e.~inverting the role of the creation and annihilation operators used for the two other modes. Only in this way do the propagating modes behave as bosons satisfying the usual commutation relations
\begin{equation}\label{abh5}
\begin{split}
& \left[\hat{b}_i(\omega),\hat{b}_j^\dagger(\omega')\right]
    =\delta_{i,j}\, \delta(\omega-\omega') \; \\
& \left[\hat{b}_i(\omega),\hat{b}_j(\omega')\right]
= \left[\hat{b}_i^\dagger(\omega),\hat{b}_j^\dagger(\omega')\right]
= 0\; ,
\end{split}
\end{equation}
for $i$ and $j\in\{0,1,2\}$. Another consequence is that the $3\times 3$ scattering matrix $S(\omega)$ whose elements are the $S_{ij}(\omega)$ obeys a skew-unitarity relation \cite{Recati2009} :
\begin{equation}\label{abh6}
    S^\dagger \eta S =\eta =S \eta S^\dagger\; ,
    \qquad
    \eta={\rm diag}(1,1,-1)\; .
\end{equation}
For $\omega>\Omega$ the situation is drastically different: the channels $2|{\rm in}$ and $2|{\rm out}$ disappear (cf.~Fig.~\ref{fig:dispersion}), as well as the operator $\hat{b}_2(\omega)$, and the $S$-matrix becomes $2\times 2$ and unitary.

We denote the $b$ modes as "incoming" since they correspond to scattering processes initiated by a single wave incident along one of the three "in" channels directed towards the horizon: $0|{\rm in}$, $1|{\rm in}$ and $2|{\rm in}$. One could equivalently choose to work with "outgoing modes" \cite{Leonhardt2003} describing processes each resulting in the emission of a single wave along one of the three "out" channels $0|{\rm out}$, $1|{\rm out}$ and $2|{\rm out}$. We denote the corresponding quantum operators as $\hat{c}_0(\omega)$, 
$\hat{c}_1(\omega)$ and $\hat{c}_2(\omega)$. They relate to the incoming operators {\it via} \cite{Recati2009}
\begin{equation} \label{eq:bog_transform-ch3}
\begin{pmatrix}
\hat{c}_{0}  \\
\hat{c}_{1} \\
\hat{c}_{2}^{\dagger} \\
\end{pmatrix}
=
\begin{pmatrix}
S_{00} & S_{01} & S_{02} \\
S_{10} & S_{11} & S_{12} \\
S_{20} & S_{21} & S_{22} \\
\end{pmatrix}
\,
\begin{pmatrix}
\hat{b}_{0}  \\
\hat{b}_{1} \\
\hat{b}_{2}^{\dagger} \\
\end{pmatrix},
\end{equation}
where for legibility we omit the $\omega$ dependence of all the
terms. The definition \eqref{eq:bog_transform-ch3}, together with the
property \eqref{abh6}, ensures that the $\hat{c}$ operators obey the
same commutation relations \eqref{abh5} as the $\hat{b}$ operators and
thus describe bosonic quasiparticles.

In the setting we consider, the analogue of the Hawking radiation
spectrum is the
number of excitations emitted per unit time and per unit frequency
into the subsonic region ($x<0$), that is, the expectation value of
$\hat{c}^\dagger_0(\omega) \hat{c}_0(\omega)$ over the state
vector. From relation \eqref{eq:bog_transform-ch3} one sees that this
current is non zero when the state vector is the vacuum $|0\rangle_b$
of incoming modes:
$_b\langle 0| \hat{c}^\dagger_0(\omega) \hat{c}_0(\omega)|0\rangle_b=
|S_{02}(\omega)|^2$; this is the analogous Hawking effect \cite{fabbri_modeling_2005,Recati2009,Macher2009,robertson2012}.  The mode
associated with operator $\hat{c}_0$ is thus denoted the Hawking
outgoing mode.  The other outgoing modes, associated with operators
$\hat{c}_1$ and $\hat{c}_2$ are denoted the Companion and the Partner,
respectively.

As can be seen from expression \eqref{eq:bog_transform-ch3}, the \textit{outgoing} operators $\hat{c}$ and $\hat{c}^{\dagger}$ are expressed as a combination of the \textit{ingoing} annihilation and creation operators $\hat{b}$ and $\hat{b}^{\dagger}$. Therefore, it is possible to associate a Bogoliubov transformation with our analogue system. This is the aim of the next section.

\section{Bogoliubov transformations}
\label{bog}

Bogoliubov transformations are linear transformations of creation and annihilation operators that preserve the canonical commutation rules \cite{balian1969}. 
In the context of quantum field theory in curved spacetime, these transformations are at the heart of the Hawking process; indeed, since they mix annihilation and creation operators they can give rise to spontaneous emission of particles from vacuum \cite{hawking1974,hawking1975,dewitt1975,davies1977,wald1975,parker1975}. This mixing of operators also occurs for analogue black holes, as is clear from Eq.~\eqref{eq:bog_transform-ch3}. This way of viewing the emergence of the analogue Hawking radiation through a Bogoliubov transformation makes a direct connection with the gravitational case: as shown by Hawking in 1974 \cite{hawking1974,hawking1975}, one of the parameters involved in the Bogoliubov transformation, the so-called $\beta$-coefficient, is directly related to the number of particles created by black holes. In our case, we can derive such a parameter and compare its properties with Hawking's $\beta$-coefficient; in particular, through this approach, we will be able to question the thermality of the analogue Hawking radiation (see \autoref{sec:thermality}).
Furthermore, identifying the Bogoliubov transformation will be an important step to understand and study the entanglement properties of the analogue Hawking radiation (see \autoref{sec:entanglement}). 

The present section is divided into two parts. First, we consider an arbitrary (but unitary) Bogoliubov transformation and derive its properties. Then, we apply these results to the particular case of analogue black holes in BECs starting from expression \eqref{eq:bog_transform-ch3}.

\subsection{General setting}
\label{sec:bogoliubov_general_case}
We start by briefly recalling some well-known facts concerning unitary  Bogoliubov transformations \cite{balian1969,blaizot1986, takayanagi2008utilizing}. Some useful intermediate results are given in Appendix \ref{appBogol}.

Let us consider $N$ boson operators $\hat{b}_1, \ldots,\hat{b}_{\sss N}$
satisfying the usual commutation relations $[\hat{b}_i,\hat{b}_j^\dagger]=\delta_{i,j}$. Defining the column vector
\begin{equation}\label{def_vct_b}
\mathbf{b}=(\hat{b}_1,\ldots,\hat{b}_{\sss N},\hat{b}_1^\dagger,\ldots,\hat{b}_{\sss N}^\dagger)^{\sss \rm T},
\end{equation}
the Bose commutation relations can be rewritten as
\begin{equation} \label{eq:comm_rel}
\left[ \mathbf{b}_i, \, \mathbf{b}_j \right] = \widetilde{\mathds{J}}_{ij},
\quad\mbox{with}\quad
\widetilde{\mathds{J}} = 
\begin{pmatrix} 0 &  \mathds{1}_{\sss N} \\
- \mathds{1}_{\sss N} & 0
\end{pmatrix},
\end{equation}
where $\mathds{1}_{\sss N}$ is the $N\times N$ identity matrix.
A (unitary) Bogoliubov transformation is a linear transformation mapping the operators $\hat{b}_i$ onto new operators $\hat{c}_i$ defined through
\begin{equation} \label{eq:BogoTr0}
\mathbf{c}_i =  \sum_{j=1}^{2N} \mathscr{T}_{ij} \, \mathbf{b}_j\; , \quad \text{or equivalently} \quad \mathbf{c}=\mathscr{T}\, \mathbf{b}\; .
\end{equation}
For unitary Bogoliubov tranformations $\mathbf{c}$ has the form
\begin{equation}\label{def_vct_c}
\mathbf{c}=(\hat{c}_1,\ldots,\hat{c}_{\sss N},\hat{c}_1^\dagger,\ldots,\hat{c}_{\sss N}^\dagger)^{\sss \rm T},
\end{equation}
i.e., $\mathbf{c}_{i+N}=\mathbf{c}_i^\dagger$.
In this case the matrix $\mathscr{T}$ admits the block decomposition
\begin{equation} \label{eq:matrix_T}
\mathscr{T} = 
\begin{pmatrix} \alpha^* &  - \beta^*\\
-\beta & \alpha
\end{pmatrix}\; ,
\end{equation}
where $\alpha$ and $\beta$ are $N\times N$ matrices.
Operators $\hat{c}_i$ and $\hat{b}_i$ can then be related by a unitary operator $T$ such that
\begin{equation} \label{eq:BogoTr}
\hat{c}_i = T^{\dagger} \, \hat{b}_i \, T,
\end{equation}
whose explicit construction from matrix $\mathscr{T}$ is detailed in Appendix \ref{appBogol}. 

 In general, the transformation $\mathscr{T}$ in \eqref{eq:BogoTr0} mixes creation and annihilation operators, so that the vacua $|0\rangle_b$ and $|0\rangle_c$, defined by
\begin{equation}
\hat{b}_{\sss i}\,|0\rangle_b = 0, \quad \text{and} \quad \hat{c}_{\sss i}\,|0\rangle_c = 0, \quad i \in \{1,\ldots,N\},
\end{equation}
differ. These vacua are related via the identity
\begin{equation}
\label{vacbvacc}
|0\rangle_b = T \, |0\rangle_c,
\end{equation}
as is clear from the fact that $\hat{b}_{\sss i}\,T|0\rangle_c=T \, \hat{c}_{\sss i}|0\rangle_c=0$. 

Defining the $N\times N$ matrix
$X = - \beta^{*} \, \alpha^{-1}$ and
using the decomposition \eqref{decompositionT},
it is possible to write Eq.~\eqref{vacbvacc} under the explicit form 
\begin{equation} \label{eq:link_vacua}
|0\rangle_b =  \frac{1}{(\det \alpha)^{\frac12}} e^{\frac{1}{2} \sum_{i,j} X_{ij} \hat{c}_i^{\dagger}\hat{c}_j^{\dagger}} |0\rangle_c .
\end{equation}
A simple example of a Bogoliubov transformation is the one leading to two-mode squeezed states \cite{schumaker1985new, agarwal2012quantum}. For a real squeezing parameter $r$, a two-mode squeezed state is obtained by applying the two-mode squeezing operator 
\begin{equation}
\label{squeezingop}
T=
\exp[r (\hat{c}_1^\dagger \hat{c}_2^\dagger-\hat{c}_1 \hat{c}_2)]
\end{equation}
to the vacuum state $|0\rangle_c$. The corresponding Bogoliubov transformation is of the form \eqref{eq:matrix_T} with $N=2$ and $\alpha, \beta$ given by
\begin{equation} \label{eq:tmss}
\alpha = 
\begin{pmatrix}
\cosh r & 0 \\
0 & \cosh r 
\end{pmatrix},\quad
\beta = -
\begin{pmatrix}
0 & \sinh r   \\
\sinh r & 0 
\end{pmatrix}.
\end{equation}
In this case \eqref{eq:link_vacua} reads 
\begin{equation}
|0\rangle_b =(\cosh r)^{-1}
\exp(\tanh r \, \hat{c}^\dagger_1 \hat{c}^\dagger_2 ) \, |0\rangle_c \; .
\end{equation}

\subsection{Bogoliubov transformation in a transonic BEC}\label{sec:Bog-BEC}

In the case described in Sec.~\ref{sec:ABH} of a transonic flow realised in a BEC, $\mathbf{b}$ and $\mathbf{c}$ correspond to sets of ingoing and outgoing modes. The associated column vectors $\mathbf{b} = (\hat{b}_{0} ,\hat{b}_{1} ,\hat{b}_{2} ,\hat{b}_{0}^\dagger , \hat{b}_{1}^\dagger,\hat{b}_{2}^\dagger )^{\sss \rm T}$ and $\mathbf{c} = (\hat{c}_{0} ,\hat{c}_{1} ,\hat{c}_{2} ,\hat{c}_{0}^\dagger , \hat{c}_{1}^\dagger,\hat{c}_{2}^\dagger )^{\sss \rm T}$ are related by Eq.~\eqref{eq:bog_transform-ch3}. One can express this relation equivalently as $\mathbf{c}=\mathscr{T}\, \mathbf{b}$, with $\mathscr{T}$ a Bogoliubov transformation of the form \eqref{eq:matrix_T} with
\begin{equation} \label{eq:alph-beta}
\alpha = 
\begin{pmatrix}
S^*_{00} & S_{01}^* & 0 \\
S_{10}^* & S_{11}^* & 0 \\
0 & 0 & S_{22} \\
\end{pmatrix},\quad
\beta = -
\begin{pmatrix}
0 & 0 & S_{02}^*  \\
0 & 0 & S_{12}^* \\
S_{20} & S_{21} & 0 \\
\end{pmatrix},
\end{equation}
where for legibility we do not write the $\omega$-dependence of the scattering amplitudes. This yields
\begin{equation} \label{eq:V-ch3}
X = \frac{1}{S_{22}}\begin{pmatrix}
0 & 0 & S_{02} \\
0 & 0 & S_{12} \\
S_{02} & S_{12} & 0 \\
\end{pmatrix}.
\end{equation}
From relation \eqref{abh6} one can show that $\det\alpha=|S_{22}|^2$, and thus
\eqref{eq:link_vacua} takes the simple form
\begin{equation} \label{eq:link_vacua_BEC-2}
|0\rangle_b = \frac{1}{|S_{22}|} \,  e^{\left(X_{02} \, \hat{c}_{0}^\dagger  + X_{12} \, \hat{c}_{1}^\dagger \right) \, \hat{c}_{2}^\dagger} \,  |0\rangle_c\, .
\end{equation}

A word of caution is in order here. The case we consider in the present section is different from the discussion of the previous \autoref{sec:bogoliubov_general_case} because, as explained in Sec.~\ref{sec:ScatteringProcess}, the modes are here continuously distributed along the energy axis [compare for instance the commutation relations \eqref{abh5} and \eqref{eq:comm_rel}]. A natural way to set up a framework encompassing both situations consists in discretizing the energies with a small mesh $\Delta \omega$ and to define coarse-grained operators
\begin{equation}
    \hat{B}_{i,p}=\frac{1}{\sqrt{\Delta\omega}} \int_{\omega_p}^{\omega_{p+1}}
  \!\!\!\! {\rm d}\omega \, \hat{b}_i(\omega),
  \end{equation}
and
\begin{equation}
  \hat{C}_{i,p}=\frac{1}{\sqrt{\Delta\omega}} \int_{\omega_p}^{\omega_{p+1}}
  \!\!\!\! {\rm d}\omega \, \hat{c}_i(\omega)\; ,
\end{equation}
where $i\in\{0,1,2\}$, $p\in\mathbb{N}$ and $\omega_p=p\, \Delta\omega$. It is easy to check that these operators obey the standard Bose commutation rules, such as $[\hat{B}_{i,p},\hat{B}^\dagger_{j,q}] = \delta_{i,j}\delta_{p,q}$ for instance. If $\Delta\omega$ is small compared to the typical scale of variation of the elements of the $S$-matrix, then the $\hat{C}_{i,p}$ and the $\hat{B}_{j,p}$ are related by a relation analogous to \eqref{eq:bog_transform-ch3}:
\begin{equation}
\begin{pmatrix}
\hat{C}_{0,p}  \\
\hat{C}_{1,p} \\
\hat{C}_{2,p}^{\dagger} \\
\end{pmatrix}
=
\begin{pmatrix}
S_{00} & S_{01} & S_{02} \\
S_{10} & S_{11} & S_{12} \\
S_{20} & S_{21} & S_{22} \\
\end{pmatrix}
\,
\begin{pmatrix}
\hat{B}_{0,p}  \\
\hat{B}_{1,p} \\
\hat{B}_{2,p}^{\dagger} \\
\end{pmatrix},
\end{equation}
where the $S_{ij}$ should be evaluated at $\omega_p$. 
Thus the relation \eqref{eq:link_vacua_BEC-2} should be replaced by
\begin{equation}\label{eq:link_vacua_BEC-2bis}
\begin{split}
    |0\rangle_b = & \frac{1}{\prod_{p=0}^\infty|S_{22}(\omega_p)|} \times \\  & e^{\sum_{p=0}^\infty \left(X_{02}(\omega_p) \, \hat{C}_{0,p}^\dagger  + X_{12}(\omega_p) \, \hat{C}_{1,p}^\dagger \right) \, \hat{C}_{2,p}^\dagger} \,  |0\rangle_c\, .
    \end{split}
\end{equation}
This remark being made, in the following we favor legibility over formal rigor: We will continue to write relations of the type \eqref{eq:link_vacua_BEC-2}, instead of the more rigourous but cumbersome Eq.~\eqref{eq:link_vacua_BEC-2bis}, keeping in mind that the correction of "naive" expressions -- such as Eqs.~\eqref{number-state}, \eqref{eq:our_state}, \eqref{eq:link_vacua_set_e} or \eqref{eq:bar_ni_def} below -- is straightforward.

From \eqref{eq:link_vacua_BEC-2}, if we define the Fock state basis of quasi-particles of type $c$ by 
\begin{equation}
\label{number-state}
 |n\rangle_i=\frac{1}{\sqrt{n!}}(\hat{c}_i^\dagger)^n|0\rangle_c\,,
 \end{equation}
 where $i$ is the mode number, then the explicit expansion of the vacuum $|0\rangle_b$ reads
\begin{equation} 
\label{eq:our_state}
|0\rangle_b = \frac{1}{|S_{22}|} \sum_{n,n'=0}^{\infty} 
\sqrt{\binom{n+ n'}{n}} X_{\sss 02}^n X_{\sss 12}^{n'}
|n\rangle_{\sss 0}|n'\rangle_{\sss 1}|n+n'\rangle_{\sss 2}\,.
\end{equation}
It is convenient for future use in sections \ref{sec:vac_three_mode} and \ref{sec:entanglement_localization} to introduce a new set of operators $\mathbf{e} = (\hat{e}_{0} ,\hat{e}_{1} ,\hat{e}_{2} ,\hat{e}_{0}^\dagger , \hat{e}_{1}^\dagger,\hat{e}_{2}^\dagger )^{\sss \rm T}$. By writing
\begin{equation}\label{eq:s_a_p}
S_{ij}(\omega)=v_{ij}(\omega)e^{\mathrm{i}\varphi_{ij}(\omega)},\quad v_{ij}\geq 0,\qquad 0\leq i,j\leq 2,
\end{equation}
we define the operators $\hat{e}_{0}$, $\hat{e}_{1}$ and $\hat{e}_{2}$ as
\begin{equation}
\label{eq:def_e}
\hat{e}_0 = e^{-\mathrm{i}\varphi_{02}} \hat{c}_0, \,\,
\hat{e}_1 = e^{-\mathrm{i}\varphi_{12}} \hat{c}_1, \,\,
\hat{e}_2 = e^{\mathrm{i}\varphi_{22}} \hat{c}_2
\end{equation}
(note the $+$ sign in front of $\varphi_{22}$).
This defines a local unitary Bogoliubov transformation, as it does not mix annihilation and creation operators. In particular $|0\rangle_e = |0\rangle_c$.
Using the notations of \autoref{sec:bogoliubov_general_case}, this transformation can be cast in the form
\begin{equation}
\label{eq:def_R}
\mathbf{e} = \mathscr{R} \, \mathbf{c},
\end{equation}
where
\begin{equation}
\label{eq:defR}
\mathscr{R} =
\diag \left( e^{-\mathrm{i}\varphi_{02}},
e^{-\mathrm{i}\varphi_{12}},
e^{\mathrm{i}\varphi_{22}},
e^{\mathrm{i}\varphi_{02}},
e^{\mathrm{i}\varphi_{12}},
e^{-\mathrm{i}\varphi_{22}}
\right).
\end{equation}
Then, using expression \eqref{eq:V-ch3} and this new set of creation and annihilation operators $\mathbf{e}$, Eq.~\eqref{eq:link_vacua_BEC-2} becomes
\begin{equation} \label{eq:link_vacua_set_e}
|0\rangle_b = \frac{1}{v_{22}} \,  e^{v_{22}^{-1}\left(v_{02} \, \hat{e}_{0}^\dagger  + v_{12} \, \hat{e}_{1}^\dagger \right) \, \hat{e}_{2}^\dagger} \,  |0\rangle_e\, .
\end{equation}


\section{Three-mode Gaussian states}
\label{sec:3mode}

In the context of analogue gravity, the general description of the system 
by means of a Gaussian state has been presented in the monograph \cite{fabbri_modeling_2005}. 
The importance of Gaussianity has been implicitly or explicitly assumed in 
many articles, but it has been thoroughly discussed only in Ref.~\cite{nova2015}.
In the present work we will extend in Secs.~\ref{sec:entanglement} and \ref{sec:finiteT} the analysis of \cite{nova2015}
to build {\it quantitative and monotone} measures of bipartite and tripartite entanglement.  Since Gaussianity is a central point in our approach, in the present section we 
briefly present general properties of Gaussian states, then construct the covariance matrix of the three-mode Gaussian pure state  which describes our system [$|0\rangle_b$ defined by Eq.~\eqref{eq:our_state}] and discuss 
in more detail the covariance matrix of the reduced state $\rho^{(0)}$, in connection with the determination of the Hawking temperature.

\subsection{Gaussian states}
\label{defgaussian}
In order to set up notations we start by reviewing the formalism for
Gaussian states (see \cite{weedbrook2012gaussian} for a
review). Gaussian states are states whose Wigner function is a
Gaussian. A Gaussian state $\rho$ can be entirely described by its
first and second moments. We define the covariance matrix $\sigma$ of
$\rho$ as the real symmetric positive-definite matrix
\begin{equation} 
  \sigma_{ij} \equiv \frac{1}{2} \, \langle \hat{\xi}_i \, \hat{\xi}_j +
  \hat{\xi}_j \, \hat{\xi}_i  \rangle - \langle \hat{\xi}_i \rangle
  \, \langle \hat{\xi}_j \rangle,
\label{eq:covariance_matrix}
\end{equation}
where $\hat{\xi}_i$ are components of the vector
$\boldsymbol{\xi} = \sqrt2 \,
(\hat{q}_1,\hat{p}_1,\ldots,\hat{q}_{\sss N},\hat{p}_{\sss N})^{\sss
  \rm T}$ of quadratures relative to mode $i$, defined so that
$[\hat{q}_i,\hat{p}_j]=i\, \delta_{i,j}$.  In the definition
\eqref{eq:covariance_matrix} and in all the following the averages
$\langle \cdots \rangle$ are taken over the density matrix $\rho$
characterizing the state of the system, which, in the simpler case, is
the projector onto the vacuum state $|0\rangle_b$. We shall discuss in
\autoref{sec:finiteT} how to generalize to a finite-temperature
configuration.

The commutation relations between the $\hat{\xi}_i$ can be expressed as $[\hat{\xi}_i,\hat{\xi}_j] = 2 \, i \, \mathds{J}_{ij},\,\, \forall \,i,j \in \{1, \ldots 2N\}$ 
with
\begin{equation} \label{eq:omega-ch3}
\mathds{J}= \overset{\sss N}{\underset{1}{\oplus}} \, {J}_i, \quad 
{J}_i =\begin{pmatrix}0&  1 \\-1&  0\end{pmatrix}.
\end{equation}
Entanglement properties of a quantum state are unchanged by local unitary (LU) operations, so that the mean values of position and momentum operators can be set to 0. A $N$-mode Gaussian state is then entirely specified by its $2N\times 2N$ covariance matrix, which can be rewritten in terms of $2\times 2$ blocks as
\begin{equation}\label{cmG}
\sigma=
\begin{pmatrix}
\sigma_1 & \varepsilon_{12} & \cdots & \varepsilon_{1\sss N} \\
\varepsilon_{12}^{\rm \sss T } & \ddots & \ddots & \vdots \\
 \vdots & \ddots & \ddots &  \varepsilon_{\sss N-1 N} \\
\varepsilon_{1\sss N}^{\rm \sss T } & \cdots & \varepsilon_{\sss N-1 N}^{\rm \sss T } &  \sigma_{\sss N} \\
\end{pmatrix},
\end{equation}
with
\begin{equation} \label{eq:cov_mat_n-modes2a}
\varepsilon_{ij} =
2 \, 
\begin{pmatrix}
\left\langle  \hat{q}_{i} \, \hat{q}_{j} \right\rangle 
& \left\langle   \hat{q}_{i} \, \hat{p}_{j} \right\rangle   \\
\left\langle \hat{p}_{i} \, \hat{q}_{j}  \right\rangle  
& \left\langle   \hat{p}_{i} \, \hat{p}_{j}  \right\rangle  \\
\end{pmatrix}
\end{equation}
and
\begin{equation}\label{eq:cov_mat_n-modes2b}
\sigma_{i} =
\begin{pmatrix}
\left\langle  2\hat{q}_{i}^2 \right\rangle 
& \!\!\!\!\!\!\left\langle  \{ \hat{q}_{i}, \, \hat{p}_{i} \} \right\rangle  \\
\left\langle    \{ \hat{q}_{i}, \, \hat{p}_{i} \} \right\rangle 
&\!\!\!\!\!\! \left\langle 2\hat{p}_{i}^2 \right\rangle \\
\end{pmatrix},
\end{equation}
$\{.,.\}$ denoting the anticommutator. A covariance matrix $\sigma$ satisfies the inequality
\begin{equation} \label{eq:uncertainty_relation}
\sigma+ i \, \mathds{J} \geq 0,
\end{equation}
which is a consequence of the canonical commutation relations and positivity of the density matrix \cite{simon1987gaussian, simon1994}. In particular, $\sigma$ is a positive matrix.

\subsection{Transformations of Gaussian states}
\label{sec:transfogaussian}
Partial tracing a Gaussian state is particularly simple. The covariance matrix of the reduced state is simply obtained by discarding the lines and columns corresponding to the modes over which the partial trace is done  (see, e.g., \cite{giedke2001quantum}).
For instance, the two-mode state obtained from \eqref{cmG} by tracing out all modes but $i$ and $j$ has covariance matrix
\begin{equation}\label{cmG2}
\sigma_{ij}=
\begin{pmatrix}
\sigma_i & \varepsilon_{ij} \\
\varepsilon_{ij}^{\rm \sss T } & \sigma_{j} 
\end{pmatrix},
\end{equation}
where the $2\times 2$ blocks are the same as the ones in \eqref{cmG}. In the same way, the reduced density matrix $\rho^{(i)}$ of mode $i$ obtained by tracing out all the other modes is a single-mode Gaussian state entirely specified by the covariance matrix $\sigma_i$. 

Let us now turn to the modification of the covariance matrix under a Bogoliubov transformation. It is important to stress here that we change operators but keep the same quantum state over which the averages $\langle \cdots \rangle$ are performed. For the vector of creation and annihilation operators $\mathbf{b}$, we denote by $\boldsymbol{\xi}_b$ the corresponding vector of position and momentum operators $\boldsymbol{\xi}_b = \sqrt2 \, (\hat{q}_1,\hat{p}_1,\ldots,\hat{q}_{\sss N},\hat{p}_{\sss N})^{\sss \rm T}$ with $\hat{q}_{j}=(\hat{b}_j+\hat{b}_j^\dagger)/\sqrt2$ and $\hat{p}_{j}=i(\hat{b}_j^\dagger-\hat{b}_j)/\sqrt2$. We thus have $\boldsymbol{\xi}_b = U \mathbf{b}$, with
\begin{equation}
\label{eq:Uxib}
U=\left(
\begin{array}{cccccccc}
 \frac{1}{\sqrt{2}} & 0 & 0 &\ldots& \frac{1}{\sqrt{2}} & 0 & 0&\ldots \\
 -\frac{i}{\sqrt{2}} & 0 & 0 &\ldots& \frac{i}{\sqrt{2}} & 0 & 0&\ldots \\
 0 & \frac{1}{\sqrt{2}} & 0 & \ldots&0 & \frac{1}{\sqrt{2}} & 0&\ldots \\
 0 & -\frac{i}{\sqrt{2}} & 0 &\ldots& 0 & \frac{i}{\sqrt{2}} & 0&\ldots \\
 0 & 0 & \ddots & \ldots&0 & 0 & \ddots&\ldots
\end{array}
\right)
\end{equation}
a $2N\times 2N$ unitary matrix. Similarly, $\boldsymbol{\xi}_c = U \mathbf{c}$. The Bogoliubov transformation $\mathbf{c}=\mathscr{T}\mathbf{b}$ then entails that $\boldsymbol{\xi}_c =\calst\boldsymbol{\xi}_b$ with 
\begin{equation}
    \calst=U\mathscr{T}U^\dagger\; .
\end{equation}
It can be proved that the matrix $\calst\in \text{Sp}(2N,\mathbb{R})$ is real and symplectic \cite{simon1994}.
Since this transformation is linear, we get from Eq.~\eqref{eq:covariance_matrix} that a Gaussian state with covariance matrix $\sigma_b$ in mode $b$ is a Gaussian state in mode $c$ with covariance matrix 
\begin{equation} \label{eq:transfo_sigma1}
\sigma_c = \calst\, \sigma_b \, \calst^{\sss \rm T}
\end{equation}
in mode $c$.

As guaranteed by Williamson theorem \cite{williamson1936algebraic}, it is always possible to find a symplectic transform that brings any covariance matrix $\sigma$ to a canonical diagonal matrix $\diag(\nu_1,...,\nu_N,\nu_1,...,\nu_N) $, which is unique up to the ordering of the $\nu_j$. The $\nu_j$ are called the {\it symplectic eigenvalues} of $ \sigma $. They can be directly obtained from the eigenvalues of the matrix $\mathds{J} \sigma$, which are given by $\pm i\nu_j$ \cite{serafini2006}. 
In terms of the $\nu_j $, given the uncertainty relation \eqref{eq:uncertainty_relation}, the positivity of $\rho$ is equivalent to
\begin{equation}
\nu_j\geq 1 \quad\quad  j=1,...,N\,.
\label{uncert-nu}
\end{equation}

\subsection{Thermal states}
\label{sec:thermality}
The symplectic eigenvalues have an appealing physical interpretation. Indeed, they can be related with the mean particle number of a thermal state. 

Recall that a generic (single-mode) thermal state is a state whose density matrix in the Fock space spanned by vectors $|n\rangle$ is of the form
\begin{equation} \label{eq:thermal_state}
\rho^{\rm \sss th}(a) = \frac{2}{a+1} \sum_{n=0}^{\infty} \left(\frac{a-1}{a+1}\right)^n \, |n \rangle \, \langle n |,
\end{equation}
with $a$ some parameter. Denoting as $\hat{n}$ the corresponding number operator, since $\bar{n}\equiv\langle\hat{n}\rangle=\tr(\rho^{\rm \sss th}\hat{n})=\tfrac12(a-1)$, the parameter $a$ is simply related with the mean particle number as $a=2\bar{n}+1$. The state $\rho^{\rm \sss th}(a)$ is in fact a Gaussian state with $2\times 2$ covariance matrix $\sigma^{\rm \sss th}=a\mathds{1}_2$. This means that a single-mode covariance matrix in diagonal form describes a thermal state with mean particle number $\bar{n}=(a-1)/2$ and symplectic eigenvalue $\nu=a$.  Another way of representing a thermal state is to set $\bar{n}=\sinh^2 r$, which yields
\begin{equation} \label{eq:reduced_density_matrix-BEC}
\rho^{\rm \sss th}(a) =   \frac{1}{\cosh^2 r} \sum_{n=0}^{\infty}  \left(\tanh r \right)^{2n}  \, |n\rangle\langle n|\,,\quad a=\cosh(2 r).
\end{equation}
The purity of $\rho^{\rm \sss th}(a)$ can be readily calculated from \eqref{eq:thermal_state} or \eqref{eq:reduced_density_matrix-BEC}; it reads $\tr[\rho^{\rm \sss th}(a)]^2=1/\cosh(2 r)=1/a$. The quantity $a$ being the inverse of the purity of a single-mode reduced density matrix, it is referred to as the {\it local mixedness} \cite{adesso2004a}. Note that the vacuum state is a thermal state $\rho^{\rm \sss th}(1)$ with mean occupation numbers $\bar{n}_j=0$ and local mixidness unity. 

More generally \cite{adesso2004}, a $N$-mode Gaussian state with arbitrary covariance matrix $\sigma$ can be brought to a product of thermal states
\begin{equation}
\label{eq:full_thermal_density_matrix}
\rho_{\boldsymbol{\nu}}= \overset{\sss N}{\underset{j=1}{\otimes}} \rho^{\sss \rm th}(\nu_{j}).
\end{equation}
Indeed, if $\cals$ is the symplectic transformation that diagonalizes $\sigma$ as $\diag(\nu_1,\nu_1,...,\nu_N,\nu_N)=\cals\sigma\cals^{\sss \rm T}$, then it can be realized on the Gaussian state by a unitary evolution generated by a quadratic Hamiltonian (see e.g.~\cite{simon1994}).
 Therefore Williamson's theorem ensures that any Gaussian state can be decomposed into a product of thermal states whose mean occupation number in mode $j$ is obtained from the symplectic eigenvalue $\nu_{j}$ as $\bar{n}_j=(\nu_j-1)/2$ \cite{simon1994}. The condition $\nu_j\geq 1$ in \eqref{uncert-nu} simply corresponds to the fact that the mean occupation numbers have to be positive.  Note that the purity of state \eqref{eq:full_thermal_density_matrix} is simply given in terms of the covariance matrix by $\tr(\rho_{\boldsymbol{\nu}})^2=1/\sqrt{\det\sigma}$.

\subsection{Vacuum as a three-mode Gaussian state}\label{sec.3mgs}
\label{sec:vac_three_mode}
\begin{figure}[!t]
\includegraphics[width=\linewidth]{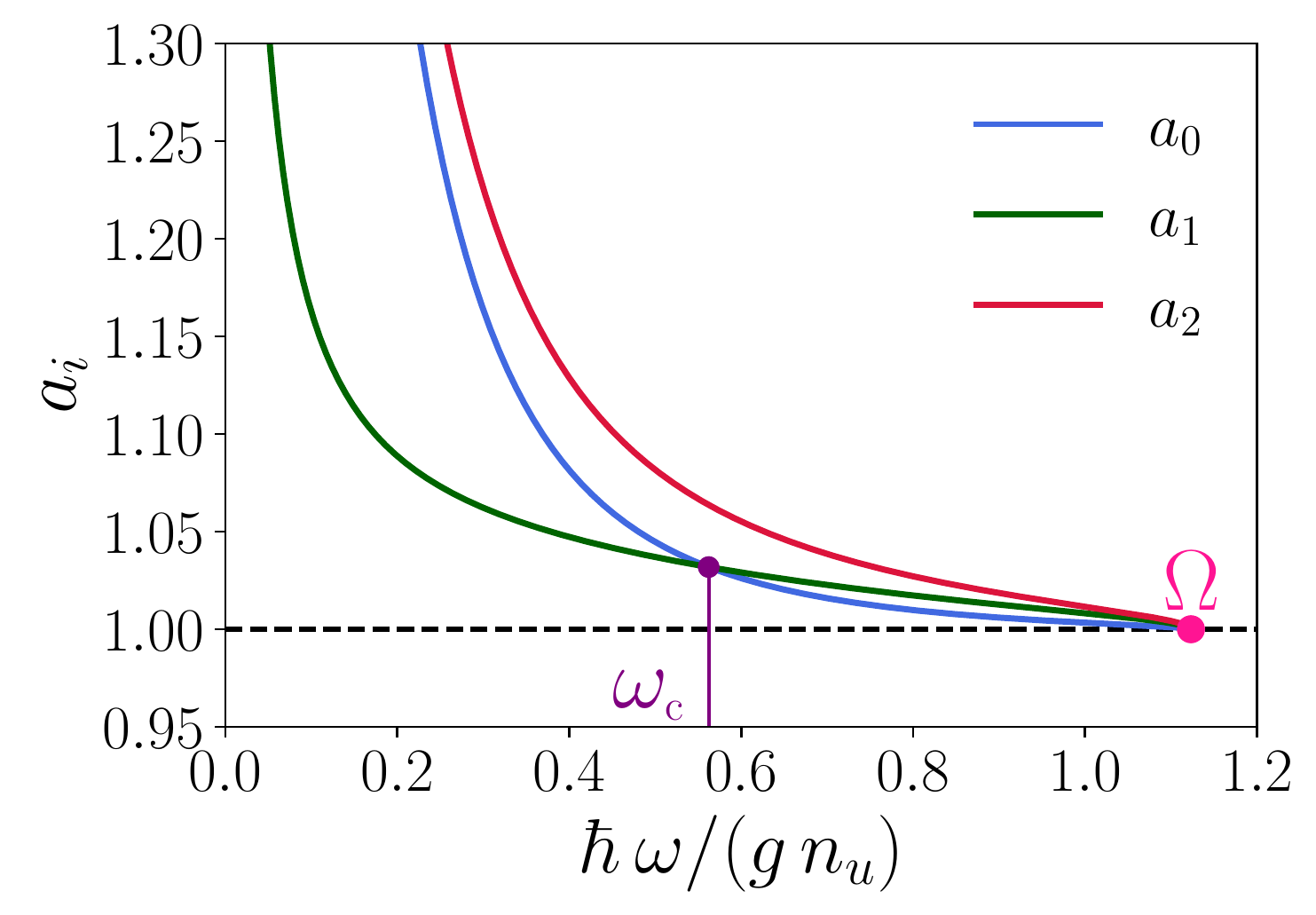}
\caption{Local mixedness $a_i(\omega)$ [see
  Eqs.~\eqref{eq:local_mixed_omega}] for each mode $0$, $1$ and $2$ as
  functions of the dimensionless quantity
  $\hbar \, \omega/ (g\, n_u)$, for a waterfall configuration with
  $m_u = 0.59$.  The frequency $\omega_{\rm c}$ indicates the turning
  point above which $a_{ 0}$ becomes lower than $a_{1}$. The
  upper-bound frequency $\Omega$ corresponds to the vanishing of the
  mode $2$ [see Eq.~\eqref{threshold}].}
\label{fig:LocalMixedness}
\end{figure}

The Bogoliubov transformation associated with the scattering process \eqref{eq:bog_transform-ch3} leads to a three-mode Gaussian pure state, given by \eqref{eq:our_state}. The covariance matrix of the vacuum $|0\rangle_b$ is the identity matrix $\mathds{1}_6$. Applying \eqref{eq:transfo_sigma1} we thus get that the covariance matrix of state \eqref{eq:our_state} is 
\begin{equation}\label{sigma_c}
\sigma_c=\calst\calst^{\sss \rm T}.
\end{equation}
Using the explicit expression of $\mathscr{T}$ derived from \eqref{eq:alph-beta}, and the explicit expression \eqref{eq:Uxib} for $U$, we obtain the $6\times 6$ matrix given in Eq.~\eqref{eq:cov-U_theta} of Appendix \ref{technicalpoints}. Note that using \eqref{eq:cov_mat_n-modes2a} and \eqref{eq:cov_mat_n-modes2b} the $2\times 2$ matrices $\sigma_i$ and $\varepsilon_{ij}$, $i \in \{0,1,2\}$, simply read
\begin{equation}
\label{eq:sigma_i_sf}
    \sigma_i = \left(1 + 2\,\langle \hat{c}_i^\dagger \hat{c}_i \rangle \right) \mathds{1}_2
\end{equation}
and (for $i\neq 2$)
\begin{equation}
\label{eq:epsilon_ij_sf}
\begin{split}
\varepsilon_{i2} & =
\begin{pmatrix}
2\, \text{Re}  \left\langle \hat{c}_i \, \hat{c}_2 \right\rangle  & 2\, \text{Im}  \left\langle \hat{c}_i \, \hat{c}_2 \right\rangle  \\
2\, \text{Im}   \left\langle \hat{c}_i \, \hat{c}_2 \right\rangle  &  -2\, \text{Re}  \left\langle \hat{c}_i \, \hat{c}_2 \right\rangle 
\end{pmatrix}, \\
\varepsilon_{01} & =
\begin{pmatrix}
2\, \text{Re}  \langle \hat{c}_0 \, \hat{c}_1^\dagger \rangle  & 2\, \text{Im}  \langle \hat{c}_0 \, \hat{c}_1^\dagger \rangle  \\
-2\, \text{Im}   \langle \hat{c}_0 \, \hat{c}_1^\dagger \rangle  &  2\, \text{Re}  \langle \hat{c}_0 \, \hat{c}_1^\dagger \rangle 
\end{pmatrix}.
\end{split}
\end{equation}
In the two-mode and three-mode cases, it is known \cite{duan2000, adesso2006a} that all pure Gaussian states can be brought by LLUBOs (local linear unitary Bogoliubov transformations) to a {\it standard form} where matrices $\sigma_{i}$ are proportional to the identity and matrices $\varepsilon_{ij}$ are diagonal. In order to get such a standard form, we use the set of operators $\hat{e}_j$ related with the $\hat{c}_j$ by $\mathbf{e}=\mathscr{R}\mathbf{c}$, where $\mathscr{R}$ has been defined in  \eqref{eq:defR}. Using the results of \ref{sec:transfogaussian} and applying Eq.~\eqref{eq:transfo_sigma1}, the covariance matrix of state \eqref{eq:link_vacua_set_e} in mode $e$ is 
\begin{equation}
\label{eq:sigma_e}
\sigma_e = \calsr \calst\calst^{\sss \rm T} \calsr^{\sss \rm T}.
\end{equation}
Note that $\calsr$ is a rotation operator. Indeed, one can easily show that $\calsr = \diag\{ R(\varphi_{02}),\, R(\varphi_{12}), \, R(-\varphi_{22})\}$, where  
\begin{equation}
R(\phi) = 
\begin{pmatrix}
 \cos\phi & \sin\phi
 \\
-\sin\phi & \cos\phi
\end{pmatrix}.
\end{equation}
Then, one proves that the covariance matrix $\sigma_e$ defined by  \eqref{eq:sigma_e} is in the standard form, with 
$\sigma_i$ given by \eqref{eq:sigma_i_sf} and
\begin{equation}
\label{eq:cov_mat_r-parameters2}
\begin{split}
\varepsilon_{ij}& = 2 \,|\langle \hat{c}_i \, \hat{c}_j^\dagger \rangle|  \mathds{1}_2 =  2 \, v_{i2} \, v_{j2} \, \mathds{1}_2 , \qquad  i,j=0, 1\\
\varepsilon_{i2}&= 2 \, |\langle \hat{c}_i \, \hat{c}_2 \rangle| \sigma_z = 2 \, v_{i2} \, v_{22} \, \sigma_z, \qquad  i=0, 1,
\end{split}
\end{equation}
where $\sigma_z$ is the third Pauli matrix.
Following the notation introduced in \autoref{sec:thermality} for thermal states, we define real parameters $r_i\geq 0$ and $a_i\geq 1$ such that
\begin{equation}\label{locmix1}
\bar{n}_i = 
\langle\hat{c}^{\dagger}_i\hat{c}_i\rangle  = \sinh^2(r_i) = \frac{a_i -1}{2}\; .
\end{equation}
To be completely accurate, we recall that the operators $\hat{c}_i = \hat{c}_i(\omega)$ all depend on the energy $\hbar \, \omega$ of the elementary excitations. Therefore, the above defined quantities $a_i$ also depend on $\omega$. They can be written explicitly as functions of the coefficients of the scattering matrix:
\begin{equation}
\label{eq:local_mixed_omega}
    \begin{split}
        a_0(\omega) & = 1 + 2\, |S_{02}(\omega)|^2 \; \\
        a_1(\omega)  & =  1 + 2\, |S_{12}(\omega)|^2 \;\\
        a_2(\omega)  & =  -1 + 2\, |S_{22}(\omega)|^2 \; \\
    \end{split}
\end{equation}
(see Appendix \ref{technicalpoints}).
From the solution of the scattering problem in the waterfall configuration, we calculate the scattering amplitudes $S_{ij}(\omega)$ following \cite{Larre2012}. This makes it possible to compute the three local mixednesses $a_{0}$, $a_{1}$ and $a_{2}$ as functions of the frequency.
In particular we have 
\begin{equation}
    \label{a0a1}
a_0(\omega)+a_1(\omega) = a_2(\omega) +1,
\end{equation}
which stems from relations \eqref{eq:local_mixed_omega} and \eqref{abh6}.
Figure \ref{fig:LocalMixedness} shows the associated curves. Here, these coefficients are computed for a waterfall configuration with downstream Mach number $m_d=2.9$, which is the one for which the experiment of \cite{de_nova_2019} has been realized. In our case, this corresponds to an upstream Mach number $m_u=0.59$. 

We can identify two regimes in Fig.~\ref{fig:LocalMixedness}: below a
frequency denoted $\omega_{\rm c}$ the lowest of the three parameters
is $a_{ 1}$; above this frequency, the minimum value becomes $a_{
  0}$. The value of this frequency is determined numerically and is
equal to $\omega_c \approx 0.56 \, g\, n_u/\hbar$ for $m_u=0.59$. We observe
that the ratio $\omega_{\rm c}/\Omega$ (where $\Omega$ is the
frequency \eqref{threshold} at which mode $2$ vanishes and also
depends on $m_u$) decreases when $m_u$ decreases. The local
mixednesses $a_0$, $a_1$ and $a_2$ go to 1 when $\omega \to \Omega$,
which means that the populations of all modes vanish.

Using Eqs.~\eqref{eq:local_mixed_omega} one may rewite
expressions 
\eqref{eq:sigma_i_sf} and \eqref{eq:cov_mat_r-parameters2} in terms of
the $a_i$ as
\begin{equation}
\begin{split}
\label{eq:cov_mat_a-parameters}
\sigma_{i} &=  a_i \,  \mathds{1}_2, \qquad\qquad\qquad\qquad \,\, i=0, 1,2\\
\varepsilon_{ij}&=  \sqrt{a_i-1} \sqrt{a_j-1} \,
\mathds{1}_2, \qquad i,j=0, 1\quad (i\neq j)\\
\varepsilon_{i2}&= \sqrt{a_i-1} \sqrt{a_2+1} \, \sigma_z, \qquad i=0, 1.
\end{split}
\end{equation}
The $6\times 6$ covariance matrix defined by
\NP{Eqs.~\eqref{cmG} and}
\eqref{eq:cov_mat_a-parameters} is no longer the covariance matrix
associated with modes $\mathbf{c}$, but the covariance matrix associated
with modes $\mathbf{e}$ defined by Eq.~\eqref{eq:def_R}; since $\mathbf{c}$ and
$\mathbf{e}$ only differ by phases, the entanglement properties are
the same. When considering entanglement in Sec.~\ref{sec:entanglement}
we will therefore use the standard form
\eqref{eq:cov_mat_a-parameters}. In the case of a pure three-mode
Gaussian state, the three local symplectic invariants $a_i$ fully
determine the entanglement content of any given bipartition
\cite{adesso2006a}. As we shall see in \autoref{sec:entanglement},
the blocks of the covariance matrix $\sigma$ in the form of
expressions \eqref{eq:cov_mat_a-parameters} are the key ingredients to compute
the amount of bipartite and tripartite entanglement.

As mentioned in \autoref{sec:transfogaussian}, $\sigma_{i}$ is the covariance matrix of the reduced state $\rho^{(i)}$ of mode $i$. Given its diagonal form, one gets from  \autoref{sec:thermality} that $\rho^{(i)}$ is a thermal state with local mixedness $a_i$. 
It can also be considered as a reduced state of a two-mode squeezed state with squeezing parameter $r_i$ \cite{agarwal2012quantum, Molmer2008,Perrier2019}. In this respect, the study of the reduced state $\rho^{(0)}$ is of particular interest in the context of analogue gravity, since the number of emitted quanta in the $0$ mode gives access to the Hawking radiation spectrum. In the context of general relativity, this spectrum is exactly Planckian\footnote{We do not consider here possible effects of s grey body factor. These will be accounted for in \autoref{sec:entanglement_localization}.} \cite{hawking1974,hawking1975}, with a temperature which is called the "Hawking temperature". For the analogue model we consider, dispersive effects significantly affect this result. Indeed, if one defines an effective temperature $T^{(0)}_{\rm eff}$ such that:
 \begin{equation}\label{eq:Teff1}
     \bar{n}_0(\omega)=\frac{1}{\exp(\hbar \omega/T^{(0)}_{\rm eff}) - 1},
 \end{equation}
one finds from \eqref{eq:reduced_density_matrix-BEC}, \eqref{locmix1} and \eqref{eq:local_mixed_omega} that $T_{\rm eff}^{(0)}$ is frequency-dependent:
\begin{equation}\label{eq:Teff2}
    T_{\rm eff}^{(0)}(\omega)=\frac{\hbar \omega}{2\, \ln[\coth (r_0(\omega))]}
    =\frac{\hbar \omega}{\ln[1+|S_{02}(\omega)|^{-2}]}
    \; .
\end{equation}
\begin{figure}
    \centering
    \includegraphics[width=\linewidth]{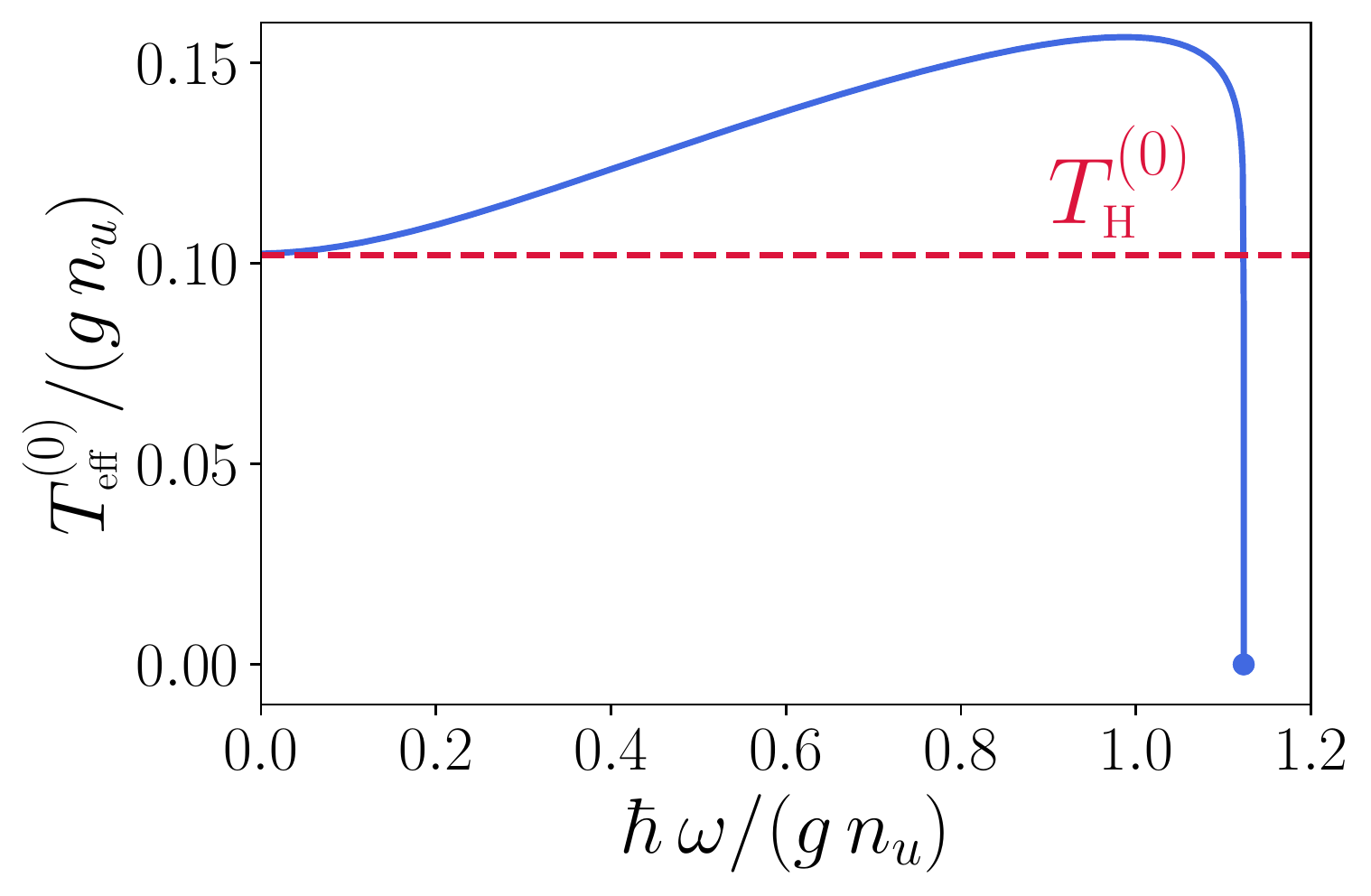}
    \caption{Blue continuous curve: Effective temperature $T_{\rm eff}^{(0)}$ defined in Eq.~\eqref{eq:Teff2} plotted as a function of the frequency $\omega$ for a waterfall configuration with
  $m_u = 0.59$. The dashed red line is the Hawking temperature $T_{\rm\sss H}^{(0)}$ given by \eqref{eq:Th0}.}
    \label{fig:T_eff}
\end{figure}
Figure \ref{fig:T_eff} represents $T_{\rm eff}^{(0)}(\omega)$  for a waterfall configuration with $m_u=0.59$. We note here that the same type of results has been obtained numerically in \cite{Macher2009}. In the long wavelength limit the effective temperature tends to a constant analogue Hawking temperature $T_{\rm\sss H}^{(0)} = 
\lim_{\omega\to 0} T^{(0)}_{\rm eff}(\omega)$. Based on the expansion \eqref{eq:Sloww}
and on the formula \eqref{eq:F02} one gets
\begin{equation}\label{eq:Th0}
    \frac{T_{\rm\sss H}^{(0)}}{g n_u} = 2\,
\frac{{m}_{u}(1-{m}_{u})^{\frac{3}{2}}
(1+{m}_{u}^{2})^{\frac{3}{2}}}{(1+{m}_{u})^{\frac{1}{2}}
(1+{m}_{u}+{m}_{u}^{2})^{2}}\; .
\end{equation}
This long wavelength determination of the analogue Hawking temperature is physically sound, since the reduced density matrix of mode 0 is indeed thermal (in the sense of \autoref{sec:thermality}). However, it has the drawback of depending of the mode considered (here the outgoing Hawking mode). The reduced density matrices of modes 1 and 2 are also thermal, and on the basis of the present reasoning there is another, different Hawking temperature for the Companion (which could be denoted as $T_{\rm\sss H}^{(1)}$), and still another one for the Partner ($T_{\rm\sss H}^{(2)}$). In \autoref{sec:entanglement_localization} we use a different reasoning and argue that $T_{\rm\sss H}^{(2)}$ gives a more satisfactory definition of the Hawking temperature, valid for the whole system.

\section{Entanglement in three-mode Gaussian states}
\label{sec:entanglement}

Entanglement detection and characterization has attracted a great deal of effort in the past two decades, as it has been identified as a key resource for quantum information processing \cite{horodecki2009quantum}. 
A quantum state is entangled if it is not separable, i.e., if it cannot be written as a convex sum of product states \cite{werner1989quantum}. 
One of the simplest necessary separability criteria is given by the positivity of the partial transpose (PPT), first proposed for discrete variables \cite{peres1996, horodecki1996} and extended to the continuous case in \cite{simon2000}.
 A wealth of entanglement measures were discussed in the literature, both for discrete and continuous variables. For bipartite pure states, quantitative measures of entanglement include the entanglement entropy (which can be shown to be unique if some additional natural requirements are imposed) \cite{popescu1997thermodynamics}, or the concurrence \cite{hill1997entanglement}. In the mixed state case, it is possible to construct 'good' entanglement measures in many different ways, which are inequivalent in the sense that they lead to different orderings of entangled states \cite{horodecki2009quantum}. A possible way is to extend measures for pure states via a convex roof construction: entanglement of a mixed state is then defined by a minimization over all its possible pure state decompositions. For instance, entanglement entropy generalizes for mixed states to the entanglement of formation \cite{bennett1996mixed}.

A striking difference between classical correlations and quantum entanglement is that the latter is monogamous \cite{Theral2004, Koashi2004}. This means that a particle which is maximally entangled with a Partner cannot be entangled with a third party, or in other words that any amount of entanglement shared with a particle limits the entanglement that can be shared with another particle. In the case of three qubits, this limitation to bipartite entanglement was expressed in \cite{coffman2000} through an inequality that must be satisfied by an entanglement measure called the concurrence, or more precisely by its square, the tangle. This monogamy inequality was later generalized to 
an arbitrary number of qubits \cite{osborne2006general}, to three qutrit systems \cite{li2017} and to continuous variables \cite{adesso2006,Zarate2017}, as we now discuss. 

In the case of continuous variables, to which the situation of black hole analogues pertains, Gaussian states are the most natural objects with which one is led to deal.
From a qualitative point of view, entanglement can be detected by the PPT criterion, which is a necessary and sufficient separability condition for $1\times N$-mode Gaussian states \cite{simon2000}; the three-mode case, which is relevant to our situation, was investigated in \cite{giedke2001separability}, and will be considered in  \autoref{subsecbipent}.
From  a  quantitative  point  of  view, entanglement can be measured by the logarithmic negativity, which quantifies by which amount the PPT criterion is violated \cite{vidal2002}.\footnote{For continuous variables it is generally highly difficult to make use of the convex roof construction, both analytically and numerically, as the optimization has to take place over all pure state decompositions. To circumvent this issue, Gaussian entanglement of formation was defined in \cite{wolf2004gaussian}, restricting the convex roof construction to Gaussian pure state decompositions. This quantity provides an upper bound for the entanglement of formation, and is more amenable to calculations. In \cite{adesso2005,hiroshima2007}
it was shown that Gaussian entanglement of formation and entanglement measured by negativity are inequivalent measures.}
In \cite{adesso2006}, it was proposed
to construct a specific measure of entanglement, the {\it contangle} (continuous tangle), defined as the convex roof extension of the square of the logarithmic negativity. In that manner, the monogamy inequality expressed by this measure also holds for Gaussian states. The amount by which both sides of the monogamy inequality differ provides an estimate for multipartite entanglement. In the present section, we will make use of this measure of entanglement to quantify tripartite entanglement in our analog black hole system. For consistency purposes we shall also quantify bipartite entanglement using the contangle.

In the domain of analogue gravity, previous approaches have already considered quantitative measures of entanglement. Using a relation \cite{klich2009quantum} between entanglement entropy and cumulants of the full counting statistics, Ref.~\cite{Giovanazzi2011} expresses the (long wave-length limit of the) entanglement entropy of a pure two-mode Gaussian state in terms of number fluctuations in a given region. Although this approach bears some similarity with the one we discuss at the end of \autoref{sec:measuring_entanglement_finiteT} we cannot directly compare it with ours because for our three-mode Gaussian state the reduced two-body state is mixed. Ref.~\cite{bruschi_2013} studies the dynamical Casimir effect in a BEC and quantifies the non-separability after a quench by means of the entanglement of formation, which takes an exact analytic expression for symmetric two-mode Gaussian states \cite{giedke2003entanglement}. Although we cannot directly compute this quantity\footnote{In our case the reduced state of modes $i$ and $j$ is non-symmetric, since in general $a_i\neq a_j$ ($i$ and $j$ in $\{0,1,2,\}$).}, the spirit of our approach is similar to theirs and to the one of Refs.~\cite{Horstmann_2011,Jacquet2020}, which use the symplectic spectrum to construct quantitative measures of entanglement in the context of ion rings and nonlinear optics analogue, respectively.

\subsection{Bipartite entanglement}
\label{subsecbipent}

The criterion usually used to detect entanglement in bipartite systems is the Peres-Horodecki (or PPT) criterion \cite{peres1996,horodecki1996}. It is a necessary and sufficient separability condition for bipartite $1\times (N-1)$-mode Gaussian states \cite{simon2000}. This corresponds to all possible bipartitions occurring in three-mode states: indeed, we will have to consider either bipartitions $i|jk$ or, after tracing out mode $k$, bipartitions $i|j$.
This criterion states that a state $\rho$ is separable if and only if its partial transpose $\rho^{\sss \rm PT}$ with respect to the first mode (mode $i$ in the above notation) is positive. Partial transposition of an $N$-mode Gaussian state is equivalent to mirror reflection in phase space for the Wigner function \cite{simon2000}. The covariance matrix of $\rho^{\sss \rm PT}$ is given by
\begin{equation}
\label{eq:transposed_sigma}
\sigma^{\rm \sss PT} = \Lambda \, \sigma \, \Lambda, 
\quad \text{with} \quad  \Lambda =  \sigma_z \oplus \mathds{1}_{2N-2}.
\end{equation}
According to the criteria \eqref{uncert-nu}, the necessary and sufficient separability criterion $\rho^{\sss \rm PT}\geq 0$ is equivalent to
\begin{equation}
\nu_j^{\sss \rm PT}\geq 1, \quad\quad  j=1,\ldots,N\; ,
\label{ppt}
\end{equation}
where  $\nu_j^{\sss \rm PT}$ are the symplectic eigenvalues of $\sigma^{\rm \sss PT}$.

In our case $N=3$. Let us investigate bipartite entanglement of two-mode states obtained by tracing out the third one. As discussed in \autoref{sec:transfogaussian}, the covariance matrix associated with the two-mode state $i,j$ obtained by tracing out mode $k$ is $\sigma_{ij}$ given by \eqref{cmG2}. Its symplectic eigenvalues $\nu_\pm$ are given by 
\begin{equation} \label{eq:nu-2mode}
2 \, \nu_{\pm}^2 = \Delta_{ij} \pm \sqrt{\Delta_{ij}^2 - 4 \, \det \, \sigma_{ij}}
\end{equation}
with $\Delta_{ij} = \det \sigma_i + \det \sigma_j + 2 \det \varepsilon_{ij}$ \cite{serafini2004}. The symplectic eigenvalues  $\nu_\pm^{\sss \rm PT}$ of $\sigma_{ij}^{\sss \rm PT}$ associated with the partial transpose are given by
\begin{equation} \label{eq:nu-2mode_transpose}
2 \, (\nu_{\pm}^{\sss \rm PT})^2 = \Delta_{ij}^{\sss \rm PT} \pm \sqrt{(\Delta^{\sss \rm PT}_{ij})^2 - 4 \, \det  \sigma_{ij}},
\end{equation}
with $\Delta_{ij}^{\sss \rm PT} = \det \sigma_i + \det \sigma_j - 2 \det \varepsilon_{ij}$.
For a two-mode state, the PPT criterion is in fact equivalent to condition $\nu_{-}^{\rm \sss PT}\geq 1$ only, since $\nu_{+}^{\rm \sss PT}$ is always larger than 1 \cite{adesso2004}. 

From \eqref{eq:cov_mat_a-parameters} one readily derives the expressions for $\nu_{-}^{\rm \sss PT}$ in our case. Note that, again, since the local mixednesses appearing in Eq.~\eqref{eq:cov_mat_a-parameters} depend on the frequency $\omega$, the lowest symplectic eigenvalue $\nu_{-}^{\rm \sss PT}$ also depends on $\omega$.
By using the fact that $a_i \geq 1$, $i = 0,1,2$ and the relation \eqref{a0a1}, one can prove easily that $\nu_{-}^{\rm \sss PT}\geq 1$ for the bipartition $0|1$, independently of the frequency. Therefore, the reduced state of modes $0|1$ is always separable: the Hawking quantum and the Companion are not entangled. On the other hand, the eigenvalues $\nu_{-}^{\sss \rm PT}$ of the reduced covariance matrices $\sigma_{02}^{\rm \sss PT}$ and $\sigma_{ 12}^{\rm \sss PT}$ are lower than 1, which implies that the reduced state of modes $0|2$ (Hawking--Partner) and $1|2$ (Companion--Partner) is entangled for all frequencies $\omega$, see for instance Fig.~\ref{fig:bip_ent_finiteT}(a), where the blue curve represents $1 - \nu_{-}^{\sss \rm PT}(\omega)$ computed for the reduced state of modes $0|2$.


The same results are obtained with the "Cauchy-Schwarz criterion" (see , e.g.,  Ref.~\cite{walls-milburn}) which has been often used in the context of analogue gravity
\cite{deNova2014,Busch2014b,Busch2014a,Boiron2015,Fabbri2018,Coutant2018}.
 According to this criterion modes $i$ and $j$ are entangled if the following inequality is verified
\begin{equation} \label{eq:CS_inequality}
\begin{cases}
|\langle \hat{c}_{i} \, \hat{c}_{j} \rangle|^2 > \langle \hat{c}_{i}^{\dagger} \, \hat{c}_{i} \rangle
\, \langle \hat{c}_{j}^{\dagger} \, \hat{c}_{j} \rangle, & \, \text{for} \; i \in \{0,1\}, \, j=2, \\[2mm]
\langle \hat{c}_{i} \, \hat{c}_{j}^{\dagger}  \rangle|^2 > \langle \hat{c}_{i}^{\dagger} \, \hat{c}_{i} \rangle
\, \langle \hat{c}_{j}^{\dagger} \, \hat{c}_{j} \rangle, & \,  \text{for} \; i\ne j \in \{0,1\}.
\end{cases}
\end{equation}
Using Eqs.~\eqref{eq:cov_mat_r-parameters2} and \eqref{eq:cov_mat_a-parameters}, one finds $\langle \hat{c}_{i}^{\dagger} \, \hat{c}_{i} \rangle
\, \langle \hat{c}_{j}^{\dagger} \, \hat{c}_{j} \rangle = \sinh^2r_i \, \sinh^2r_j$, $|\langle \hat{c}_{i} \, \hat{c}_{2} \rangle|^2 = \sinh^2r_i \, \cosh^2r_2$ ($i\neq 2$) and $|\langle \hat{c}_{0} \, \hat{c}_{1}^\dagger \rangle|^2 = \sinh^2r_0 \, \sinh^2r_1$. Therefore, when considering the bipartition $0|1$, one concludes immediately that the second inequality of \eqref{eq:CS_inequality} is never true; one has instead the equality $|\langle \hat{c}_{0} \, \hat{c}_{1}^\dagger \rangle|^2 =\langle \hat{c}_{0}^{\dagger} \, \hat{c}_{0} \rangle
\, \langle \hat{c}_{1}^{\dagger} \, \hat{c}_{1} \rangle$ for all frequencies $\omega$. Therefore, the reduced state $0|1$ is separable. For bipartitions $0|2$ and $1|2$, since $\tanh(r_2)<1$ (with $r_2>0$, finite), the first inequality of \eqref{eq:CS_inequality} is always true. The criterion of violation of the Cauchy-Schwarz inequality thus leads to the same conclusion as the PPT criterion for the reduced states $0|2$ and $1|2$: these states are always entangled.

However, the Cauchy-Schwarz criterion does not give any clue about the amount of entanglement shared by each bipartition. Indeed, as will be discussed in \autoref{sec:measuring_entanglement_finiteT}, in an experimental setup for which the temperature of the system cannot be exactly equal to zero, a stronger violation of the Cauchy-Schwarz inequality does not necessarily imply a greater amount of entanglement.



\subsection{Tripartite entanglement}
\label{sec:tripartite_entanglement}

\subsubsection{Monogamy inequality}
Monogamy is a fundamental property of entanglement correlations. It can be described by monogamy inequalities, which in the case of a
tripartite system with subsystems labelled by $(i,j,k)$ takes the form
\begin{equation}
E^{(i|jk)} - E^{(i|j)} - E^{(i|k)}\geq 0
\label{mineq}
\end{equation}
where $E^{(A|B)}$ is a proper measure of bipartite entanglement between subsystems $A$ and $B$ (nonnegative on separable states and monotonic under (G)LOCC). 
This inequality expresses the fact that the total amount of entanglement that can be shared between $i$ and $j$ and between $i$ and $k$ is upper bounded by the amount of entanglement between $i$ and $jk$ taken as a whole.
The left-hand side of inequality \eqref{mineq} provides a quantifier of genuine tripartite entanglement. 

Not all entanglement measures satisfy a monogamy inequality. However, it is possible to find and construct proper measures of entanglement which satisfy these relations, both in the qubit case and in the continuous-variable case. In the case of qubits, the monogamy inequality holds for entanglement measured by the square of the concurrence. 
For Gaussian states a measure satisfying \eqref{mineq} was constructed in \cite{adesso2006}; it is called the contangle $E_\tau$ and it corresponds to the squared logarithmic negativity. For an arbitrary pure state $\rho=\vert \psi \rangle\langle \psi\vert$ with covariance matrix $\sigma^p$ ($ p $ stands for pure), it is defined as
\begin{equation}
\label{eq:def_log_neg}
E_\tau(\sigma^p)=\left(\ln \Vert \rho^{\sss \rm PT}\Vert_1\right)^2,
\end{equation}
where $ \Vert \hat{O}\Vert_1 = \mathrm{tr}\sqrt{\hat{O}^{\dagger}\hat{O}}$ is the trace norm.

The state considered in our case is a pure three-mode Gaussian state; thus, any bipartition $i|jk$ is a pure state, for which the term $E^{(i|jk)}$ in \eqref{mineq} can be computed easily (see \autoref{psc} below). On the contrary, the two other terms of \eqref{mineq} correspond to reduced two-mode states, which are mixed. The squared logarithmic negativity can be extended to mixed states by taking the infimum over all convex decompositions of $ \rho $ in terms of pure states $ \lbrace \vert\psi_i\rangle\rbrace$. In order to get a quantity more amenable to computations, the Gaussian contangle $G_{\tau}$ was defined by restricting this convex-roof construction to decompositions over pure Gaussian states only. The Gaussian contangle can be expressed as
\begin{equation}
\label{eq:G_tau}
G_{\tau}(\sigma)=\underset{\sigma^p\leq \sigma}{\inf}E_\tau(\sigma^{p}),
\end{equation}
where the notation $ \sigma^p\leq \sigma $ means that the matrix $ \sigma-\sigma^p $ is positive semidefinite. It is an upper bound to the true contangle $ E_\tau $ obtained from unrestricted pure-state decompositions, but for pure states  both coincide.

For three qubits the residual tangle (or three-way tangle) $E^{(i|jk)} - E^{(i|j)} - E^{(i|k)}$ provides a measure of tripartite entanglement. It has an explicit expression \cite{coffman2000}, which is symmetric in the three qubits. The corresponding quantity in the continuous case is no longer symmetric in the three modes. One can however define a permutation-invariant quantity by minimizing it over all permutations of the modes \cite{adesso2006}. This measure of tripartite entanglement shared among Gaussian modes was called residual contangle \cite{adesso2006a}. Its explicit expression reads
\begin{equation}
G_{\tau}^\rmres=G_{\tau}^{(i\vert j\vert k)}= \underset{i,j,k}{\min} \left(G_{\tau}^{(i\vert jk)}-G_{\tau}^{(i\vert j)}-G_{\tau}^{(i\vert k)}\right).
\label{gctres}
\end{equation}

\subsubsection{Pure-state contangle}
\label{psc}
Let us consider first a bipartition $i|jk$. For a pure state, the Gaussian contangle $G_\tau^{(i|jk)}$ coincides with the true contangle $E_{\tau}^{(i|jk)}$. In general, for a multimode Gaussian state $\vert \psi \rangle$ with covariance matrix $\sigma^{p}$ and generic bipartition $ i_1 \ldots i_{N-1}\vert i_N $ ($N=3$ in our case), the squared logarithmic negativity can be written as \cite{adesso2006}
\begin{equation}
\label{eq:log_neg_sympl_eigen}
 E_\tau^{i_1 \ldots i_{N-1}\vert i_N}(\sigma^{p})=   \left( \sum_{j:\nu_j^{\rm \sss PT}<1}  \ln \nu_j^{\rm \sss PT} \right)^2,
\end{equation}
where $\nu_j^{\rm \sss PT}$ are the symplectic eigenvalues associated with the partial transpose state $\rho^{\rm \sss PT}$. 

It is actually possible to write Eq.~\eqref{eq:log_neg_sympl_eigen} in terms of the local mixedness $a_{i_N}$ associated with mode $i_N$. Indeed, for any covariance matrix $\sigma$ associated with a pure multimode Gaussian state and generic bipartition $ i_1 \ldots i_{N-1} \vert i_N $, there exists a local symplectic transformation $\cals$ such that
\cite{botero2003modewise} 
\begin{equation}
\label{eq:sympl_transformation_Botero}
\cals \, \sigma  \, \cals^{\rm \sss T}= \mathds{1}_{2\, (N-2)} \oplus \sigma_{ \rm \sss sq} , 
\end{equation}
where $\sigma_{ \rm \sss sq}$ is the covariance matrix of a two-mode squeezed state and reads
\begin{equation} \label{eq:cov-squeezed-BEC}
\footnotesize
\begin{pmatrix}
a_{i_N} & 0 & \sqrt{a_{i_N}^2-1} & 0\\
0 & a_{i_N} & 0  & -\sqrt{a_{i_N}^2-1}   \\
\sqrt{a_{i_N}^2-1} & 0 & a_{i_N} & 0\\
0 & -\sqrt{a_{i_N}^2-1} & 0  & a_{i_N}   \\
\end{pmatrix}.
\end{equation}
In the case of a tripartite system ($N=3$), a direct proof of Eqs.~\eqref{eq:sympl_transformation_Botero}--\eqref{eq:cov-squeezed-BEC}, as well as explicit expressions of the symplectic matrix $\cals$ for each bipartition , $12|0$ and $02|1$ and $01|2$, can be found in Appendix \ref{app:tripartite_eigenvalues}. 

The symplectic eigenvalues of $\sigma^{\rm \sss PT}$ (corresponding to taking the partial transpose with respect to mode $i_N$) are 
then readily obtained from \eqref{eq:nu-2mode_transpose}, using the form \eqref{eq:cov-squeezed-BEC}; the symplectic eigenvalue 1 has degeneracy $2\, (N-2)$, while the ones associated with \eqref{eq:cov-squeezed-BEC} are $e^{\pm 2 r_{i_N}}$, with twofold degeneracy.
They can be related to the local mixedness $a_{i_N}$ through the relations $a_{i_N} = \cosh (2 \, r_{i_N})$. 
Equation \eqref{eq:log_neg_sympl_eigen} then gives
\begin{equation}
\label{eq:log_neg}
E_\tau^{i_1 \ldots i_{N-1}\vert i_N}(\sigma^{p}) =\arsinh^2\left(\sqrt{a_{i_N}^2-1}\,\right)= 4 \, r_{i_N}^2,
\end{equation}
which only depends on the local mixedness of mode $i_N$ and has a simple expression in terms of $r_{i_N}$. We will perform explicit calculations for our system in the next section.

\subsubsection{Residual contangle}\label{sec:res_contangle}

Equation \eqref{eq:log_neg} provides an explicit expression
for the first term $G_{\tau}^{(i\vert jk)}$ in \eqref{gctres}. For a pure three-mode Gaussian state, an explicit expression
of $G_\tau^{(i|j)}$ and $G_\tau^{(i|k)}$ can also be obtained. Indeed, in this specific case, any reduced two-mode state saturates the uncertainty relation \eqref{eq:uncertainty_relation} and belongs to a class of states called Gaussian least entangled mixed states (GLEMS). For GLEMS, one has \cite{adesso2006a}
\begin{equation}
\label{gtauij}
    G_\tau^{(i|j)} = \arsinh^2 \, \left[\sqrt{m^{\rm \sss GLEMS}(a_i,a_j,a_k)-1}\right],
\end{equation}
where $m^{\rm \sss GLEMS}$ can be explicitly calculated as a function of the three local mixednesses, as shown in Appendix \ref{app:app_C}. In our case [see Eqs.~\eqref{eq:G_appD}--\eqref{eq:m_theta_star_appD}], we obtain
\begin{equation} \label{eq:G_01_02_12}
\begin{split}
G_\tau^{(0|1)} & = 0, \\
 G_\tau^{(j|2)} & = \arsinh^2 \left( \frac{2 }{1+a_k} \sqrt{ (a_2+1)\, (a_j-1)} \right) \\
 & = \arsinh^2 \left( \frac{2 \,|\langle \hat{c}_j \, \hat{c}_2 \rangle| }{1+ \langle \hat{c}_k^\dagger \, \hat{c}_k \rangle}  \right),
\end{split}
  \end{equation}
with $j=0, \, k=1$ or $j=1, \, k=0$. 

Let us now introduce the quantity
\begin{equation}
\label{eq:G_res_tau_i}
    G_\tau^{\rmres(i)} = G_{\tau}^{(i\vert jk)}-G_{\tau}^{(i\vert j)}-G_{\tau}^{(i\vert k)},
\end{equation}
such that the residual contangle is given by
\begin{equation}\label{eq:Gresmin}
 G_\tau^\rmres= \underset{i\in \{0,1,2\}}{\min} \left[ G_\tau^{\rmres(i)} \right].
\end{equation}
Using \eqref{eq:log_neg} and \eqref{eq:G_01_02_12},  Eq.~\eqref{eq:G_res_tau_i} yields
\begin{equation} \label{eq:G_res_tau_0}
\begin{split}
    G_\tau^{\rmres(0)} & = \arsinh^2\left(\sqrt{a_{0}^2-1}\,\right) \\ 
   & - \arsinh^2 \left( \frac{2 }{1+a_1} \sqrt{ (a_2+1)\, (a_0-1)} \right),
    \end{split}
\end{equation}
\begin{equation} \label{eq:G_res_tau_1}
\begin{split}
    G_\tau^{\rmres(1)} & = \arsinh^2\left(\sqrt{a_{1}^2-1}\,\right) \\ 
   & - \arsinh^2 \left( \frac{2 }{1+a_0} \sqrt{ (a_2+1)\, (a_1-1)} \right),
    \end{split}
\end{equation}
and
\begin{equation} \label{eq:G_res_tau_2}
\begin{split}
    G_\tau^{\rmres(2)} & = \arsinh^2\left(\sqrt{a_{2}^2-1}\,\right) \\ 
   & - \arsinh^2 \left( \frac{2 }{1+a_1} \sqrt{ (a_2+1)\, (a_0-1)} \right) \\
   & - \arsinh^2 \left( \frac{2 }{1+a_0} \sqrt{ (a_2+1)\, (a_1-1)} \right).
    \end{split}
\end{equation}
The residual contangle only depends on the three local mixednesses $a_0$, $a_1$ and $a_2$ (this is no longer true at finite temperature, see \autoref{sec:detec_ent_th} and Appendix \ref{technicalpoints}). The minimum over all possible permutations of $i,\, j$ and $k$ in Eq.~\eqref{eq:Gresmin} can be obtained by choosing as reference mode $i$ the one with smallest local mixedness \cite{adesso2006a}.

\begin{figure}[!t]
\includegraphics[width=\linewidth]{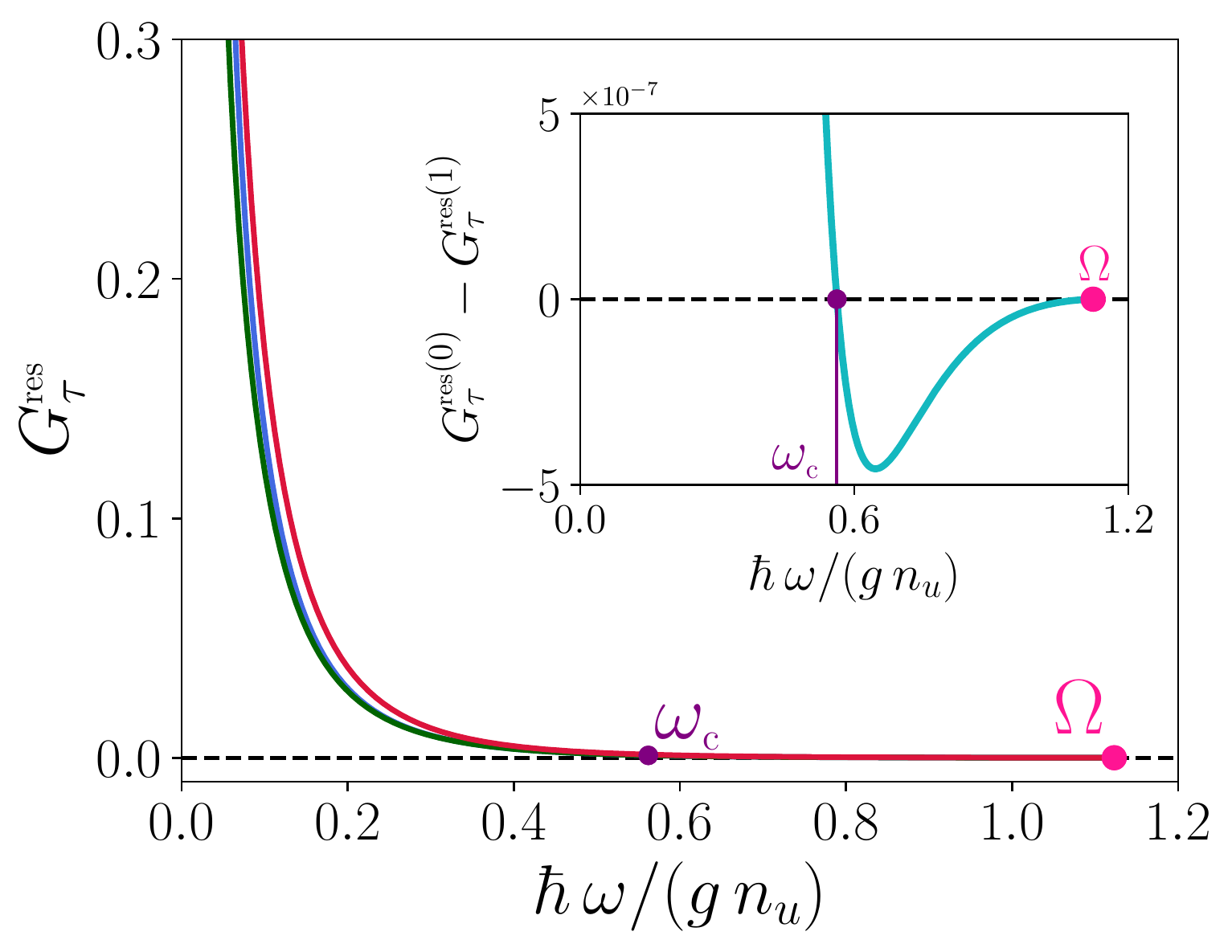}
\caption{Residual contangles $G_\tau^{\rmres(0)}$ (blue), $G_\tau^{\rmres(1)}$ (green), $G_\tau^{\rmres(2)}$ (red). The upper-bound frequency $\Omega$ corresponds to the vanishing of the mode $2$. The frequency $\omega_{\rm c}$ is the value above which $a_{ 0}$ becomes lower than $a_{1}$ (see Fig.~\ref{fig:LocalMixedness}), and coincides with the point above which $G_\tau^{\rmres(0)} < G_\tau^{\rmres(1)}$. The inset displays the difference $G_\tau^{\rmres(0)} - \, G_\tau^{\rmres(1)}$ (cyan). }
\label{fig:TripartiteEntanglement}
\end{figure}

We can then compute the residual Gaussian contangle for our three-mode Gaussian state using the expression of the $a_i$'s given in \eqref{eq:local_mixed_omega}. The results for $m_u=0.59$ are shown\footnote{These results have been previously presented in \cite{Isoard_PhD}.} in Fig.~\ref{fig:TripartiteEntanglement}. 
Tripartite entanglement naturally emerges from quantum fluctuations around a sonic horizon and diverges when the energy goes to zero. This divergence always comes from the first term in Eqs.~\eqref{eq:G_res_tau_0}, \eqref{eq:G_res_tau_1} and \eqref{eq:G_res_tau_2}. Indeed, this term diverges as $\ln^2 \omega$ (see discussion in Sec.~\ref{sec:exp_persp}).
On the other hand, it may be proven that $G_\tau^{(j|2)}$ given by expressions \eqref{eq:G_01_02_12} for $j=0,1$ is bounded at zero energy for any $m_u<1$. Indeed, for $j=0, \, k=1$ or $j=1, \, k=0$,
\begin{equation}\label{eq:lowwg}
    G_\tau^{(j|2)} \underset{\omega \to 0}{=} \arsinh^2 \left(\frac{2\, |F_{22} \, F_{j2}|}{|F_{k2}|^2} \right),
\end{equation}
where the explicit expressions of the constant coefficients $|F_{i2}|^2, \, i \in\{0,1,2\}$ are given in Appendix \ref{sec:lowwS}. It means in particular  that the entanglement of bipartitions $j|2, \, j \in\{0,1\}$ remains finite at zero energy, while the tripartite entanglement becomes infinite.
Then, for higher frequencies, the residual contangle decreases rapidly to zero and vanishes at the upper-bound frequency $\Omega$.

Moreover, we show in the inset of Fig.~\ref{fig:TripartiteEntanglement} (cyan curve) that while at low frequency $G_\tau^{\rmres(1)} < G_\tau^{\rmres(0)}$ the situation is reversed for $\omega > \omega_{\rm c}$, i.e., when $a_{ 0} < a_{1}$ (the difference is anyway quite small). We note that this result may be different for Mach numbers different from the value $m_u=0.59$ we consider here. In particular, based on the estimate \eqref{eq:gresapprox} below, one can show that, when $m_u<0.17$, at low frequency $G_\tau^{\rmres(0)}$ becomes the contribution which minimizes \eqref{eq:Gresmin}.


\subsubsection{Experimental perspectives}\label{sec:exp_persp}

The waterfall model we use has proven to provide a fairly good description of the experimental setting \cite{Isoard2020}. In this section we use the relevance of our model to assess what is the best choice of parameters for an experimental measure of tripartite entanglement.

\begin{figure}
\includegraphics[width=\linewidth]{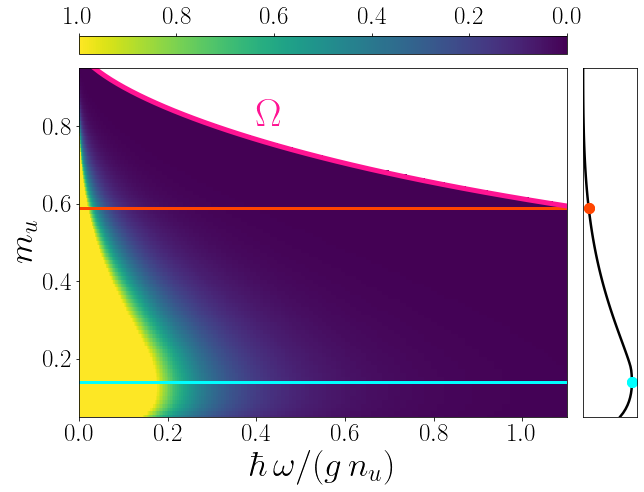}
\caption{Measure of tripartite entanglement $G_\tau^{\rm res}$  as a function of the (dimensionless) energy $\hbar \, \omega/ (g \, n_u)$ and of the upstream Mach number $m_u \in [0.05,0.95]$ defined in Eq.~\eqref{eq:Mach_nb}. The pink curve corresponds to the upper bound frequency $\Omega$ \eqref{threshold}. For a fixed value $m_u$ mode 2 only exists for a frequency $\omega$ lower than $\Omega(m_u)$. Beyond this value the tripartite system $\{0,1,2\}$ no longer exists; this is the reason why the corresponding area is left blank. The horizontal red line corresponds to the value $m_u\simeq 0.59$, corresponding to $m_d=2.9$ as realized in the experiment of  \cite{de_nova_2019}. The right plot shows the integral $\int G_\tau^{\rmres}{\rm d}\omega$  over frequencies $\omega\in[0,\Omega]$ for each value of $m_u$. The red dot pinpoints the numerical estimate of the integral for the specific value $m_u\simeq 0.59$, while the blue dot locates the maximum of the black curve reached for $m_u=0.14$. The light blue horizontal line on the left graph corresponds to this value.}
\label{fig:tripartite_mu}
\end{figure}
Figure \ref{fig:tripartite_mu} displays the
amount of genuine tripartite entanglement $G_\tau^{\rm res}$ expected in our one-dimensional analogue black hole as a function of frequency (horizontal axis) and upstream Mach number $m_u$ (vertical axis). For most cases, as proved by this two-dimensional graph, the entanglement is indeed shared among the Hawking, the Partner and the Companion quanta. Therefore, the Companion plays an important role in the distribution of entanglement within the emitted quanta. 

From Fig.~\ref{fig:tripartite_mu} one sees that the amount of tripartite entanglement is maximal for $m_u=0.14$, in the sense that the integral of $G_\tau^{\rm res}(\omega)$ over all frequencies is maximal for this value of upstream Mach number. This specific value of $m_u$ is indicated by the light blue horizontal cut on the graph. It has been determined by numerical integration of the residual contangle \eqref{eq:Gresmin}. One can also obtain an analytic estimate of this value of $m_u$, as we now explain. 
From the low-frequency behavior \eqref{eq:Sloww} of the components of the $S$-matrix involved [through Eq.~\eqref{eq:local_mixed_omega}] in \eqref{eq:G_res_tau_0}, \eqref{eq:G_res_tau_1} and \eqref{eq:G_res_tau_2},
one obtains the following expression for the low-frequency residual contangle:
\begin{equation}\label{eq:gresapprox}
G_\tau^{\rm res}(\omega) \underset{\omega \to 0}{\simeq} \underset{i\in \{0,1,2\}}{\min} 
\left[
\ln^2\left(\frac{4\, |F_{i2}|^2}{\hbar\omega/g\, n_u}\right)
\right].
\end{equation}
The value of $m_u$ for which $G_\tau^{\rm res}(\omega)$ in \eqref{eq:gresapprox} is the largest is thus simply the value for which the mimimum of $|F_{02}|$, $|F_{12}|$ and $|F_{22}|$ reaches a maximum. From the analytic expressions  
\eqref{eq:F02}, \eqref{eq:F12} and \eqref{eq:F22} of these coefficients one obtains $m_u=0.17$. Although this value has been determined using a different criterion than the numerical estimate $m_u=0.14$ plotted in Fig.~\ref{fig:tripartite_mu} (the former is based on the low $\omega$ behavior and the latter on the integrated signal)
the fact that both are quite close confirms their relevance.


\subsection{Entanglement localization}
\label{sec:entanglement_localization}
The tripartite entanglement of our system can be concentrated in a two-mode state by applying a local linear Bogoliubov transformation \cite{serafini2005, daems2010}; this is called entanglement localization. 
This transformation can be obtained by means of the symplectic transformation $\cals$ given by \eqref{eq:sympl_transformation_Botero}. To the mapping \eqref{eq:sympl_transformation_Botero} between $\sigma$ and its three-mode localized version $\mathds{1}_{2} \oplus \sigma_{ \rm \sss sq}$ one can associate the Bogoliubov transformation $\mathscr{T}=U^\dagger\cals U$ [see Eq.~\eqref{eq:transfo_sigma1}]. The modes $\mathbf{e}$ defined in \eqref{eq:def_e} (which coincide with the modes $\mathbf{c}$ up to a phase) are mapped through this Bogoliubov transformation to new modes $\mathbf{f}$. The Bogoliubov transformation from $\mathbf{e}$ to 
$\mathbf{f} = ( \hat{f}_{0},\hat{f}_{1}, \hat{f}_{2},\hat{f}_{0}^\dagger,\hat{f}_{1}^\dagger, \hat{f}_{2}^\dagger )^{\rm \sss T}$ is denoted $\mathscr{T}_{\mathbf{e} \rightarrow \mathbf{f}}$, and thus we have $\mathbf{f} = \mathscr{T}_{\mathbf{e} \rightarrow \mathbf{f}} \, \mathbf{e} $.
This transformation is such that the tripartite entanglement $e_0|e_1|e_2$ 
gets completely localized in a two-mode squeezed state.

Let us consider in turn the different cases.
If we consider bipartitions $ij|k=12|0$ and $02|1$ for modes $\mathbf{e}$, as derived explicitly in Appendix \ref{app:tripartite_eigenvalues} [see in particular Eq.~\eqref{eq:bogo_transfo_appB}], the new operators $\hat{f}_i$ and $\hat{f}_2$ correspond to a mixing of annihilation and creation operators $\hat{e}_{i},\hat{e}_{2}$ and $\hat{e}_{i}^\dagger,\hat{e}_{2}^\dagger$. In the case of bipartition $ij|k=01|2$ of modes $\mathbf{e}$, entanglement can also be localized but without mixing annihilation and creation operators. The corresponding Bogoliubov transformation is given by Eq.~\eqref{bogef012} and corresponds to a change of basis from $\{\hat{e}_{0},\hat{e}_{1},\hat{e}_{2}\}$ to $\{\hat{f}_0,\hat{f}_1,\hat{f}_{2}\}$ given by
\begin{subequations}
 \label{eq:expressions_e_theta}
\begin{align}
 \hat{f}_0  &= -\sin \theta\,  \hat{e}_{0} + \cos \theta \, \hat{e}_{1}, \label{eq:expressions_e_theta1}\\
  \hat{f}_1  &= \cos \theta \, \hat{e}_{0} + \sin \theta \, \hat{e}_{1}, \label{eq:expressions_e_theta2}\\
\hat{f}_2&=\hat{e}_{2} \label{eq:expressions_e_theta3},
\end{align}
\end{subequations}
 where (see Appendix \ref{sec:01|2})
 \begin{equation}\label{eq:deftheta}
   \cos\theta=\frac{\sinh r_{0}}{\sinh r_{2}} \quad\mbox{and}\quad
   \sin \theta=\frac{\sinh r_{1}}{\sinh r_{2}}.
\end{equation}
 The transformation leading to entanglement localization is thus particularly simple in the case of bipartition $01|2$. Inserting \eqref{eq:expressions_e_theta} into Eq.~\eqref{eq:link_vacua_set_e} leads to 
 \begin{equation}
|0\rangle_b = T \, |0\rangle_{f}, \quad\mbox{where}\quad 
      T = \exp[r_2 ( \hat{f}_1^\dagger \, \hat{f}_2^\dagger - \hat{f}_1 \, \hat{f}_2 )].
 \end{equation}
  The operator $T$ is a two-mode squeezing operator [compare with the generic form \eqref{squeezingop}] between $\hat{f}_1$ and $\hat{f}_2$, with squeezing parameter $r_2(\omega)$ defined in \eqref{locmix1}. Note that the modes $\hat{e}_0$ and $\hat{e}_1$ that are combined in \eqref{eq:expressions_e_theta1} and \eqref{eq:expressions_e_theta2} are those of positive norm; this leads to a squeezed state between the only mode of negative norm (mode 2) and a combination of the modes of positive norm (modes 0 and 1), exactly as occurs in the gravitational case \cite{birrell1982}.

\begin{figure}
    \centering
    \includegraphics[width=\linewidth]{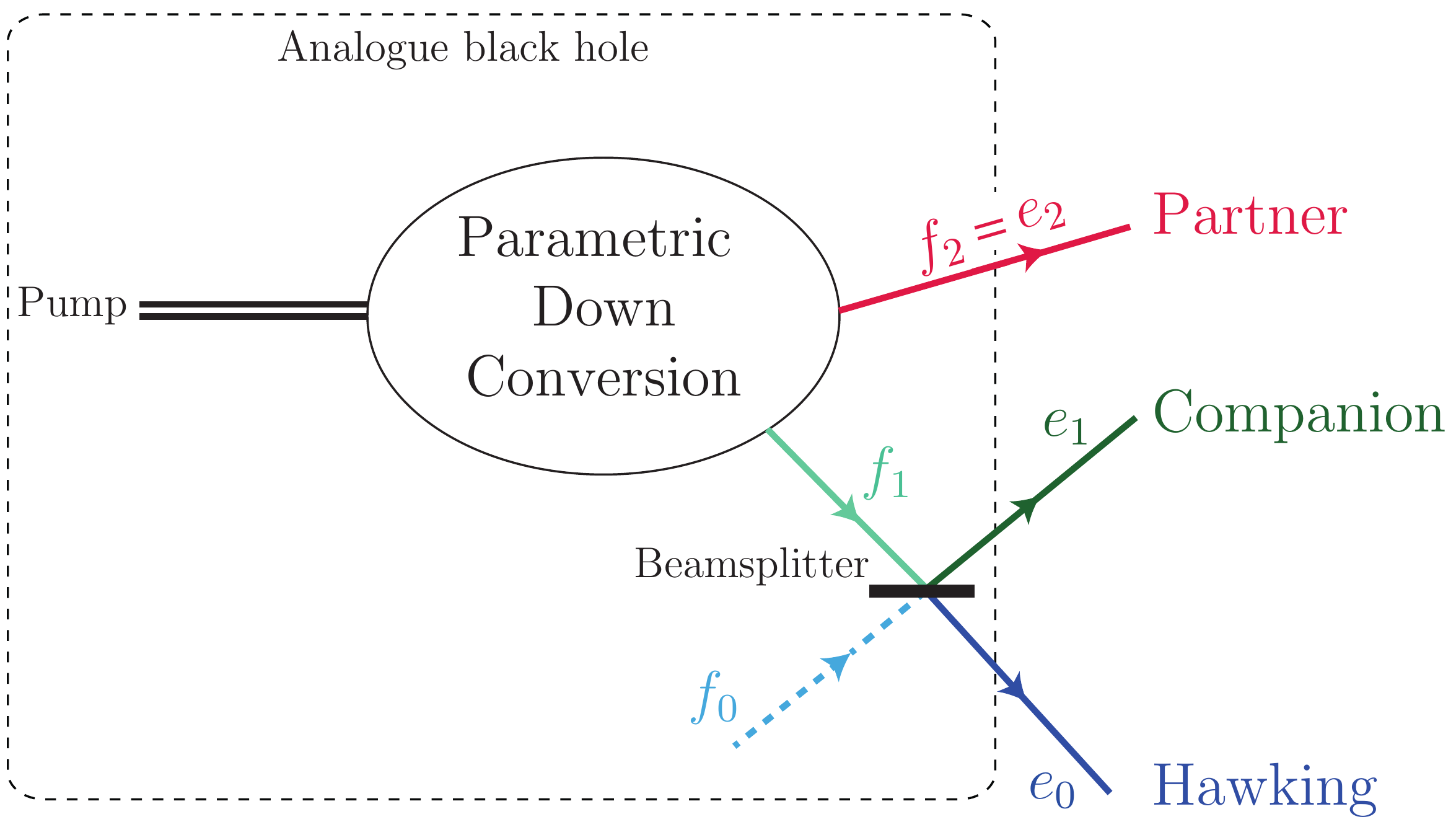}
    \caption{Schematic representation of an optical process equivalent to the Hawking emission in the transonic BEC system we consider. Entanglement is localized in the two-mode squeezed state $f_1|f_2$. The mode $f_0$ being empty is represented by a dashed line.}
    \label{fig:PDC}
\end{figure}

To summarize, the tripartite entanglement in our system can be unitarily localized by linearly combining modes $\hat{e}_0$ and $\hat{e}_1$ as in Eqs.~\eqref{eq:expressions_e_theta1} and \eqref{eq:expressions_e_theta2} to obtain mode $\hat{f}_1$, which forms a two-mode squeezed state with $\hat{f}_2=\hat{e}_2$.
Besides, using the definition \eqref{eq:deftheta} 
and noticing that $\langle \hat{e}_1^\dagger\hat{e}_0\rangle
= |\langle \hat{c}_1^\dagger\hat{c}_0\rangle|$
one obtains
\begin{equation}\label{eq:occuf0}
\begin{split}
    \langle \hat{f}_0^\dagger\hat{f}_0\rangle = & \sin^2\theta \,
    \langle \hat{c}_0^\dagger\hat{c}_0\rangle + \cos^2\theta \, \langle \hat{c}_1^\dagger\hat{c}_1\rangle \\
& -2\sin\theta\cos\theta\, 
|\langle \hat{c}_1^\dagger\hat{c}_0\rangle| 
\\
=& \left(\sin\theta |S_{02}| -\cos\theta |S_{12}|\right)^2
    = 0.
    \end{split}
\end{equation}
This means that mode $f_0$ is not occupied. This comes as no surprise since the corresponding local mixedness is equal to 1 in the transformed covariance matrix given by \eqref{eq:sympl_transformation_Botero}, which entails from \eqref{locmix1} that the mean particle number is equal to 0.
One can thus schematically describe the Bogoliubov transformation \eqref{eq:expressions_e_theta} operating in our analogue black hole by means of the equivalent
optical setup represented in Fig.~\ref{fig:PDC}: non-degenerate parametric down-conversion in a nonlinear crystal creates a two-mode squeezed state\footnote{We note here that the relevance of a non-degenerate parametric amplifier model has already been pointed out in Ref.~\cite{nova2015}.}. 
One of the modes is the Partner $\hat{f}_2=\hat{e}_2$. The other one, $\hat{f}_1$, is directed to a beam-splitter that generates the two other outgoing channels $\hat{e}_{1}$ and  $\hat{e}_{0}$ which are, up to a phase, the Companion and the Hawking mode, respectively.

We note that various theoretical proposals and experimental works have addressed the issue of generating and measuring tripartite entangled states for continuous variables, based on different setups of nonlinear optical parametric oscillators \cite{Aoki2003, Villar2006, Coelho2009, daems2010}. The analogue black hole we consider here is another such setup. It is quite peculiar in the sense that genuine tripatite entanglement is realized although two of the outgoing modes (0 and 1) are not entangled.

An interesting outcome of the present study is a redefinition of the analogue Hawking temperature, associated to a so-called grey-body factor. Redoing for the $f_1$ and $f_2$ modes at the output of the parametric down conversion process the analysis done for the Hawking mode at the end of \autoref{sec:vac_three_mode}, it is clear that these two modes have the same 
occupation number
\begin{equation}
    \langle \hat{f}_2^\dagger \hat{f}_2\rangle =
    \langle \hat{f}_1^\dagger \hat{f}_1\rangle = \sinh^2 r_2\; ,
\end{equation}
and the same effective temperature
\begin{equation}\label{eq:Teff3}
\begin{split}
    T_{\rm eff}^{(2)}(\omega)= & \frac{\hbar \omega}{2\, \ln[\coth (r_2(\omega))]} \\
    = & \frac{\hbar \omega}
    {\ln\left[\frac{\displaystyle |S_{22}(\omega)|^{2}}{\displaystyle |S_{22}(\omega)|^{2}-1}\right]}
    \; .
    \end{split}
\end{equation}
The $f_1$ mode being sent to the beam splitter is transmitted onto the Hawking mode with a transmission coefficient $\cos^2\theta$, and indeed one can easily check that 
\begin{equation}\label{1to0}
    \langle \hat{e}_0^\dagger \hat{e}_0\rangle = \cos^2\theta \, 
    \langle \hat{f}_1^\dagger \hat{f}_1\rangle \; .
\end{equation}
We saw in \autoref{sec:vac_three_mode} that the Hawking mode could be considered as a thermal state with temperature $T_{\rm eff}^{(0)}(\omega)$. Equation \eqref{1to0} shows that it can also be considered as a thermal state of temperature $T_{\rm eff}^{(2)}(\omega)$ affected by a grey-body factor $\Gamma(\omega)=\cos^2\theta$. Such a factor is invoked in general relativity for explaining that the Hawking radiation is subject to an effective potential at the horizon which affects its thermal character \cite{Page_emission_rate_1976}. 
The introduction of a grey-body term in the present analysis has the advantage to ascribe a single, global effective temperature to the analogue system: $T_{\rm eff}^{(2)}$. In this framework, the difference in population of the modes is explained by the transmission coefficients $\cos^2\theta$ and $\sin^2\theta$ of the beam splitter, not by a difference in temperature. In the long wavelength limit it yields an analogue Hawking radiation $T_{\rm\sss H}^{(2)}=\lim_{\omega\to 0}
T_{\rm eff}^{(2)}(\omega)$ which explicit expression in the waterfall configuration reads (from Appendix \ref{sec:lowwS})
\begin{equation}\label{eq:Th2}
    \frac{T_{\rm\sss H}^{(2)}}{g n_u}=  \frac{1}{2}\frac{(1-{m}_{u}^{4})^{\frac{3}{2}}}
{(1+{m}_{u}+{m}_{u}^{2})^{2}},
\end{equation}
and a grey-body factor
\begin{equation}
    \Gamma_0=\lim_{\omega\to 0} \Gamma(\omega) = \lim_{\omega\to 0} \frac{|S_{02}|^2}{|S_{22}|^2-1} = \frac{4\, m_u}{(1+m_u)^2}\; .
\end{equation}
It is satisfactory to note that the present approach yields a result for $\Gamma_0$ identical to the universal limit obtained in Refs. \cite{Anderson_loww_rig2015,Fabbri_grey_body2016} by means of a different technique.
\begin{figure}
    \centering
    \includegraphics[width=\linewidth]{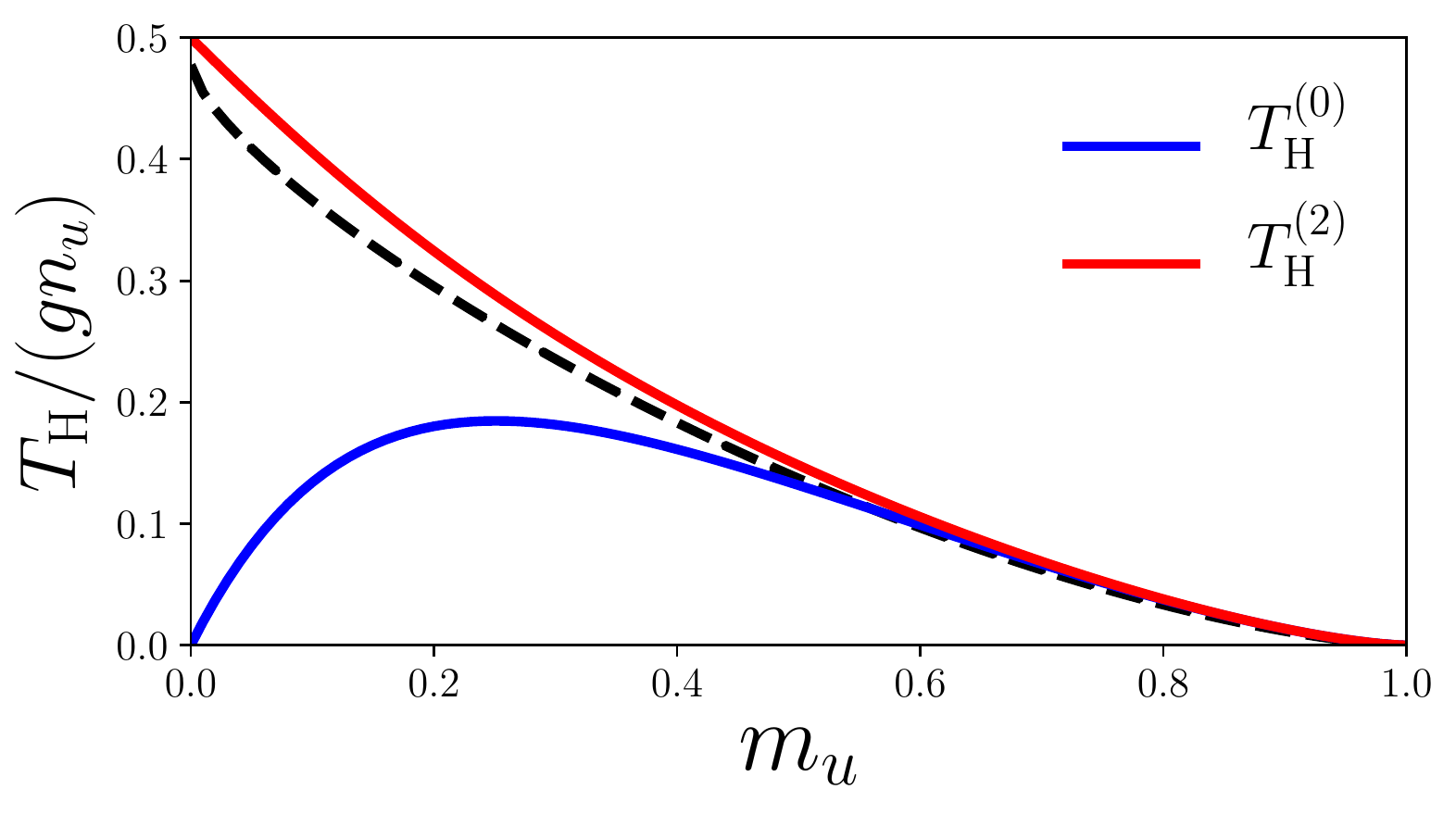}
    \caption{Hawking temperature as a function of the upstream Mach number in the waterfall configuration. The blue solid line is the result \eqref{eq:Th0} and the red solid line comes from expression \eqref{eq:Th2}. The dashed line is the semi-classical expectation \eqref{eq:Th_SC}.}\label{fig:Th}
\end{figure}

Another advantage of the present definition of the Hawking temperature over the one introduced at the end of \autoref{sec:vac_three_mode}, is that $T_{\rm\sss H}^{(2)}$ defined in Eq.~\eqref{eq:Th2} is in good agreement with the semi-classical result \eqref{eq:Th_SC}. This is to be contrasted with $T_{\rm\sss H}^{(0)}$, defined in Eq.~\eqref{eq:Th0} by studying the thermal character of the reduced mode 0 state. The discrepancy between the two behaviors is illustrated in Fig.~\ref{fig:Th}.  $T_{\rm\sss H}^{(0)}$ has the unpleasant property of vanishing at $m_u\to 0$, although in this limit the "surface gravity" is the largest. For the model sketched in Fig.~\ref{fig:PDC} instead, the disappearance of Hawking radiation when $m_u\to 0$ is due to a vanishing grey-body factor, which is physically more satisfactory. One sees also in Fig.~\ref{fig:Th} that in the limit $m_u\to 1$ all definitions of the Hawking temperature coalesce to zero: in this regime $\Gamma_0\to 1$ hence $T_{\rm\sss H}^{(0)}=T_{\rm\sss H}^{(2)}$. Also, in this limit the density profile is smoother, the semi-classical approach is more legitimate (cf.~the discussion at the end of App.~\ref{sec:lowwS}), and the surface gravity vanishes: the semiclassical estimation of the Hawking temperature thus vanishes as $T_{\rm\sss H}^{(2)}$ does.

\section{Finite temperature}\label{sec:finiteT}

We previously considered the zero-temperature case, where all the averages $\langle \cdots \rangle$ in Sec.~\ref{defgaussian} are taken over the vacuum state $|0 \rangle_b$. In the present section we study a finite temperature system.

\subsection{Finite-temperature states}

Because of the
existence of negative-energy modes, the transonic flow we consider is energetically unstable  and  cannot support a  thermal state. However, one can define a finite temperature configuration \cite{Recati2009,Macher2009,Fabbri2018} as follows: One considers a uniform BEC (with density $n_u$) initially flowing at constant velocity $V_u$, at thermal equilibrium at temperature $T_{\rm \sss BEC}$ in the frame moving along with the fluid. Then the potential $U(x)$ of Eq.~\eqref{abh2} is slowly ramped up until the system reaches the configuration described in Sec.~\ref{sec:background} and Fig.~\ref{fig:waterfall}. At the end of this adiabatic  branching process one can define an occupation number $ \bar{n}_i(\omega, T_{\rm \sss BEC})$ for each of the incoming modes $\hat{b}_i$.
 As explained, e.g., in \cite{Fabbri2018}, for a fixed frequency $\omega$ these occupation numbers are given by
\begin{equation}
\label{eq:bar_ni_def}
\langle \hat{b}^\dagger_i(\omega) \hat{b}_i(\omega) \rangle = 
    \bar{n}_i(\omega, T_{\rm \sss BEC}) = n_{\rm \sss th} \left[ \omega_{\sss \rm B, \alpha} \left(  q_{i|\rm in}(\omega)  \right) \right],
\end{equation}
where $n_{\rm \sss th}(\varpi)=[\exp(\hbar \varpi/T_{\rm \sss BEC})-1]^{-1}$ is the thermal Bose occupation distribution. In this expression
$\omega_{\sss \rm B, \alpha}(q_{i|\rm in})$ is the Bogoliubov dispersion relation \eqref{eq:Bogoliubov_dr}, with $\alpha = u$ if $i=0$ and $\alpha=d$ if $i=1$ or 2, and the functions $q_{i|\rm in}(\omega)$ are defined above [just after Eq.~\eqref{threshold}].

The regime in which the separation \eqref{abh1} between a classical field and quantum fluctuations is valid and where the Bogoliubov treatment of the fluctuations applies has been denoted as the ``weakly interacting quasicondensate regime'' in \cite{Deuar2009}. It is valid up to a temperature $T_{\rm \sss BEC} \simeq  g \, n_u$ \cite{Isoard2020}, where $g$ is the coefficient of the nonlinearity in the Gross-Pitaevskii equation \eqref{abh2}. For typical experimental parameters $g \, n_u \simeq 3 $ nK \cite{de_nova_2019}. While it is difficult to precisely determinate the experimental temperature,  
 we note that the agreement between the experimental results of \cite{de_nova_2019} and the theoretical expectations \cite{Isoard2020} suggests that the temperature of the condensate in the analogue black hole realized by J.~Steinhauer and collaborators is possibly lower than 3 nK. 

At a finite temperature $T_{\rm \sss BEC}$, the vacuum state $|0 \rangle_b$ is replaced by a product of thermal states of ${b}$ modes given by 
\begin{equation}
\label{eq:rho_nu_finiteT}
\rho_{\boldsymbol{a}^b}= \overset{\sss 2}{\underset{i=0}{\otimes}} \rho^{\sss \rm th}({a}^b_i),\qquad {a}^b_i=1+2\bar{n}_i(\omega, T_{\rm \sss BEC})
\end{equation}
where $\bar{n}_i$ is given by Eq.~\eqref{eq:bar_ni_def}. 
We recall that we use the term "thermal" to designate that state in a loose sense, since, as explained in the beginning of this section, the occupation numbers  \eqref{eq:bar_ni_def} do not correspond to an equilibrium distribution in the transonic configuration we consider.
The covariance matrix associated with this state is given by $\sigma^{\sss \rm th}_b = \text{diag}({a}^b_0, {a}^b_0, {a}^b_1, {a}^b_1, {a}^b_2, {a}^b_2)$. After the Bogoliubov transformation $\mathbf{c} = \mathscr{T} \, \mathbf{b}$, the covariance matrix becomes $\sigma_c^{\sss \rm th} = \calst  \, \sigma_b^{\sss \rm th} \, \calst^{\sss \rm T}$ [see Eq.~\eqref{eq:transfo_sigma1}]. The $2\times 2$ matrices $\sigma_i$ and $\varepsilon_{ij}$ in the block decomposition \eqref{cmG} of $\sigma_c^{\sss \rm th}$ are given by expressions \eqref{eq:sigma_i_sf} and \eqref{eq:epsilon_ij_sf} where the averages $\langle \ldots \rangle$ should be replaced by  
\begin{equation}
  \langle \ldots \rangle_{\rm \sss th} = \text{Tr} \left[ \rho_{\boldsymbol{a}^b} \, \ldots \right]. 
\end{equation}
In particular, 
\begin{equation}\label{eqsith}
\sigma_i=a_{i,{\sss\rm th}} \mathds{1}_2 ,
\end{equation} 
 where
\begin{equation}\label{eqsith2}
    a_{i, {\sss \rm th}} = 1 + 2\, \langle \hat{c}_i^\dagger \, \hat{c}_i \rangle_{\sss \rm th}\; , \quad i \in \{0,1,2 \}\; ,
\end{equation} 
is the corresponding local mixedness [compare to \eqref{eq:sigma_i_sf} and to the first of Eqs.~\eqref{eq:cov_mat_a-parameters}].

We conclude this short section by noting that the optical analogue proposed in Fig.~\ref{fig:PDC} remains relevant at finite temperature. The difference with the zero-temperature case is just the occupation number of the $f$-modes: they now acquire an incoherent contribution. In particular the occupation $\langle \hat{f}_0^\dagger \hat{f}_0\rangle$
is no longer zero as in Eq.~\eqref{eq:occuf0}. This suggests a possible 
experimental study of the effects of temperature on tripartite entanglement: one could realize the optical setup of Fig.~\ref{fig:PDC},
send a non-coherent beam along the mode $f_0$, and evaluate the associated effect on entanglement in the system.

\subsection{Detection of entanglement}\label{sec:detec_ent_th}
Contrary to the zero-temperature case, $\sigma_c^{\sss \rm th }= \calst\, \sigma_b^{\rm th} \, \calst^{\sss \rm T}$ is associated with a  mixed state with no special symmetry, and  it cannot be put in a standard form where the matrices $\varepsilon_{ij}$ are all diagonal \cite{adesso2006a}. In this section we thus restrict our study to bipartite entanglement. In this case, the $4 \times 4$ covariance matrix associated with the reduced two-mode state $ij$ can always be brought by LLUBOs to its standard form \cite{duan2000}. One easily proves that matrices $\varepsilon_{ij}$ have, {\it mutatis mutandis}, the same form as those in the zero-temperature case, namely
\begin{equation}
\label{eq:epsilon_ij_th}
\begin{split}
\varepsilon_{ij}& = 2 \,|\langle \hat{c}_i \, \hat{c}_j^\dagger \rangle_{\sss \rm th}| \, \mathds{1}_2 , \qquad  i,j=0, 1,\quad i\neq j\\
\varepsilon_{i2}&= 2 \, |\langle \hat{c}_i \, \hat{c}_2 \rangle_{\sss \rm th}|\, \sigma_z, \qquad  i=0, 1.
\end{split}
\end{equation}
As a consequence, the lowest symplectic eigenvalue associated with the partial-transposed reduced two-mode state  $ij$ takes the same form as in the zero-temperature case. Eq.~\eqref{eq:nu-2mode_transpose} still holds, and in particular
\begin{equation} 
\label{eq:nu_finiteT1}
2 \, (\nu_{-}^{\sss \rm PT})^2 = \Delta_{ij}^{\sss \rm PT} - \sqrt{(\Delta^{\sss \rm PT}_{ij})^2 - 4 \, \det  \sigma_{ij}},
\end{equation}
with here
\begin{equation}
\label{eq:nu_finiteT2}
\begin{split}
\det  \sigma_{01} & = \left( a_{0, {\sss \rm th}} \, a_{1, {\sss \rm th}} - 4 \, |\langle \hat{c}_0 \, \hat{c}_1^\dagger \rangle_{\sss \rm th}|^2 \right)^2, \\
\Delta_{01}^{\sss \rm PT} & = a_{0, {\sss \rm th}}^2 + a_{1, {\sss \rm th}}^2  - 8 \, |\langle \hat{c}_0 \, \hat{c}_1^\dagger \rangle_{\sss \rm th}|^2, \\
  \det  \sigma_{i2} & = \left( a_{i, {\sss \rm th}} \, a_{2, {\sss \rm th}} - 4 \, |\langle \hat{c}_i \, \hat{c}_2 \rangle_{\sss \rm th}|^2 \right)^2, \! \! \! \qquad i =0,1 \\
  \Delta_{i2}^{\sss \rm PT}  & = a_{i, {\sss \rm th}}^2 + a_{2, {\sss \rm th}}^2  + 8 \, |\langle \hat{c}_i \, \hat{c}_2 \rangle_{\sss \rm th}|^2, \qquad i =0,1.
\end{split}
\end{equation}
Note that the above expressions only involve moduli of mean values, so that we could equivalently use operators $\hat{e}_i$ instead of $\hat{c}_i$ since the transformation defined by Eqs.~\eqref{eq:def_R} and \eqref{eq:defR} is diagonal.
The explicit form of the quantities appearing in  Eqs.~\eqref{eq:nu_finiteT2} is given in Eqs. \eqref{finiteTcc}. At variance with the zero-temperature case they do not depend only on the local mixednesses.
They should be experimentally accessible through the measurement of the structure form factor and of real space
density correlations \cite{steinhauer2015}, meaning that the PPT criterion can be used to experimentally detect entanglement (cf.~the discussion at the end of \autoref{sec:measuring_entanglement_finiteT}).

\begin{figure*}[h!]
    \centering
    \includegraphics[width=\textwidth]{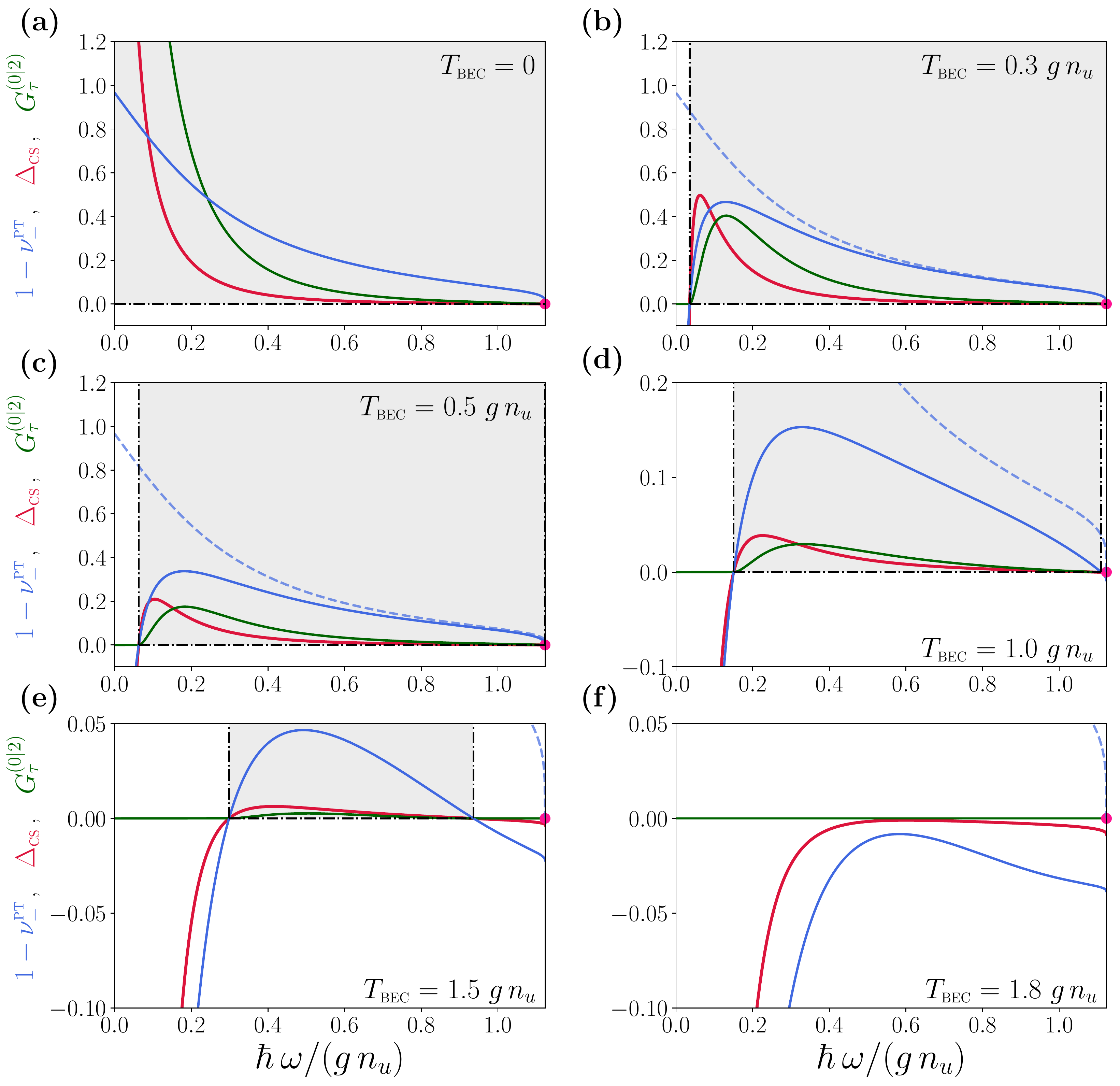}
    \caption{Evolution of the PPT measure $1-\nu_-^{\rm \sss PT}$ (blue), of the Cauchy-Schwarz parameter $\Delta_{\sss \rm CS}$ (red) and of the Gaussian contangle $G_{\tau}^{(0|2)}$ (green) for the bipartite system $0|2$ (i.e., the analogue Hawking pair) as functions of the dimensionless frequency $\hbar \omega/(g\, n_u)$ and for different temperatures of the system, denoted by $T_{\rm \sss BEC}$, ranging from 0 (a) to $1.8 \, g\,n_u$ (f). All the plots are obtained for an upstream Mach number $m_u = 0.59$. The dashed blue curves in graphs (b)-(f) correspond to the zero-temperature value of $1-\nu_-^{\rm \sss PT}$. The grey areas indicate the range of frequencies for which the bipartite system is entangled [see text and Eqs.~\eqref{eq:PPT_finiteT} and \eqref{eq:CS_finiteT}]. The purple dots locate the upper-bound frequency $\Omega$ at which mode 2 vanishes.  }
    \label{fig:bip_ent_finiteT}
\end{figure*}

The PPT criterion asserts that the bipartite state is entangled iff
\begin{equation}
\label{eq:PPT_finiteT}
    1 - \nu_-^{\rm \sss PT} > 0.
\end{equation}
In the following we denote this quantity as the "PPT measure". It is of particular interest to focus on the bipartition $0|2$ since it corresponds to the Hawking--Partner pair. In this case expression \eqref{eq:nu_finiteT1} leads to
\begin{equation}
\begin{split}\label{nu02}
    \nu_-^{\sss\rm PT}= & 
    \frac{a_{0, {\sss \rm th}}+a_{2, {\sss \rm th}}}{2}\\
    & -
    \sqrt{
    \left(\frac{a_{0, {\sss \rm th}}-a_{2, {\sss \rm th}}}{2}\right)^2 +
    4 |\langle \hat{c}_0 \, \hat{c}_2 \rangle_{\sss \rm th}|^2
    }\; .
\end{split}
\end{equation}
The corresponding value of the PPT measure \NP{$1-\nu_-^{\rm\sss PT}$} is represented in 
Fig.~\ref{fig:bip_ent_finiteT} as a function of the frequency $\omega$ of the elementary excitations and for different temperatures ranging from 0 to 1.5 $g \, n_u$ (blue curves).
In each plot the dashed blue curves display the same quantity at zero-temperature for comparison. 

It is instructive to compare the conclusions drawn from the study of the PPT measure with those obtained using the criterion of violation of the Cauchy-Schwarz inequality \eqref{eq:CS_inequality}. According to this criterion, the analogue Hawking--Partner pair $0|2$ is entangled iff
\begin{equation}
\label{eq:CS_finiteT}
    \Delta_{\rm \sss CS} \equiv |\langle \hat{c}_0 \, \hat{c}_2 \rangle_{\sss \rm th}|^2 - \frac{(a_{0, {\sss \rm th}} -1) \, (a_{2, {\sss \rm th}}-1)}{4} > 0.
\end{equation}
In the following we denote $\Delta_{\rm \sss CS}$ as the "Cauchy-Schwarz parameter". It is represented by the red curves in Figs.~\ref{fig:bip_ent_finiteT} which
confirm the results obtained with the PPT criterion: the blue and red curves are positive in the same region and
cross zero exactly at the same frequency. This means, as expected, that both criteria lead to the same qualitative result for entanglement detection. However, as we shall see in \autoref{sec:measuring_entanglement_finiteT}, they lead to different quantitative estimation of the amount of entanglement. 

The analogue Hawking pair is entangled in the range of frequencies for which inequalities \eqref{eq:PPT_finiteT} 
and \eqref{eq:CS_finiteT}
hold. This corresponds to the grey shaded regions bounded by two vertical black dot-dashed lines in Figs.~\ref{fig:bip_ent_finiteT}. The range of parameters over which entanglement can be observed decreases when the temperature of the Bose gas increases. In agreement with the findings of Refs. \cite{bruschi_2013,Finazzi2014}, we observe that when $T_{\rm\sss BEC}$ increases entanglement first
disappears at low $\omega$. It eventually
completely disappears when $T_{\rm \sss BEC} \gtrsim 1.8 \, g \, n_u$, cf.~Fig.~\ref{fig:bip_ent_finiteT}(f). Therefore the temperature of the experimental system should not exceed this limiting value to be able to observe entanglement. It is interesting to compare this value to the one obtained in Ref.~\cite{Finazzi2014}, which studies an analogue black hole configuration different from the waterfall we consider here (it had been denoted as "flat profile" in Ref.~\cite{Larre2012}) with values of the upper and lower Mach numbers not significantly different from ours\footnote{They have $m_u=0.75$ and $m_d=1.5$, whereas here $m_u=0.59$ and $m_d=2.9$.}. The authors of Ref. \cite{Finazzi2014} find a disappearance of entanglement for $T_{\rm \sss BEC} \gtrsim 0.195 \, g \, n_u$, i.e., at much lower temperature than what is observed here. This is in agreement with the findings of Ref.~\cite{Fabbri2018} where entanglement was shown to be much less resilient to temperature in the flat profile configuration than in the waterfall configuration.

\begin{figure}
    \centering
    \includegraphics[width=\linewidth]{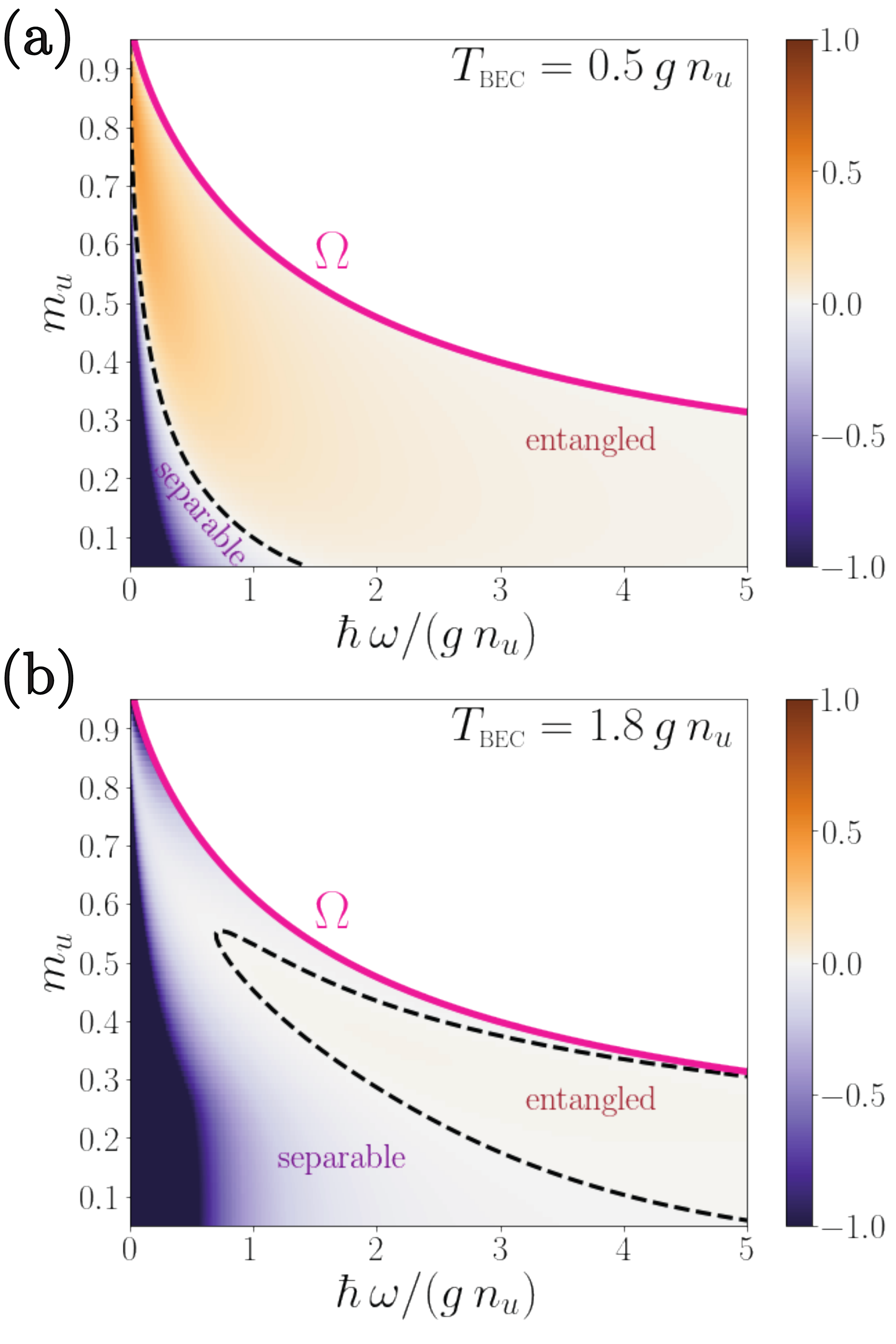}
    \caption{PPT measure $1-\nu_-^{\rm \sss PT}$ of the Hawking pair $0|2$ plotted as a function of the upstream Mach number $m_u$ and of the frequency $\omega$, for temperatures (a) $T_{\rm \sss BEC}=0.5 \, g\, n_u$ and (b) $T_{\rm \sss BEC}=1.8 \, g\, n_u$.
    The pink curve corresponds to the upper-bound frequency $\Omega$ \eqref{threshold}. For a fixed value $m_u$, i.e., along a horizontal cut on the graph, mode 2 only exists for a frequency $\omega$ lower than $\Omega(m_u)$ (see Fig.~\ref{fig:dispersion}). The dashed black curve corresponds to $1-\nu_-^{\rm \sss PT} = 0$ and thus delimits the region where the analogue Hawking pair is entangled. }
    \label{fig:bipartite_HP_T05}
\end{figure}

In order to perform a more detailed discussion of the effects of temperature
on entanglement, we represent in
Fig.~\ref{fig:bipartite_HP_T05}(a) the PPT measure $1-\nu_-^{\rm \sss PT}$ of the Hawking pair $0|2$ at temperature $T_{\rm \sss BEC}=0.5 \, g \, n_u$ for different configurations parameterized by the upstream Mach number $m_u$. 
As already seen in Fig.~\ref{fig:bip_ent_finiteT}, which corresponds to the specific case $m_u=0.59$, a finite temperature reduces the range of frequencies for which entanglement occurs. One
observes in this new plot that the entanglement of the Hawking pair persists for a larger fraction of the available frequency domain when the parameter $m_u$ is closer to unity. This is in agreement with the results obtained in \cite{Busch2014a}; it was noticed that not only the temperature $T_{\rm \sss BEC}$ destroys the entanglement of the analogue Hawking pair, but also that a strong ``coupling'' of mode 1 with the other modes can  affect their entanglement. This coupling is measured through the squared modulus of the scattering matrix coefficients $|S_{01}(\omega)|^2$ and $|S_{21}(\omega)|^2$. One finds numerically (and analytically in the low-$\omega$ sector \cite{Larre2012}) that these two quantities decrease when $m_u$ increases. This exactly corresponds to the results presented in Fig \ref{fig:bipartite_HP_T05}(a): when $m_u$ increases, the coupling between 0-1 and 1-2 decreases, and indeed leads to a stronger violation of PPT criterion for a larger fraction of frequencies. However, it is important to note that this phenomenon is only valid at low enough temperatures. This is illustrated in Fig.~\ref{fig:bipartite_HP_T05}(b): for a temperature as large as $T=1.8 \, g \, n_u$ the region where the pair is entangled greatly diminishes and entanglement only survives at moderate values of $m_u$ (at variance with the conclusion of the above discussion). Likewise, at this temperature, even in the region where entanglement is present, the PPT measure is significantly lower than in the equivalent regions in Fig.~\ref{fig:bipartite_HP_T05}(a).

\begin{figure}
    \centering
    \includegraphics[width=\linewidth]{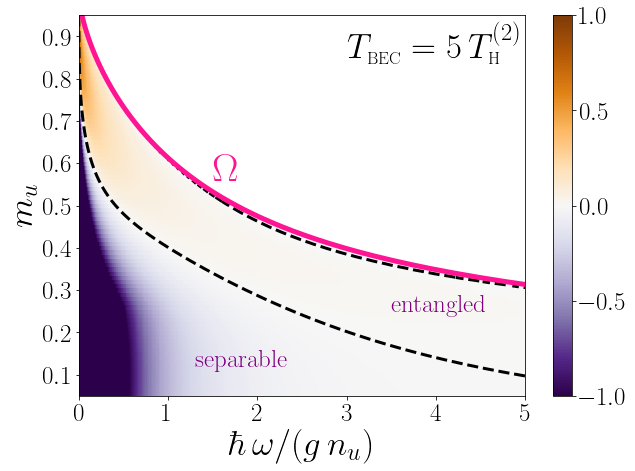}
    \caption{Same as Fig.~\ref{fig:bipartite_HP_T05} for $T_{\rm \sss  BEC}=5 \, T_{\rm \sss H}^{(2)}$.}
    \label{fig:bipartite_HP_10TH}
\end{figure}

It is also interesting to study the entanglement of the Hawking pair, not as a function of the absolute temperature, but as a function of the Hawking temperature $T_{\rm \sss H}^{(2)}$ \eqref{eq:Th2}. 
There is no obvious reason why entanglement between modes should disappear when the temperature of the system exceeds the Hawking temperature. This is indeed what is observed in Fig.~\ref{fig:bipartite_HP_10TH}: entanglement persists in sizeable regions even when $T_{\rm \sss BEC}=5\, T_{\rm \sss H}^{(2)}$. 

We can conclude from the above discussion that whereas entanglement persists for temperatures noticeably larger than $T_{\rm \sss H}^{(2)}$, it is significantly reduced when $T_{\rm \sss BEC}$ becomes larger than the chemical potential  $g \, n_u$.

\subsection{Measurement of entanglement}
\label{sec:measuring_entanglement_finiteT}
The violation of the Cauchy-Schwarz inequality is often used to study the entanglement between the elementary excitations in the context of analogue gravity  \cite{Busch2014b,deNova2014,Busch2014a,Boiron2015,Fabbri2018,Coutant2018}. However, while this criterion tells us whether the bipartite system is entangled or not, the Cauchy-Schwarz parameter $\Delta_{\rm \sss CS}$ it is not a good measure of the amount of entanglement at finite temperature. 

To clarify this point, we compute the amount of entanglement at finite temperature for the bipartition $0|2$, as measured by the Gaussian contangle $G_\tau^{(0|2)}$ defined in Eq.~\eqref{eq:G_tau}. This computation is slightly more difficult here than in the zero-temperature case, where it is given by Eq.~\eqref{eq:G_01_02_12}. In the presence of temperature the reduced two-mode state $0|2$ is not a GMEMMS, a GMEMS or a GLEMS, for which analytic expressions hold \cite{adesso2005}. Nevertheless, the Gaussian contangle can be put under the form \cite{adesso2005}
\begin{equation}
\label{eq:G_tau_finiteT}
    G_\tau^{(0|2)} =  \arsinh^2 \, \left\{\sqrt{\underset{\theta}{\text{min}} [m(\theta)]-1}\right\},
\end{equation}
where $m(\theta)$ is explicitly given by Eq.~\eqref{eq:m_theta}.
In Figs.~\ref{fig:bip_ent_finiteT} we represent by a green solid line the value of $G_\tau^{(0|2)}$ in the range of frequencies for which the system is entangled (the minimum over the angle $\theta$ in Eq.~\eqref{eq:G_tau_finiteT} is obtained numerically). 
The results for $G_\tau^{(0|2)}$
confirm that the PPT and Cauchy-Schwarz criteria correctly determine the region where entanglement exists.

As  expected,  entanglement decreases  as  the  temperature  increases. In  the  zero-temperature case [Fig.~\ref{fig:bip_ent_finiteT}(a)], both $1-\nu_-^{\rm \sss PT}$ and $\Delta_{\sss \rm CS}$ vary in the same way as $G_{\tau}^{(0|2)}$. The situation at finite temperature is different: while the quantities $1 - \nu_-^{\rm \sss PT}$ and $G_\tau^{(0|2)}$ appear to behave similarly, being increasing and decreasing in the same regions and having a maximum at the same value of $\omega$, this is not the case for $\Delta_{\rm \sss CS}$ whose maximum is shifted with respect to the two others, see Figs.~\ref{fig:bip_ent_finiteT}(b)-(f).

\begin{figure}
    \centering
    \includegraphics[width=\linewidth]{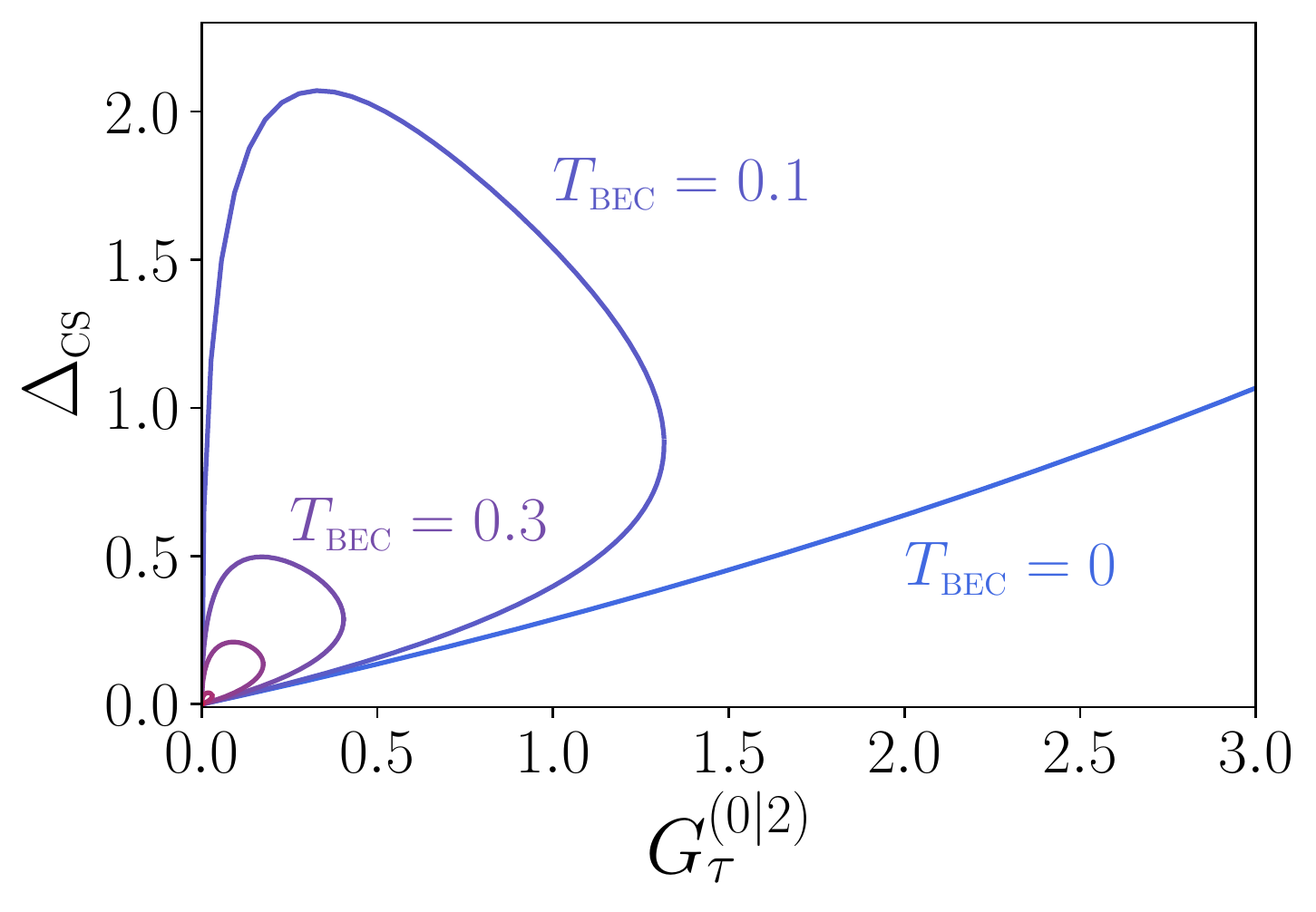}
    \caption{Evolution of $\Delta_{\sss \rm CS}$ given by Eq.~\eqref{eq:CS_finiteT} as a function of the measure of bipartite entanglement $G_\tau^{(0|2)}$ given by expressions \eqref{eq:G_tau_finiteT} and \eqref{eq:m_theta}, for the same set of temperatures as in Fig.~\ref{fig:bip_ent_finiteT}, ranging from $T_{\rm \sss BEC} = 0$ (blue curve) to $T_{\rm \sss BEC} = 1.5 \, g\, n_u$ (red curve), with $m_u=0.59$. When possible, the corresponding temperature for each curve is indicated on the graph (we dropped the factor $g\, n_u$ for readability). Note that for $T_{\rm \sss BEC} > 0$, the curves describe a loop.}
    \label{fig:bip_ent_violation_finiteT_CS}
\end{figure}

In order to illustrate this phenomenon, in Fig.~\ref{fig:bip_ent_violation_finiteT_CS}  we plotted $\Delta_{\sss \rm CS}$ as a function of $G_\tau^{(0|2)}$ for several temperatures.
These are parametric curves obtained from expressions \eqref{eq:CS_finiteT} and \eqref{eq:G_tau_finiteT}, $\omega$ playing the role of the parameter. Except at $T_{\rm \sss BEC}=0$, $\Delta_{\sss \rm CS}$ is not a monotonous function of 
$G_\tau^{(0|2)}$, as demonstrated by the closed loops with regions of negative slope observed for each finite temperature. Another way to note the same point is to remark that the maximal violation of Cauchy-Schwarz inequality ($\Delta_{\sss \rm CS}$ maximal) is not reached when  $G_\tau^{(0|2)}$ is maximal.
This confirms that the parameter $\Delta_{\sss \rm CS}$ is not an entanglement monotone.

\begin{figure}[h]
    \centering
    \includegraphics[width=\linewidth]{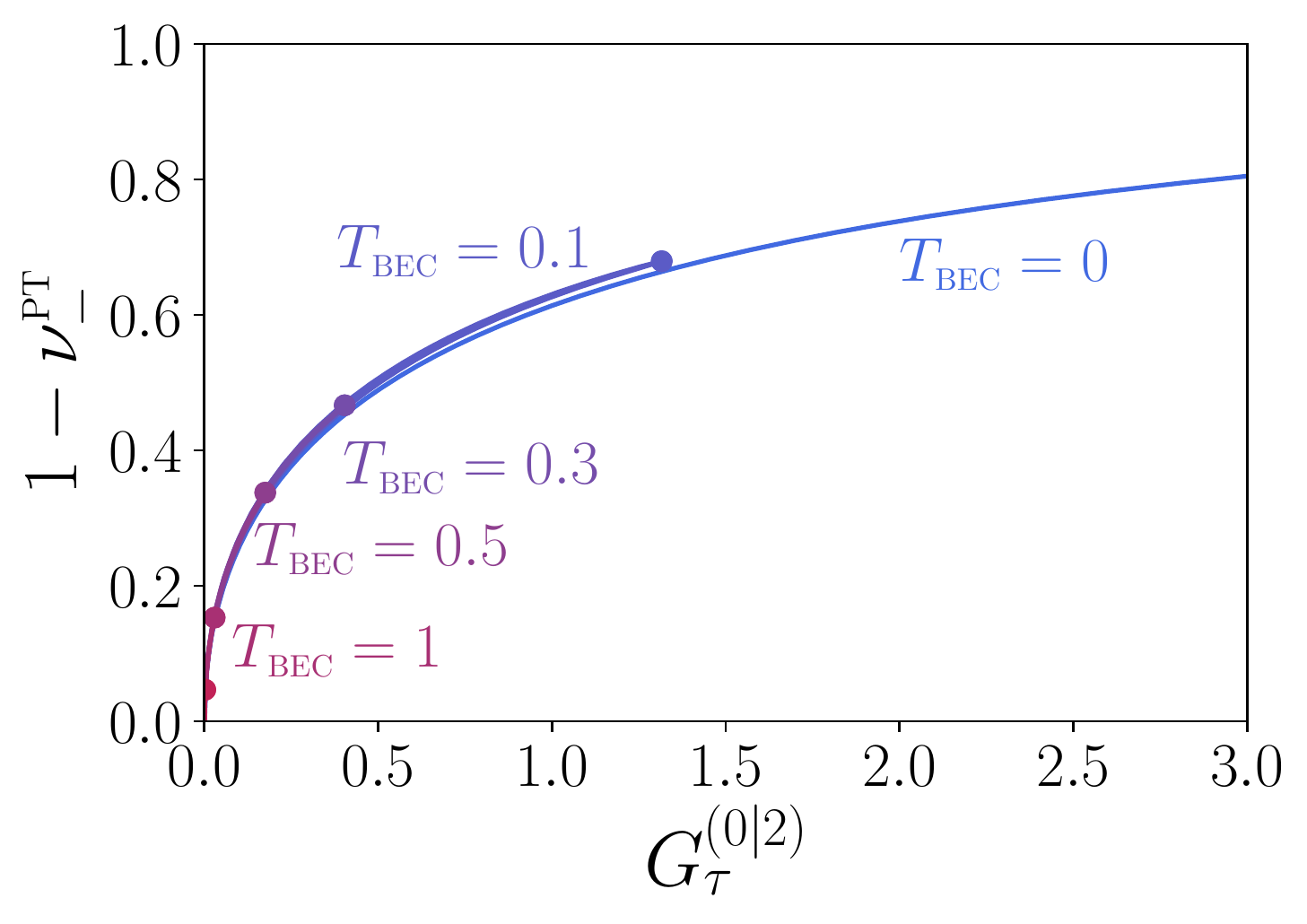}
    \caption{  Evolution of $1 - \nu_-^{\rm \sss PT}$ given by Eq.~\eqref{nu02} as a function of the measure of bipartite entanglement $G_\tau^{(0|2)}$ given by expressions \eqref{eq:G_tau_finiteT} and \eqref{eq:m_theta}, for the same set of temperatures as in Fig.~\ref{fig:bip_ent_finiteT}, ranging from $T_{\rm \sss BEC} = 0$ (blue curve) to $T_{\rm \sss BEC} = 1.5 \, g\, n_u$ (red curve), with $m_u=0.59$. When possible, the corresponding temperature for each curve is indicated on the graph (we dropped the factor $g\, n_u$ for readability). For each temperature the common maximal value of  $1 - \nu_-^{\rm \sss PT}$ and $G_\tau^{(0|2)}$ is marked with a point.}
    \label{fig:bip_ent_violation_finiteT_nu}
\end{figure}

In Fig.~\ref{fig:bip_ent_violation_finiteT_nu} we underline the difference between the behaviors of the Cauchy-Schwarz parameter and the PPT measure by plotting $1 - \nu_-^{\rm \sss PT}$ as a function of $G_\tau^{(0|2)}$. 
The difference with Fig.~\ref{fig:bip_ent_violation_finiteT_CS} is striking. For each temperature, $1 - \nu_-^{\rm \sss PT}$ is a monotonous increasing function of $G_\tau^{(0|2)}$. It is not easily seen in the figure, but for finite $T_{\sss\rm BEC}$ the relation between the two quantities is not one to one: for each $G_\tau^{(0|2)}$ there are two (close) values of $1 - \nu_-^{\rm \sss PT}$ which coalesce at the common maximum of the two quantities, marked with a point on Fig.~\ref{fig:bip_ent_violation_finiteT_nu}. This confirms without ambiguity that the PPT measure is still an entanglement monotone at finite temperature. We also note that all the curves in Fig.~\ref{fig:bip_ent_violation_finiteT_nu} almost superimpose, meaning that relation between the two quantities $1 - \nu_-^{\rm \sss PT}$ and $G_\tau^{(0|2)}$ is very weakly dependent on temperature, which makes the PPT measure an even better candidate for quantifying entanglement.

\begin{figure}
    \centering
\includegraphics[width=\linewidth]{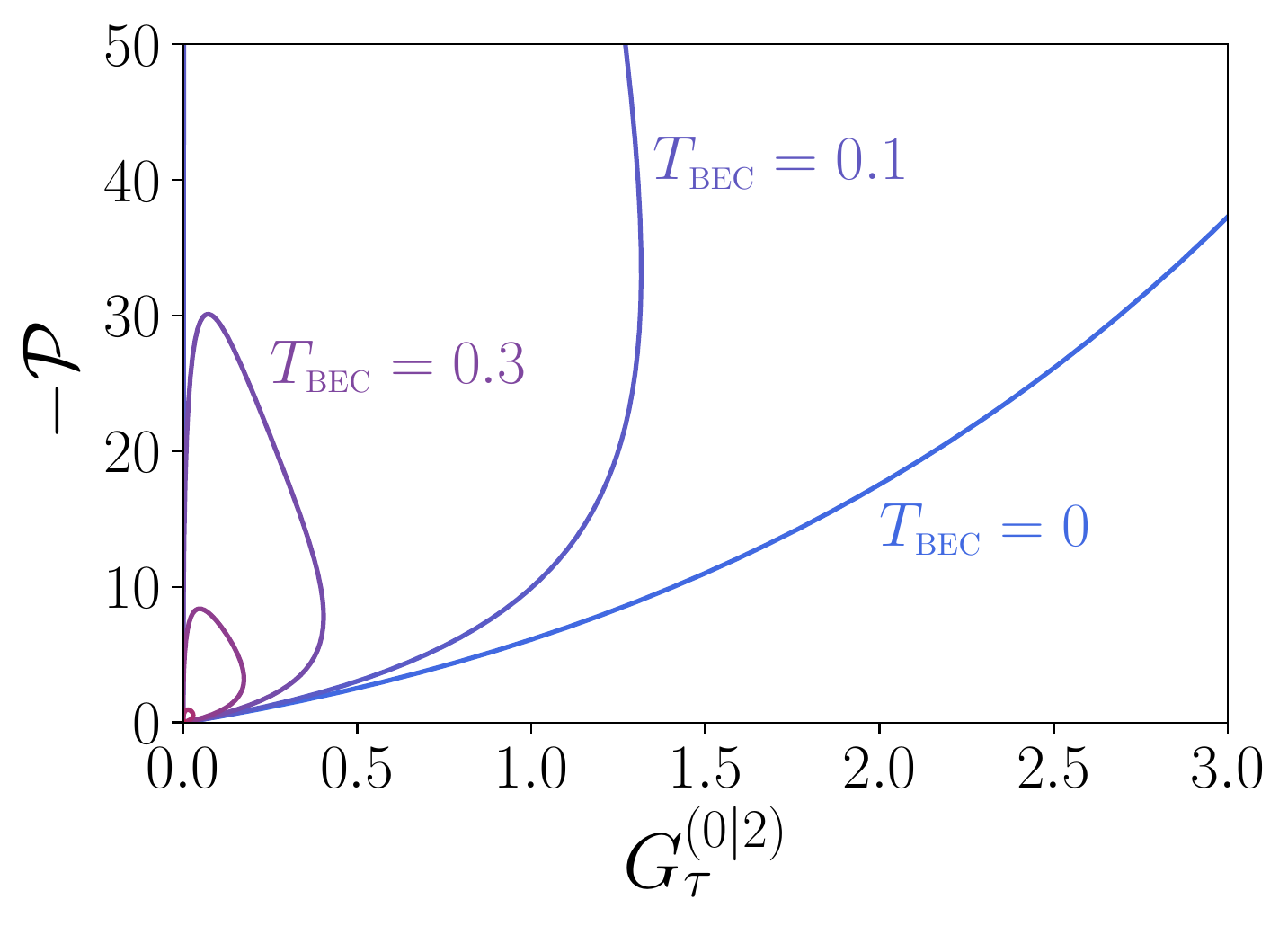}
    \caption{Same as Figs.~\ref{fig:bip_ent_violation_finiteT_CS} and \ref{fig:bip_ent_violation_finiteT_nu} for the GPH parameter $-{\cal P}$ defined in Eqs.~\eqref{GPH1} and \eqref{exprP}.}
    \label{fig:bip_ent_violation_finiteT_P}
\end{figure}

In view of the results presented in Figs.~\ref{fig:bip_ent_violation_finiteT_CS} and \ref{fig:bip_ent_violation_finiteT_nu}, it is of interest to also discuss the generalized Peres-Horodecki (GPH) parameter which has been used in \cite{Finazzi2014,nova2015} for witnessing entanglement in analogue systems. As shown by Simon \cite{simon2000}, non-separability of modes 0 and 2 can be defined as ${\cal P}< 0$, where, using our conventions, the GPH parameter reads
\begin{equation}
\begin{split}\label{GPH1}
    {\cal P} = & \det\sigma_0 \det\sigma_2 + (1 -|\det \varepsilon_{02}|)^2 \\
    & - \tr (\sigma_0 \, {J}\,  \varepsilon_{02} \, {J} \, \sigma_2 \, {J} \, \varepsilon_{02}^{\sss\rm T}\, {J} )
    - \det\sigma_0-\det\sigma_2\; ,
\end{split}
\end{equation}
and the matrix ${J}$ is defined in \eqref{eq:omega-ch3}. This yields
\begin{equation}
    {\cal P} = 
(1-4|\langle \hat{c}_0 \, \hat{c}_2 \rangle_{\sss \rm th}|^2 +a_{0, {\sss \rm th}} a_{2, {\sss \rm th}})^2
    -(a_{0, {\sss \rm th}} + a_{2, {\sss \rm th}})^2\; .
    \label{exprP}
\end{equation}
As is clear from expressions \eqref{nu02}, \eqref{eq:CS_finiteT} and \eqref{exprP}, negativity of ${\cal P}$ is equivalent to the positivity of $1-\nu_-^{\sss\rm PT}$ and to that of $\Delta_{\rm \sss CS}$: these three criteria are equivalent in terms of qualitative assessment of non-separability. This being ascertained, we want to check if $-{\cal P}$ is a good {\it quantitative} measure of entanglement. To this end, we plot it as a function of $G_\tau^{(0|2)}$ in Fig.~\ref{fig:bip_ent_violation_finiteT_P}.
It appears clearly that, as $\Delta_{\rm\sss CS}$, $-{\cal P}$ is not an entanglement monotone at finite temperature. 

We would like to insist on the positive aspects of using the PPT measure in future experimental studies of analogue black hole configurations: (i) as just seen, contrary to the Cauchy-Schwarz and GPH parameters, the PPT measure is a good quantitative measure of entanglement, whatever the temperature of the system is; (ii) from Fig.~\ref{fig:bip_ent_violation_finiteT_nu} it appears that 
$1-\nu_-^{\sss \rm PT}$ is almost as good a measure of entanglement as
$G_\tau^{(0|2)}$, but it
has a much simpler expression in terms of the local mixednesses and mode correlation functions [compare Eq. \eqref{nu02} with Eqs. \eqref{eq:G_tau_finiteT} and \eqref{eq:m_theta}];
(iii) the calculations of \autoref{sec:detec_ent_th} show that the computation of the lowest symplectic eigenvalue requires essentially the knowledge of the the same quantities \eqref{eq:nu_finiteT2} as $\Delta_{\rm \sss CS}$ and
${\cal P}$
[compare Eqs.~\eqref{nu02}, \eqref{eq:CS_finiteT} and \eqref{exprP}]. This means that the value of $\nu_-^{\sss \rm PT}$ is experimentally accessible and can be measured, for instance, from the density correlations along the acoustic black hole, as we now demonstrate. 

An experimental evaluation of the quantities used in the present work for characterising bipartite and tripartite entanglement necessitates 
to experimentally determine the coefficients of the  covariance matrix \eqref{cmG}. For our three mode Gaussian state this matrix is $6\times 6$  and its coefficients are all expressed in terms of correlation functions of the $c$-operators [see, e.g., Eqs.~\eqref{eqsith}, \eqref{eqsith2} and \eqref{eq:epsilon_ij_th}]. Steinhauer \cite{steinhauer2015} has devised a clever method for determining such quantities from the knowledge of the static structure factor and the density-density correlation function, which are both experimentally accessible quantities.
This technique has been used in Refs. \cite{steinhauer_2016,de_nova_2019} and can be in principle extended for evaluating all the relevant averages of $c$-operators. Note however that there are potential practical difficulties:
the method necessitates the computation of windowed Fourier transforms of the real space density-density correlation function and this quantity has to be accurately determined 
over a large spatial range in order to correctly perform all the necessary Fourier transforms. Also the windows used to evaluate these Fourier transforms have to be selected with special care, as discussed in Refs. \cite{nova2015,Fabbri2018,Isoard2020}.

\begin{figure}
    \centering
    \includegraphics[width=\linewidth]{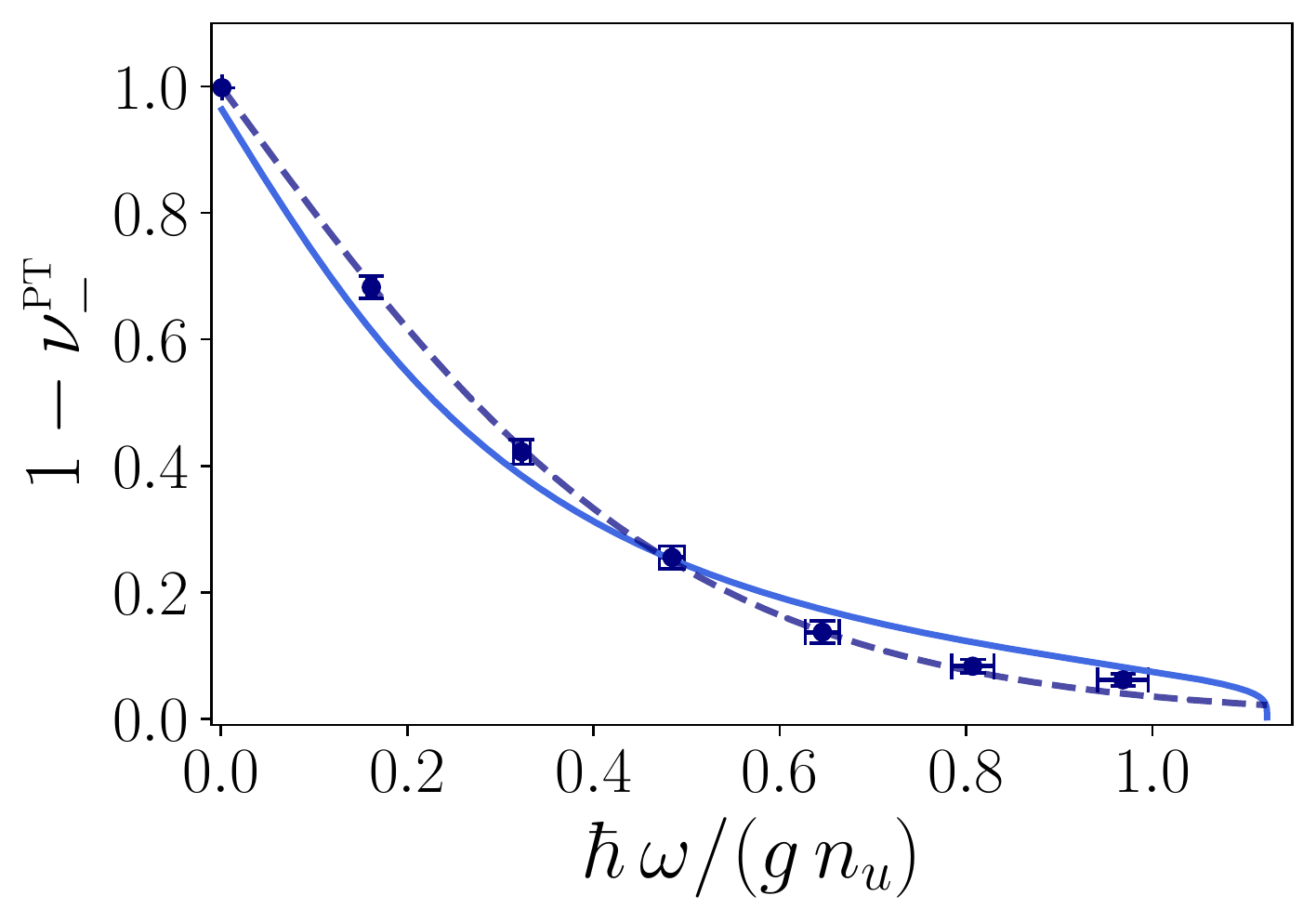}
    \caption{Zero temperature PPT measure as a function of energy in a waterfall configuration with $m_u=0.59$. The blue solid curve is the same as in Fig.~\ref{fig:bip_ent_finiteT}(a). The points with error bars are evaluated from Eq.~\eqref{nuapprox}, extracting the value of $|\langle \hat{c}_0\hat{c}_2\rangle|$ from the experimental data of Ref.~\cite{de_nova_2019} as discussed in the text. The dashed line represents the result of Eq.~\eqref{nuapprox} obtained by assuming that $|\langle \hat{c}_0\hat{c}_2\rangle|\simeq \tfrac12 \sqrt{a_0^2-1}\simeq \sqrt{\bar{n}_0(\bar{n}_0+1)}$, where $\bar{n}_0(\omega)$ 
    is a thermal Bose occupation evaluated at the Hawking temperature $T_{\rm\sss H} = 0.124\, g n_u$ determined in \cite{de_nova_2019}.}
    \label{fig:POC}
\end{figure}

In order to give a proof of concept of the method, we performed the following computation: considering a zero temperature system we neglected the occupation of the Companion mode, which, from Eq.~\eqref{a0a1} yields $a_0\simeq a_2$. This makes it possible, through \eqref{eq:cov_mat_r-parameters2}
and \eqref{eq:cov_mat_a-parameters}, to express the symplectic eigenvalue \eqref{eq:nu-2mode_transpose} as
\begin{equation}\label{nuapprox}
    \nu_-^{\rm\sss PT}\simeq 
    \sqrt{1+4|\langle \hat{c}_0\hat{c}_2\rangle|^2} - 
    2\, |\langle\hat{c}_0\hat{c}_2\rangle| \; .
\end{equation}
The corresponding value of $1-\nu_-^{\rm\sss PT}$ is reported in Fig.~\ref{fig:POC}.
We determined the quantity $|\langle \hat{c}_0\hat{c}_2\rangle|$ appearing in \eqref{nuapprox} by combining the results of the experimental analysis of \cite{de_nova_2019} with the theoretical value of the structure form factor.
A self-contained experimental analysis should resort to the experimentally determined value of this quantity. Also, as discussed in Ref.~\cite{Isoard2020}, (i) neglecting the occupation number of the Companion mode is too crude an approximation, or at least necessitates an independent experimental confirmation and (ii) the windowed Fourier analysis of the experimental density-density correlation function deserves a careful analysis. This is the reason why Fig.~\ref{fig:POC} does not provide an experimental signature of bipartite entanglement in the BEC analogue realised in Ref.~\cite{de_nova_2019}, but is rather a proof of concept, demonstrating that the theoretical techniques employed in the present work provide valuable tools for analyzing experimental data.


\section{Conclusion}\label{ccl}
In the present work, we have investigated entanglement properties of modes emitted from an analogue black hole realized in the flow of a Bose-Einstein condensate. The ground state of the system is seen by an external observer as a three-mode Gaussian state. Gaussian states are entirely characterized by their first and second moments. Thus, their entanglement properties can be expressed in terms of their covariance matrix.
We have characterized bipartite and tripartite entanglement in the system 
using tools developed in the field of continuous-variable entanglement.
We identified the best configuration for the experimental measurement of 
tripartite entanglement: the Gaussian residual entanglement $G^{\rm res}_\tau$ is larger for waterfall configurations with moderate upstream Mach number ($m_u\simeq 0.15$) and at small frequencies. 
An interesting result is the finiteness of bipartite entanglement (for instance, between the Hawking and the Partner) at zero energy, while the tripartite entanglement diverges. This point sheds new light on the importance of the Companion particle, which is sometimes discarded when studying entanglement in analogue black holes.
 We also showed that, quite counter-intuitively, while there is no bipartite entanglement between two of the outgoing modes (tracing out the third one), there is nevertheless genuine tripartite entanglement between the three modes.

Our detailed investigation of the distribution of entanglement in the system in \autoref{sec:entanglement} enabled us to propose a table top optical setup modeling the physical process we study. This, in turn, suggested a new manner to define the analogue Hawking temperature and the associated grey-body factor, in better agreement with the gravitational paradigm.

In \autoref{sec:finiteT}, we studied the effect of temperature on bipartite entanglement and obtained several new results. The Cauchy-Schwarz and the GPH criteria which have been studied in previous studies of analogue systems merely give a qualitative assessment of whether the system is entangled or not. In this paper, we go beyond this qualitative approach by evaluating the amount of entanglement in the Hawking pair using the Gaussian contangle. We assess the capability of several parameters to correctly quantify the amount of entanglement between the Hawking pair by comparing them with our measure of entanglement. Our results should be relevant in future experiments: as a main message, we advise to use the PPT measure $1-\nu_-^{\rm\sss PT}$ instead of the Cauchy-Schwarz or the Generalyzed Peres-Horodecki parameters as a good quantifier of entanglement. We have also observed that the connection between the PPT measure and the contangle is weakly affected by thermal effects, which strengthens even more its relevance in the context of analogue black holes in BEC.
     
Extensions of the present work include the investigation of zero-norm modes, which were shown in \cite{Isoard2020} to play an important role in the correct quantum description of Bogoliubov excitations. The study of the influence of thermal effects on the amount of tripartite entanglement also constitutes a natural continuation of the present study. 
 
During the completion of this work, we became aware of the preprint \cite{nambu2021tripartite} which studies tripartite entanglement in an analogue system thanks to the residual contangle, as done in the present paper. The model in \cite{nambu2021tripartite} corresponds to a "subluminal" dispersion relation, whereas in the BEC case we consider, the dispersion is rather "superluminal". 

\begin{acknowledgments}
We thank M. Jacquet for inspiring discussions on entanglement in analogue gravity.
We also acknowledge fruitful exchanges with A. Aspect, T. Bienaim\'e, A. Fabbri, Q. Glorieux, F. Sols and C. Westbrook. We thank J. Steinhauer for
providing us with experimental data. We acknowledge financial support from the DIM SIRTEQ (Science et Ing\'enierie en R\'egion \^Ile-de-France pour les Technologies Quantiques), project HydroLive.
LPTMS is member of the QUANTUM center of Universit\'e Paris-Saclay.
\end{acknowledgments}

\appendix

\section{Bogoliubov transformations}
\label{appBogol}

In this Appendix we detail some intermediate steps useful for establishing the results presented in Sec.~\ref{sec:bogoliubov_general_case}.

The form \eqref{def_vct_b} and \eqref{def_vct_c} of vectors $\mathbf{b}$ and $\mathbf{c}$  implies that the
$2N\times 2N$ matrix $\mathscr{T}$ defining the unitary Bogoliubov transformation \eqref{eq:BogoTr0} has a block structure given by \eqref{eq:matrix_T}. In order that the $\hat{c}_i$ defined in Eq.~\eqref{eq:BogoTr0} satisfy bosonic commutation relations, the matrix $\mathscr{T}$ must verify
\begin{equation} \label{eq:sympl_cond}
\mathscr{T} \, \widetilde{\mathds{J}} \, \mathscr{T}^{\sss \rm T} = \widetilde{\mathds{J}},
\end{equation}
where $\widetilde{\mathds{J}}$ is defined in Eq.~\eqref{eq:comm_rel}. Eq.~\eqref{eq:sympl_cond} means that $\mathscr{T}$ belongs to the symplectic group Sp$(2 N, \mathbb{C})$. As a consequence, one has
\begin{equation} \label{eq:T_inverse}
\mathscr{T}^{-1} = - \widetilde{\mathds{J}} \, \mathscr{T}^{\rm \sss T} \, \widetilde{\mathds{J}} =
\begin{pmatrix} \alpha^{\rm \sss T} &   \beta^\dagger \\
\beta^{\rm \sss T} & \alpha^\dagger
\end{pmatrix}.
\end{equation}
Condition \eqref{eq:sympl_cond} can be reexpressed in terms of the two $N \times N$ matrices $\alpha$ and $\beta$ as
\begin{equation}
\begin{split}
& \alpha \, \alpha^\dagger - \beta \, \beta^\dagger = \mathds{1}_{\sss N}, \quad \alpha \, \beta^{\rm \sss T} - \beta \, \alpha^{\rm \sss T} = 0 \\
&  \alpha^\dagger \,  \alpha - \beta^{\rm \sss T} \, \beta^* = \mathds{1}_{\sss N}, \quad \alpha^{\rm \sss T} \, \beta^* - \beta^\dagger \, \alpha = 0.
\end{split}
\end{equation}
The matrix $\mathscr{T}$ being symplectic, it can be written as
\begin{equation} \label{eq:matrix_Tprime}
\mathscr{T} =  \exp(\widetilde{\mathds{J}} \, Q),
\end{equation}
with $Q$ a $2N \times 2N$ symmetric matrix. 

The unitary operator $T$ relating operators $\hat{c}_i$ and $\hat{b}_i$ according to \eqref{eq:BogoTr} is defined as
\begin{equation} \label{eq:quadratic_operator}
T = \exp(\frac{1}{2} \mathbf{b}^{\sss \rm T} \, Q \, \mathbf{b}),
\end{equation}
as can be shown by using the Baker-Campbell-Hausdorff formula \cite{blaizot1986}. 
Note that using Eqs.~\eqref{eq:BogoTr0} and \eqref{eq:matrix_Tprime} one has $\mathbf{b}^{\sss \rm T} \, Q \, \mathbf{b} = \mathbf{c}^{\sss \rm T} \, (\mathscr{T}^{-1})^{\sss \rm T} \, Q \, \mathscr{T}^{-1} \,  \mathbf{c}= \mathbf{c}^{\sss \rm T} \, Q \, \mathbf{c}$. This indicates that $T$ has the same expression in term of the $c$'s and in term of the $b$'s.

\begin{widetext}
It is possible to show \cite{balian1969,ma1990, takayanagi2008utilizing} that $T$ can be uniquely decomposed into the product 
\begin{equation}
\label{decompositionT}
 T =  (\det \alpha)^{-1/2}  \exp\left[ \displaystyle  \frac{1}{2} \, \sum_{i,j=1}^{\sss N} X_{ij} \,  \hat{c}_i^\dagger \, \hat{c}_j^\dagger\right] 
 \exp\left[ \displaystyle   \sum_{i,j=1}^{\sss N} Y_{ij} \, \hat{c}_i^\dagger \, \hat{c}_j\right] \, \exp\left[ \displaystyle  \frac{1}{2} \, \sum_{i,j=1}^{\sss N} Z_{ij} \,  \hat{c}_i \, \hat{c}_j\right],
\end{equation} 
where $X$, $Y$, $Z$ are $N \times N$ matrices defined by
\begin{equation}
\label{defXYZ}
X = - \beta^{*} \, \alpha^{-1}, \quad e^{-Y^{\sss \rm T}} = \alpha, \quad Z= \alpha^{-1} \,  \beta.
\end{equation}
The interest of the decomposition \eqref{decompositionT} lies in the fact that all annihilation operators have been put to the right. Therefore, when applied to the vacuum $|0\rangle_c$, $T$ only acts through matrix $X$. This directly yields Eq.~\eqref{eq:link_vacua}.

\section{Explicit expression of the covariance matrix}
\label{technicalpoints}

In this Appendix we give a useful formula for the covariance matrix, present explicit expressions necessary for evaluating its finite-temperature form, and discuss their zero-temperature limit.

The decomposition \eqref{eq:s_a_p} makes it possible to write the covariance matrix $\sigma_c$ of Eq. \eqref{sigma_c} under the form 
\begin{equation} 
\label{eq:cov-U_theta}
\begin{footnotesize}
\hspace{-1cm}\sigma_c=\left(
\begin{array}{cccccc}
 1 + 2 \, v_{02}^2 & 0 & 2 v_{02} v_{12} \cos (\nota_{01}) & -2 v_{02} v_{12} \sin (\nota_{01}) & 2 v_{22} v_{02} \cos (\nota_{02}) & 2 v_{22} v_{02} \sin (\nota_{02}) \\
 0 & 
  1 + 2 \, v_{02}^2
 & 2 v_{02} v_{12} \sin (\nota_{01}) & 2 v_{02} v_{12} \cos (\nota_{01}) & 2 v_{22} v_{02} \sin (\nota_{02}) & -2 v_{22} v_{02} \cos (\nota_{02}) \\
 2 v_{02} v_{12} \cos (\nota_{01}) & 2 v_{02} v_{12} \sin (\nota_{01}) & 
  1 + 2 \, v_{12}^2
 & 0 & 2 v_{22} v_{12} \cos (\nota_{12}) & 2 v_{22} v_{12} \sin (\nota_{12})
   \\
 -2 v_{02} v_{12} \sin (\nota_{01}) & 2 v_{02} v_{12} \cos (\nota_{01}) & 0 & 
  1 + 2 \, v_{12}^2
 & 2 v_{22} v_{12} \sin (\nota_{12}) & -2 v_{22} v_{12} \cos (\nota_{12}) \\
 2 v_{22} v_{02} \cos (\nota_{02}) & 2 v_{22} v_{02} \sin (\nota_{02}) & 2 v_{22} v_{12} \cos
   (\nota_{12}) & 2 v_{22} v_{12} \sin (\nota_{12}) & 
   -1 + 2 \, v_{22}^2
   & 0 \\
 2 v_{22} v_{02} \sin (\nota_{02}) & -2 v_{22} v_{02} \cos (\nota_{02}) & 2 v_{22} v_{12} \sin
   (\nota_{12}) & -2 v_{22} v_{12} \cos (\nota_{12}) & 0 & 
   -1 + 2 \, v_{22}^2
   \\
\end{array}
\right),
\end{footnotesize}
\end{equation} 
where $v_{ij}$ and $\varphi_{ij}$ are defined in Eq.~\eqref{eq:s_a_p} and $\nota_{ij}=\varphi_{i2}-\varphi_{j2}$.

Also, for explicitly computing the finite temperature entanglement properties studied in \autoref{sec:detec_ent_th} 
[see Eqs.~\eqref{eq:nu_finiteT2}  and \eqref{eq:CS_finiteT}]
one uses the formulae:
\begin{equation}\label{finiteTcc}
    \begin{split}
\langle \hat{c}_0 \hat{c}_1^\dagger \rangle_{\rm th}=& S_{00}S_{10}^* (1+\bar{n}_0) + S_{01}S_{11}^* (1+\bar{n}_1) + S_{02}S_{12}^* \bar{n}_2,
        \\
\langle \hat{c}_i^\dagger \hat{c}_i\rangle_{\rm th}=& |S_{i0}|^2 \bar{n}_0 + |S_{i1}|^2 \bar{n}_1 + |S_{i2}|^2 (1+\bar{n}_2), \hspace{1.7cm} i=0,1,
        \\
\langle \hat{c}_i \hat{c}_2\rangle_{\rm th}=& S_{i0}S_{20}^* (1+\bar{n}_0) + S_{i1}S_{21}^* (1+\bar{n}_1) + S_{i2}S_{22}^* \bar{n}_2,  \quad i=0,1,
        \\
\langle \hat{c}_2^\dagger \hat{c}_2\rangle_{\rm th}=& |S_{20}|^2 (1+\bar{n}_0) + |S_{21}|^2 (1+\bar{n}_1) + |S_{22}|^2 \bar{n}_2,
    \end{split}
\end{equation}
where the quantities $\bar{n}_0$ $\bar{n}_1$ and $\bar{n}_2$ are defined in Eq.~\eqref{eq:bar_ni_def}, and, as in Eq.~\eqref{eq:cov-U_theta}, we do not write the explicit $\omega$ dependences for legibility.
\end{widetext}
 At zero temperature the above equations reduce to
\begin{equation}\label{eq:appb3}
   \begin{split}
\langle \hat{c}_0 \hat{c}_1^\dagger \rangle=& S_{00}S_{10}^*  + S_{01}S_{11}^*=S_{12}^* S_{02},
        \\
\langle \hat{c}_i^\dagger \hat{c}_i\rangle=&  |S_{i2}|^2, \hspace{3.3cm} i=0,1,
        \\
\langle \hat{c}_i \hat{c}_2\rangle=& S_{i0}S_{20}^*  + S_{i1}S_{21}^*=S_{i2} S_{22}^* ,  \quad i=0,1,
        \\
\langle \hat{c}_2^\dagger \hat{c}_2\rangle= & |S_{20}|^2 + |S_{21}|^2=-1+|S_{22}|^2,
    \end{split}
\end{equation}
where use has been made of property \eqref{abh6}. Using Eqs.~\eqref{eq:appb3} and
expressions \eqref{eq:local_mixed_omega}, one may show that the finite-temperature components \eqref{eqsith} and \eqref{eq:epsilon_ij_th} of the covariance matrix reduce at $T_{\rm \sss BEC}=0$ to the form \eqref{eq:cov_mat_a-parameters} as they should. However, at finite temperature, Eq.~\eqref{finiteTcc} holds instead of \eqref{eq:appb3}, implying that, contrarily to the zero-temperature case, the covariance matrix, its symplectic eigenvalues, and thus the entanglement properties of the system, do not depend only on the local mixednesses.

\section{Long wavelength limit of the scattering amplitudes}\label{sec:lowwS}

In the long wavelength limit, the $S_{i2}$ coefficients of the $S$-matrix \eqref{eq:bog_transform-ch3} behave as
\begin{equation}\label{eq:Sloww}
    S_{i2}(\omega) = F_{i2} \, \sqrt{\frac{g n_u}{\hbar \omega}} +{\cal O}(\omega^{1/2}),
\quad i\in\{0,1,2\},
\end{equation}
where the $F_{i2}$ are dimensionless constant coefficients. For the waterfall configuration we consider here, analytic expressions of their moduli have been determined in \cite{Larre2012}:
\begin{equation}\label{eq:F02}
|F_{02}|^2=2
\frac{{m}_{u}(1-{m}_{u})^{\frac{3}{2}}
(1+{m}_{u}^{2})^{\frac{3}{2}}}{(1+{m}_{u})^{\frac{1}{2}}
(1+{m}_{u}+{m}_{u}^{2})^{2}},
\end{equation}

\begin{equation}\label{eq:F12}
|F_{12}|^2=\frac{1}{2}
\frac{(1-{m}_{u})^{\frac{7}{2}}
(1+{m}_{u}^{2})^{\frac{3}{2}}}{(1+{m}_{u})^{\frac{1}{2}}
(1+{m}_{u}+{m}_{u}^{2})^{2}},
\end{equation}
and
\begin{equation}\label{eq:F22}
|F_{22}|^2=\frac{1}{2}\frac{(1-{m}_{u}^{4})^{\frac{3}{2}}}
{(1+{m}_{u}+{m}_{u}^{2})^{2}},
\end{equation}
where $m_u$ is the upstream Mach number. 

From the low-frequency behavior of the scattering coefficients it is possible in to evaluate the analogue Hawking temperature of the waterfall configuration, see Eqs.~\eqref{eq:Th0} and \eqref{eq:Th2} in the main text. An alternative way to evaluate the Hawking temperature is to use the semi-classical analogue surface gravity expression \cite{Unruh_1981,visser_acoustic_1998}
\begin{equation}\label{surf_grav}
    T_{\rm \sss H}=\frac{\hbar}{2 \pi} 
    \left( \frac{{\rm d}v}{{\rm d}x}-\frac{{\rm d}c}{{\rm d}x}
    \right)_{x_{\rm\sss H}},
\end{equation}
where $v(x)$ is the velocity of the flow, $c(x)=\sqrt{g n(x)/m}$ is the local sound velocity and $x_{\rm \sss H}$ is the position of the horizon, defined as the point at which 
\begin{equation}\label{horizon_loc}
v(x_{\rm \sss H})=c(x_{\rm \sss H})\; .
\end{equation}
However, as argued in 
Sec.~\ref{sec:background} the definition \eqref{surf_grav} is not expected to apply in the case we consider because, strictly speaking, the local sound velocity is ill-defined for the waterfall profile around $x=0$. A blindfolded use of Eqs.~\eqref{horizon_loc} and \eqref{surf_grav} leads to
\begin{equation}
    \frac{x_{\rm \sss H}}{\xi_u}=-\frac{1}{\sqrt{1-m_u^2}}\arcosh \sqrt{\frac{1-m_u^2}{1-m_u^{2/3}}}\; ,
\end{equation}
and
\begin{equation}\label{eq:Th_SC}
\frac{T_{\rm \sss H}}{g n_u} 
=\frac{3}{2 \pi}\left(1-m_u^{2/3}\right)\sqrt{1-m_u^{4/3}}\; .
\end{equation}
This expression is compared with alternative definitions of the Hawking temperature in Fig.~\ref{fig:Th}. Note that when $m_u$ increases, $x_{\rm \sss H}$ goes deeper in a region of smooth density profile where the concept of local sound velocity becomes relevant: $x_{\rm\sss H}\ll -\xi_u$ when $m_u\to 1$. In this regime expression \eqref{surf_grav} and the corresponding result \eqref{eq:Th_SC} are mathematically sound.

\section{Entanglement localization in a tripartite system}
\label{app:tripartite_eigenvalues}

In this Appendix  we present the specifics of the process of entanglement localization discussed in \autoref{sec:entanglement_localization}.
Let $\sigma$ be a covariance matrix associated with a pure three-mode Gaussian state. We want to determine the explicit form of the symplectic matrix $\cals$ which transforms $\sigma$ according to
\begin{equation}
\label{eq:sympl_transformation_Botero_appB}
\cals \, \sigma  \, \cals^{\rm \sss T}= \mathds{1}_{2} \oplus \sigma_{ \rm \sss sq} , 
\end{equation}
where $\sigma_{ \rm \sss sq}$ is the covariance matrix of a two-mode squeezed state [see Eq.~\eqref{eq:cov-squeezed-BEC}].

\subsection{General form of the symplectic matrix}
Consider a bipartition $ij|k$. The covariance matrix associated with the subsystem $k$ is denoted as $\sigma_{k}$ and the one associated with subsystem $ij$ reads
\begin{equation}
    \sigma_{ij} = 
    \begin{pmatrix}
    \sigma_i & \varepsilon_{ij} \\
    \varepsilon_{ij}^{\rm \sss T} & \sigma_j \\
    \end{pmatrix}.
\end{equation}
The whole covariance matrix associated with the tripartite system is then
\begin{equation}
\label{eq:cov_matrix_appB}
    \sigma = 
    \begin{pmatrix}
    \sigma_i & \varepsilon_{ij} &   \varepsilon_{ik}\\
    \varepsilon_{ij}^{\rm \sss T} & \sigma_j &  \varepsilon_{jk}  \\
    \varepsilon_{ik}^{\rm \sss T} & \varepsilon_{jk}^{\rm \sss T} &  \sigma_k
    \end{pmatrix}.
\end{equation}
Consider the case where the covariance matrix is in its standard form \eqref{eq:cov_mat_a-parameters}, i.e.~$\sigma_k = a_k \, \mathds{1}_2$ and either
 \begin{equation}
\label{eq:cov_matrices_appBa}
    \sigma_{ij} = 
    \begin{pmatrix}
    a_i & 0 & c & 0 \\
    0 & a_i & 0 & - c \\
    c & 0 & a_j & 0 \\
    0 & - c & 0 & a_j 
    \end{pmatrix},\quad c = \sqrt{a_i-1} \, \sqrt{a_j+1}
  \end{equation}
  for bipartitions $ij|k=02|1$  and $12|0$ (for which $\varepsilon_{i2} = c\sigma_z$) or 
\begin{equation}
  \sigma_{ij} = 
    \begin{pmatrix}
    a_i & 0 & c & 0 \\
    0 & a_i & 0 & c \\
    c & 0 & a_j & 0 \\
    0 & c & 0 & a_j
    \end{pmatrix},\quad c = \sqrt{a_i-1} \, \sqrt{a_j-1}
\label{eq:cov_matrices_appBb}
\end{equation}
for the bipartition $01|2$ (for which $\varepsilon_{01}=c\mathds{1}_2$). The difference in the sign in front of $c$ between  \eqref{eq:cov_matrices_appBa} and \eqref{eq:cov_matrices_appBb} is actually of great importance and leads to two different types of symplectic transformations in Eq.~\eqref{eq:sympl_transformation_Botero_appB}. Note that we also impose $a_i<a_j$ in \eqref{eq:cov_matrices_appBa}; in fact, the order of the local mixednesses does not matter in \eqref{eq:cov_matrices_appBb}, as shall be clear at the end of this section.

The symplectic eigenvalues $\sigma_k = a_k \, \mathds{1}$ are $\nu_k=a_k$. Using Williamson theorem, we can bring $\sigma_{ij}$ to a diagonal matrix
\begin{equation}
\label{eq:definition_diag_matrices_appB}
  (\sigma_{ij})^{\prime} =  \cals_{ij} \, \sigma_{ij} \, (\cals_{ij})^{\rm \sss T} = \diag\{\nu_i, \nu_i,\nu_j, \nu_j\},
\end{equation}
where we ordered the symplectic eigenvalues such that $\nu_i < \nu_j$.
Easy calculations lead to 
\begin{equation}
\label{eq:sympl_matrix_jk_appB}
\cals_{ij} = 
\begin{pmatrix}
a & 0 & b & 0 \\
0 & a & 0 & \eta b \\
\eta b & 0 & -a & 0 \\
0 & b & 0 & -a
\end{pmatrix},
\end{equation}
with 
\begin{equation} \label{eq:exp_a_b_appB}
         a =  - \sqrt{\frac{a_j \, \nu_j-a_i \,\nu_i }{\nu_j^2 - \nu_i^2}}, \quad
         b=  \sqrt{\frac{ a_i \, \nu_j - a_j \,\nu_i }{\nu_j^2 - \nu_i^2}}.
\end{equation}
and $\eta=-1$ for bipartitions $ij|k=02|1$ and $12|0$, and $\eta=1$ for bipartition $01|2$. The coefficients $a$ and $b$ satisfy the identity
\begin{equation} \label{eq:exp_a_b_appBnorm}
 a^2 +\eta b^2 = \frac{a_i +\eta a_j }{\nu_i +\eta \nu_j} = 1.
\end{equation}
The last equality is valid only if $\cals_{ij}$ is a symplectic matrix. 

Expressions \eqref{eq:sympl_matrix_jk_appB} and \eqref{eq:exp_a_b_appB} are valid for any covariance matrix $\sigma_{ij}$ of the form \eqref{eq:cov_matrices_appBa}-\eqref{eq:cov_matrices_appBb}. In our case, we can further simplify these expressions using the purity constraint of the three-mode Gaussian state under consideration. Indeed, one can easily prove that for any reduced two-mode states $ij$ of a pure three-mode Gaussian state, $\Delta_{ij} = \det \sigma_{ij} + 1= \det \sigma_{k}+1$ \cite{adesso2006a}. Therefore, considering the reduced state $jk$, Eq.~\eqref{eq:nu-2mode} immediately gives $\nu_i=1$ and $\nu_j = \sqrt{\det \sigma_{ij}}=a_k$, which imply from the last equality of \eqref{eq:exp_a_b_appBnorm} that $a_i +\eta a_j = \nu_i +\eta \nu_j = 1 +\eta a_k$. This expression is true for the case \eqref{eq:cov_matrices_appBa} iff $a_i < a_j$, because $\eta = -1$. For \eqref{eq:cov_matrices_appBb}, the order is not important because $\eta=1$.  Thus,  \eqref{eq:exp_a_b_appB} simplifies to 
\begin{equation} \label{eq:exp_a_b_appBai}
    \begin{split}
         a = & - \sqrt{\frac{(a_j -\eta) \,(a_k +\eta )  }{a_k^2 - 1}}, \\
         & \phantom{(a_j -\eta)(a_j -\eta)} b =   \sqrt{\frac{ (a_i-1) \,(a_k +\eta ) }{a_k^2 - 1}}.
    \end{split}
\end{equation}

\subsection{Standard form}

The symplectic matrix defined by 
\begin{equation}
\label{eq:expression_S_appB}
    \cals =  \cals_{ij} \oplus \cals_k,
\end{equation}
with $\cals_k=a_k\mathds{1}_2$, transforms the covariance matrix \eqref{eq:cov_matrix_appB} to
\begin{equation} \label{eq:transfo_sigma_appB}
 \sigma^{\prime} = \cals \, \sigma \cals^{\rm \sss T} = 
\begin{pmatrix}
  \sigma_{ij}^{\prime} & \rm K \\
  \rm K^{\rm \sss T} & \sigma_k^{\prime}
\end{pmatrix}, 
\end{equation}
with $\rm K$ some matrix and
\begin{equation}
 \sigma_{ij}^{\prime} =
\begin{pmatrix}
  1 & 0 & 0 & 0 \\
 0  & 1 & 0 & 0\\
0 & 0 & a_k & 0 \\
0 & 0 & 0 & a_k \\
 \end{pmatrix}, \qquad  
 \sigma_k^{\prime}=
 \begin{pmatrix}
  a_k & 0 \\
  0 & a_k
  \end{pmatrix}.
\end{equation}
Following \cite{botero2003modewise}, we first notice that 
\begin{equation}
\label{eq:relation_0_appB}
    - (\mathds{J} \, \sigma)^2 = \mathds{1}_6,
\end{equation}
where we recall that $\mathds{J}$ is given by Eq.~\eqref{eq:omega-ch3}. The previous expression is not difficult to prove: the Williamson theorem ensures the existence of a symplectic matrix $\mathcal{O}$ mapping the covariance matrix $\sigma$ to the identity (for a pure state all the symplectic eigenvalues are equal to one); thus, one obtains $- (\mathds{J} \, \sigma)^2  = - \mathds{J} \, \mathcal{O} \, \mathcal{O}^{\rm \sss T}\,  \mathds{J} \,  \mathcal{O} \, \mathcal{O}^{\rm \sss T}= \mathds{1}_6$. Note that one also has $- (\mathds{J} \, \sigma^{\prime})^2 = \mathds{1}_6$.
Then, inserting expression \eqref{eq:transfo_sigma_appB} in Eq.~\eqref{eq:relation_0_appB} and using the fact that $\sigma_{ij}^{\prime}$ and $\sigma_{k}^{\prime}$ are diagonal, it is easy to prove the following conditions for the matrix $\rm K$:
\begin{equation}
\label{eq:conditions_K_appB}
    \begin{cases}
       (\sigma_{ij}^{\prime})^2 - {J}_{ij} \, {\rm K} \, {J}_k \,
       {\rm K^{\rm \sss T}} = \mathds{1}_4, \\
        (\sigma_{k}^{\prime})^2 - {J}_{k} \, {\rm K^{\rm \sss T}} \, 
        {J}_{ij} \, {\rm K} = \mathds{1}_2, \\
        - \sigma_{ij}^{\prime} \, {\rm K} +  {J}_{ij} \, {\rm K} \,  {J}_{k} \, \sigma_k^{\prime} =0,
    \end{cases}
\end{equation}
where ${J}_k$ is defined in Eq.~\eqref{eq:omega-ch3} and ${J}_{ij} = {J}_i \oplus {J}_j$. The last condition in \eqref{eq:conditions_K_appB} implies that a given coefficient ${\rm K}_{mn} \neq 0$ iff $(\sigma_{ij}^{\prime})_{mm} = (\sigma_k^{\prime})_{nn}$, i.e., if and only if the symplectic eigenvalue on the row $m$ of the subsystem $ij$  matches with the one on the column $n$ of the subsystem $k$. Therefore, one can rewrite expression \eqref{eq:transfo_sigma_appB} in the form
\begin{equation}
    \sigma^{\prime}= 
    \begin{pmatrix}
    \mathds{1}_2 & 0 \\
    0 & \widetilde{\sigma}
    \end{pmatrix},
    \quad \text{with} \quad
   \widetilde{\sigma} =
   \begin{pmatrix}
  a_k \, \mathds{1}_2 & \widetilde{\rm K} \\
\widetilde{\rm K}^{\rm \sss T} & a_k \, \mathds{1}_2 \\
 \end{pmatrix},
\end{equation}
where we introduced a new 2$\times$2 matrix $\rm \widetilde{K}$. Then, noticing that $-({J}_{ij} \, \widetilde{\sigma})^2 = \mathds{1}_4$, one obtains
\begin{equation}
\begin{cases}
    \widetilde{\rm K} \, {J}_k \, \widetilde{\rm K}^{\rm \sss T} = (1-a_k^2) \, {J}_k, \\
    {J}_k \, \widetilde{\rm K} \, {J}_k = \widetilde{\rm K}.
\end{cases}
\end{equation}
The above conditions lead to 
\begin{equation}
   \widetilde{\rm K} = 
   \begin{pmatrix}
   a & \sqrt{\lambda^2 - a^2} \\
   -  \sqrt{\lambda^2 - a^2} & -a
   \end{pmatrix},
\end{equation}
where $\lambda = \sqrt{a_k^2 - 1}$. Then, given that $\sigma$ is in its standard form [meaning that $\varepsilon_{ik}$ and $\varepsilon_{jk}$ are diagonal matrices, see Eq.~\eqref{eq:cov_matrix_appB}] and remembering that $\cals_{ij}$ is given by expression \eqref{eq:sympl_matrix_jk_appB}, one can see easily that $\widetilde{K}$ must be diagonal; therefore, $a= \lambda$. As a result, one finds that
\begin{equation}
\widetilde{\sigma} =
   \begin{pmatrix}
  a_k \, \mathds{1}_2 & \sqrt{a_k^2-1} \, \sigma_z \\
\sqrt{a_k^2-1} \, \sigma_z & a_k \, \mathds{1}_2 \\
 \end{pmatrix}, 
\end{equation}
which exactly corresponds to the covariance matrix of a squeezed state with squeezing parameter $r_k >0$, with $\cosh (2 \,r_k) = a_k$ (see Eq.~\eqref{eq:cov-squeezed-BEC}). This statement ends the proof: the symplectic matrix $\cals$ given by expression \eqref{eq:expression_S_appB}, with  $\cals_{ij}$ explicitly written in equations \eqref{eq:sympl_matrix_jk_appB} and \eqref{eq:exp_a_b_appB} and $\cals_{k} = \mathds{1}_2$, indeed lead to the transformation \eqref{eq:sympl_transformation_Botero_appB}.

\subsection{Bipartitions $02|1$ and $12|0$}

For bipartitions $ij|k=02|1$ and $12|0$ of modes $e$, the symplectic transformation \eqref{eq:expression_S_appB} involves the matrix $\cals_{ij}$ (with $j=2$) given by Eq.~\eqref{eq:sympl_matrix_jk_appB} with $\eta=-1$. Using the identity \eqref{eq:exp_a_b_appBnorm}, we introduce a parameter $\gamma$, such that $a = \cosh \gamma$ and $b = \sinh \gamma$, where $a$ and $b$ are the coefficients of the $\cals_{ij}$ matrix. In this case, one finds
\begin{equation}
\label{eq:sympl_matrix_bogo_appB}
\cals_{i2}=\begin{pmatrix}
-\cosh \gamma & 0 & \sinh \gamma & 0 \\
0 & -\cosh \gamma & 0 & -\sinh \gamma \\
 -\sinh \gamma & 0 & \cosh \gamma & 0 \\
0 & \sinh \gamma & 0 & \cosh \gamma
\end{pmatrix},
\end{equation}
with $\cosh \gamma = \cosh r_2/  \cosh r_k $, $\sinh \gamma = \sinh r_i/  \cosh r_k $, computed from expressions \eqref{eq:exp_a_b_appBai}.

To this symplectic transformation $\cals \sigma \cals^{\rm \sss T}$ at the level of the covariance matrix corresponds a Bogoliubov transformation
\begin{equation}
    \mathbf{f} = \mathscr{T}_{\mathbf{e \to f}} \,  \mathbf{e}, 
    \quad \text{with} \quad  
    \mathscr{T}_{\mathbf{e \to f}} = U^\dagger \, \cals \, U,
\end{equation}
(see Eqs.~\eqref{eq:Uxib} and \eqref{eq:transfo_sigma1}). Using the explicit expression of the symplectic matrix \eqref{eq:sympl_matrix_bogo_appB}, one finds
\begin{equation} \label{eq:bogo_transfo_appB}
\begin{cases}
    \hat{f}_{i} =- \cosh \gamma \,\,\,\hat{e}_i + \sinh \gamma \,\, \, \hat{e}_j^{\dagger}, \\
    \hat{f}_{j} = - \sinh \gamma \,\,\, \hat{e}_i^{\dagger}+ \cosh \gamma \, \,\,\hat{e}_j, \\
    \hat{f}_{k} = \hat{e}_k. \\
\end{cases}
\end{equation}
Therefore, entanglement can be localized in the subsystem $f_2|f_k$ with $k=0$ or $1$ only through a Bogoliubov transformation which mixes annihilation and creation operators.

\subsection{Bipartition $01|2$}\label{sec:01|2}

For the bipartition $ij|k=01|2$ the matrix $\cals_{01}$ is given by Eq.~\eqref{eq:sympl_matrix_jk_appB} with $\eta=1$. One finds
\begin{equation}
\cals_{01} = 
\begin{pmatrix}
-\sin \theta & 0 & \cos \theta & 0 \\
0 & -\sin \theta & 0 & \cos \theta \\
 \cos \theta & 0 & \sin \theta & 0 \\
0 & \cos \theta & 0 & \sin \theta
\end{pmatrix},
\end{equation}
with $\cos\theta=\sinh r_{0}/\sinh r_{2}$, $\sin \theta=\sinh r_{1}/\sinh r_{2}$, 
 using again the identity \eqref{eq:exp_a_b_appBnorm} and expressions \eqref{eq:exp_a_b_appBai}. 
The associated Bogoliubov transformation
\begin{equation}
\mathscr{T}_{\mathbf{e \to f}} = U^\dagger \,  \left( \cals_{01} \oplus  \cals_2 \right)\, U,
\end{equation}
leads to the new set of operators
\begin{equation}
\label{bogef012}
\begin{cases}
    \hat{f}_{0} = - \sin \theta  \, \hat{e}_0 + \cos \theta \, \hat{e}_1, \\
    \hat{f}_{1} = \cos \theta \, \hat{e}_0+\sin \theta  \, \hat{e}_1,  \\
    \hat{f}_{2} = \hat{e}_{2},
    \end{cases}
\end{equation}
where, as in the previous subsection, $f_0$ and $f_1$ are new combinations of modes $e_0$ and $e_1$, and $f_2 = e_2$.
Here, there is no mixing of annihilation and creation operators and the matrix $\cals_{01} \oplus  \cals_{2}$ is unitary.

\section{Computation of the finite-temperature Gaussian contangle}
\label{app:app_C}
In this Appendix we explain how to obtain expression \eqref{eq:m_theta}
used in Eq.~\eqref{eq:G_tau_finiteT} for evaluating the Gaussian contangle at finite temperature. We could not find a derivation of this formula
in the literature, and since the explicit form given in \cite{adesso2005} appears to contains some missprints, we find it useful to give the whole proof, following the same path as in \cite{adesso2005}. For a general (mixed or pure) two-mode Gaussian state, a measure of bipartite entanglement is given by the Gaussian contangle $G_\tau(\sigma)$ defined in Eq.~\eqref{eq:G_tau}. It has been proven in \cite{adesso2005} that finding the infinimum over pure Gaussian states amounts to minimize 
\begin{equation}
\label{eq:m1}
m(x_0,x_1,x_3) =1 +  \frac{x_1^2}{\det \Gamma},
\end{equation}
with $\det \Gamma = x_0^2 - x_1^2 - x_3^2$, where $x_0,\, x_1,$ and $x_3$ must belong to the following cones
\begin{equation}
\label{eq:eq_cones_appD}
 \begin{cases}
 x_0 =  \displaystyle 
 \frac{a+b}{2} - \sqrt{   \left(x_1-c_+\right)^2 +  \left(x_3- \frac{a-b}{2}\right)^2 }, \\[2mm]
 x_0 =  \displaystyle \frac{a+b}{2 \, d} + \sqrt{  \left(x_1+\frac{c_-}{ d}\right)^2 +  \left(x_3 + \frac{a-b}{2 \, d}\right)^2 },
\end{cases}
\end{equation}
where $a$, $b$, $c_+$ and $c_-$ are the coefficients of the covariance matrix $\sigma$ written in the standard form and associated with a given two-mode Gaussian state:
\begin{equation} \label{eq:sigma_appD}
    \sigma = 
    \begin{pmatrix}
    a & 0 & c_+ & 0 \\
    0 & a & 0 & c_- \\
    c_+ & 0 & b & 0 \\
    0 & c_- & 0 & b \\
    \end{pmatrix}.
\end{equation}
In Eqs.~\eqref{eq:eq_cones_appD}, $d = a \, b - c_-^2$. The minimum of expression \eqref{eq:m1} is located at the intersection of both cones \eqref{eq:eq_cones_appD} \cite{adesso2005}. Therefore, in the following, we aim at finding this intersection, which corresponds to an ellipse. To find the equation of this ellipse, we first make a change of coordinates (Lorentz boost):
\begin{equation}
\label{eq:change1}
\begin{cases}
x_0^\prime & = \gamma (x_0 - v \, x_3), \\
x_3^\prime & = \gamma (x_3- v \, x_0), \\
x_1^\prime & = x_1,
\end{cases}
\end{equation}
with
\begin{equation}
\begin{cases}
\displaystyle v = \frac{a-b}{a+b} \, \frac{d+1}{d-1} \quad (<1 \quad \text{for} \quad d>1), \\[2mm]
\displaystyle \gamma = \frac{(a+b) \, (d-1)}{  2 \, \sqrt{(a\, d - b )\, (b \, d - a )}  }.
\end{cases}
\end{equation}
We find after simplifications:
\begin{equation}
\begin{cases}
\left(x_0^\prime-\alpha_1 \right)^2  - \left(x_1^\prime- \beta_1\right)^2 -  \left(x_3^\prime-  \gamma_1 \right)^2 = 0, \\
\left(x_0^\prime-\alpha_2 \right)^2  - \left(x_1^\prime- \beta_2\right)^2 -  \left(x_3^\prime-  \gamma_2 \right)^2 = 0,
\end{cases}
\end{equation}
with
\begin{equation}
\begin{split}
& \alpha_1 = \frac{\gamma \, (a-b)}{2} \, \left(  \frac{a+b}{a-b} - v  \right), \quad \beta_1 = c_+, \\
 &\alpha_2 = \frac{\gamma \, (a-b)}{2\, d} \, \left(  \frac{a+b}{a-b} + v \right), \quad \beta_2 = -\frac{c_-}{d},
 \end{split}
\end{equation}
and
\begin{equation}
\gamma_1 = \gamma_2 =  -\frac{\gamma \, (a-b)}{d-1}.
\end{equation}
Note that $\alpha_1$ and $\alpha_2$ simplify to
\begin{equation}
\begin{split}
 \alpha_1 & = \frac{2 \, a \, b\, d  - a^2 - b^2}{2\,  \sqrt{  (a\, d -b ) \, (b \, d - a )  } } , \\
 \alpha_2 & = \frac{d \left(a^2 + b^2 \right)-  2\, a \, b}{  2\, d\,\sqrt{  (a\, d -b ) \, (b \, d - a )  } }.
 \end{split}
\end{equation}
Let us now make another change of variables 
\begin{equation}
\label{eq:change2}
\begin{cases}
x_0^{\prime \prime}& = x_0^\prime - L_+, \\
x_1^{\prime \prime}& = x_1^\prime -H_+, \\
x_3^{\prime \prime}& = x_3^\prime - \gamma_1 = x_3^\prime - \gamma_2,
\end{cases}
\end{equation}
with
\begin{equation}
\label{eq:L+}
\begin{split}
L_+ & = \frac{ \alpha_1 +  \alpha_2}{2} = \frac{a\, b\, (d^2 -1)}{2\,d \,  \sqrt{  (a\, d -b ) \, (b \, d - a )  }}, \\
H_+ & = \frac{ \beta_1 + \beta_2}{2} = \frac{ c_+ \, d - c_-  }{2 \, d}.
\end{split}
\end{equation}
This leads to
\begin{equation}
\label{eq:cones_sym}
\begin{cases}
\left(x_0^{\prime \prime}-L_-\right)^2  - \left(x_1^{\prime \prime}-H_-\right)^2 -  \left. x_3^{\prime \prime}\right.^2 = 0, \\
\left(x_0^{\prime \prime}+L_- \right)^2  - \left(x_1^{\prime \prime}+H_-\right)^2 - \left. x_3^{\prime \prime} \right.^2 = 0,
\end{cases}
\end{equation}
with 
\begin{equation}
\label{eq:L-}
\begin{split}
L_- & = \frac{ \alpha_1 -  \alpha_2}{2} =   \frac{  \sqrt{  (a\, d -b ) \, (b \, d - a )  } }{2 \, d},  \\
H_-  & =  \frac{ \beta_1 -  \beta_2}{2} = \frac{ c_+ \, d + c_-  }{2 \, d}.
\end{split}
\end{equation}
Looking at Eqs.~\eqref{eq:cones_sym}, one sees that the changes of coordinates \eqref{eq:change1} and \eqref{eq:change2} make it possible to eliminate one variable ($x_3^{\prime \prime}$) and to symmetrise the equations. Note that both cone tops belong to the plane $x_3^{\prime \prime} = 0$.

The intersection of the cones \eqref{eq:cones_sym} is now simple to find. By combining equations \eqref{eq:cones_sym}, one can first eliminate $x_3^{\prime \prime}$ to find the relation between $x_0^{\prime \prime}$ and $x_1^{\prime \prime}$:
\begin{equation}
x_0^{\prime \prime} = \frac{H_-}{L_-} \, x_1^{\prime \prime}.
\end{equation}
Inserting this relation in one of the equations \eqref{eq:cones_sym} yields 
\begin{equation}
\left( 1 -  \frac{H_-^2}{L_-^2} \right) \left. x_1^{\prime \prime}\right.^2 + \left. x_3^{\prime \prime}\right.^2  = L_-^2 - H_-^2,
\end{equation}
which exactly corresponds to the equation of an ellipse. Let us define the angle $\theta$ such that
\begin{equation}
\label{eq:rel_angle}
\begin{cases}
x_0^{\prime \prime} = H_- \, \cos \theta, \\
x_1^{\prime \prime} = L_- \,  \cos \theta, \\
x_3^{\prime \prime} =\sqrt{ L_-^2 - H_-^2}  \, \sin \theta.
\end{cases}
\end{equation}
At this stage, we have everything needed to express Eq.~\eqref{eq:m1} only in terms of the parameter $\theta$ and coefficients of the covariance matrix. Since the Lorentz boost preserves the relations between both cones, one can find the minimum of the function $m$ in the basis $(x_0^\prime, \, x_1^\prime, \, x_3^\prime)$, that is to say
\begin{equation}
\label{eq:m}
m =1 +  \frac{\left. x_1^{\prime}\right.^2}{ \left. x_0^{\prime}\right.^2 -  \left. x_1^{\prime}\right.^2 - \left. x_3^{\prime}\right.^2}.
\end{equation}
Using Eqs.~\eqref{eq:change2},  \eqref{eq:L+},  \eqref{eq:L-} and  \eqref{eq:rel_angle}, one finds
\begin{equation}
\begin{cases}
x_0^{\prime} = H_- \, \cos \theta + L_+, \\
x_1^{\prime} = L_- \, \cos \theta + H_+  \\
\phantom{x_1^{\prime}} = \frac{1}{2\, d} \left[ c_+\, d - c_{-} 
+  \sqrt{(a\, d -b ) \, (b \, d - a )  } \, \cos \theta   \right], \\
x_3^{\prime} = \sqrt{ L_-^2 - H_-^2}  \, \sin \theta + \gamma_1.
\end{cases}
\end{equation}
This gives
\begin{equation}
\label{eq:det}
\begin{split}
 \left. x_0^{\prime}\right.^2 -  \left. x_1^{\prime}\right.^2 - \left. x_3^{\prime}\right.^2 & = \alpha_1 \, \alpha_2 - \beta_1 \, \beta_2 - \gamma_1^2 \\
 & - 2 \left(  L_- \, H_+ - H_-\, L_+  \right) \cos \theta  \\
&  - 2\, \gamma_1 \,  \sqrt{ L_-^2 - H_-^2}  \, \sin \theta.
 \end{split}
\end{equation}
After some simplifications, the first right-hand side term of Eq.~\eqref{eq:det} reads
\begin{equation}
\label{eq:term1_appD}
 \alpha_1 \, \alpha_2 - \beta_1 \, \beta_2 - \gamma_1^2 = \frac{a^2 +b^2 + 2 \, c_- \, c_+}{2 \, d}.
\end{equation}
Expanding the coefficient of $-\cos \theta$ in the second right-hand side term of Eq.~\eqref{eq:det} leads to
\begin{equation}
\label{eq:term2_appD}
\begin{split}
&2  \left(  L_- \, H_+ - H_-\, L_+  \right) =  \alpha_1 \, \beta_2 - \beta_1 \, \alpha_2 \\
& = \left\{ 2\, a\, b \, c_-^3 + (a^2 +b^2) \, c_+ \, c_-^2  \right. \\
& \left. + c_- \left[ a^2 (1 - 2\, b^2) +b^2  \right] - a\, b \, c_+ (a^2 +b^2 - 2) \right\} \\
& \times \left[2 \, d \,  \sqrt{  (a\, d -b ) \, (b \, d - a )  }\right]^{-1}.
\end{split}
\end{equation}
The coefficient of $-\sin \theta$ in the last right-hand side term of Eq.~\eqref{eq:det} reads
\begin{equation}
\label{eq:term3_appD}
\begin{split}
 2\, \gamma_1 \,  \sqrt{ L_-^2 - H_-^2} = & -\frac{a^2 - b^2}{2\, d} \\
 & \times \sqrt{1 - \frac{(c_+ \, d + c_-)^2}{(a\, d -b ) \, (b \, d - a ) }}.
 \end{split}
\end{equation}
The last step consists of inserting expressions \eqref{eq:term1_appD}, \eqref{eq:term2_appD} and \eqref{eq:term3_appD} in Eq.~\eqref{eq:det}; then, Eq.~\eqref{eq:det} in expression \eqref{eq:m}. This leads to the final result 
\begin{widetext}
\begin{equation}
\label{eq:m_theta}
\begin{split}
m(\theta) = 1 & + \frac{1}{2\, d} \left[ \sqrt{  (a\, d -b ) \, (b \, d - a )  } \, \cos \theta + c_+\, d - c_- \right]^2 \\
  & \times \left\{  (a^2 +b^2 + 2 \, c_- \, c_+)  -  \cos \theta \,\frac{2\, a\, b \, c_-^3 + (a^2 +b^2) \, c_+ \, c_-^2 + c_- \left[ a^2 (1 - 2\, b^2) +b^2  \right]  - a\, b \, c_+ (a^2 +b^2 - 2)}{\sqrt{  (a\, d -b ) \, (b \, d - a )  } }     \right. \\    
  & + \left.   (a^2 - b^2) \, \sin \theta \, \sqrt{1 - \frac{(c_+ \, d + c_-)^2}{(a\, d -b ) \, (b \, d - a ) }}    \right\}^{-1}.
\end{split}
\end{equation}
\end{widetext}
The explicit expressions of $a$, $b$, $c_+$, $c_-$ and $d = a \, b - c_-^2$ used for evaluating
$G_\tau^{(0|2)}$ in Eq.~\eqref{eq:G_tau_finiteT} are $a = a_{2,{\rm \sss th}}$, $b = a_{0,{\rm \sss th}}$, $c_+ = 2 \, |\langle \hat{c}_0 \, \hat{c}_2 \rangle_{\sss \rm th}|$,  $c_- = - c_+$ and $d = a_{0,{\rm \sss th}} \, a_{2,{\rm \sss th}} - 4 \, |\langle \hat{c}_0 \, \hat{c}_2 \rangle_{\sss \rm th}|^2$. 

Let us finally consider the case of a pure state (i.e., the zero temperature case) and compute the explicit expressions of the Gaussian contangles $G_{\tau}^{(j|2)}$ given by \eqref{gtauij}. We need to evaluate
\begin{equation}
\label{eq:G_appD}
   G_\tau^{(j|2)} = \arsinh^2 \, \left\{\sqrt{\underset{\theta}{\text{min}} [m(\theta)]-1}\right\},
\end{equation}
where $m(\theta)$ is given by Eq.~\eqref{eq:m_theta} and $j=0,1$.
First, by noticing that any reduced two mode state of a pure three mode Gaussian state belongs to the class of GLEMS \cite{adesso2006a}, expression \eqref{eq:m_theta} simplifies to \cite{adesso2005}
\begin{equation}
\label{eq:m_theta_glems_appD}
    m^{\rm \sss GLEMS}(\theta) = 1 + \frac{\left(A \, \cos \theta + B\right)^2}{2 \, d [ (g^2 -1) \, \cos \theta + g^2 + 1  ]},
\end{equation}
where $g = \sqrt{\det \sigma}$, 
with $\sigma$ given by \eqref{eq:sigma_appD}, 
$A = c_+\, d + c_-$ and $B= c_+ \, d - c_-$. 
Using our notations and the explicit expression of the covariance matrix written in the standard form \eqref{eq:cov_mat_a-parameters}, for a given bipartition $j|2$, one has $a = a_2$, $b=a_j$, $c_+ = - c_- = \sqrt{a_j -1} \sqrt{a_2 +1}$, $d = g = a_k$; we recall that $j=0$ or 1 and that the remaining (third) mode (1 or 0) is denoted as $k$.  One proves in this case that the minimum over $\theta$ in expression \eqref{eq:m_theta_glems_appD} is reached when $\theta=\theta^\star$, with \cite{adesso2005}
\begin{equation}
    \cos \theta^\star  = - 1 + \frac{2}{1+ a_k}.
\end{equation}
Inserting this expression in Eq.~\eqref{eq:m_theta_glems_appD} leads to 
\begin{equation}
\label{eq:m_theta_star_appD}
     m^{\rm \sss GLEMS}(\theta^\star) = \left( \frac{-1 + 2\, a_j + a_k}{1 + a_k} \right)^2.
\end{equation}
Using this result in Eq.~\eqref{eq:G_appD} and remembering that $a_j + a_k = a_2 +1$ yields immediately expressions \eqref{eq:G_01_02_12}.

\bibliography{bib_file}

\end{document}